\documentclass[english,11pt,reqno]{article}
\usepackage{natbib}
\usepackage{lmodern}

\usepackage[T1]{fontenc}
\usepackage{babel}
\usepackage{array}
\usepackage{float}
\usepackage{booktabs}
\usepackage{multirow}
\usepackage{todonotes}
\usepackage{changepage}
\usepackage{amsmath}
\usepackage{amsthm}
\usepackage{amssymb}
\usepackage{graphicx}
\usepackage{varwidth}
\usepackage{doi}
\usepackage[]{hyperref}
\hypersetup{
	colorlinks = true,
	linkcolor = blue,
	anchorcolor = blue,
	citecolor = blue,
	filecolor = blue,
	urlcolor = blue
}
\newtheorem{prop}{Proposition}
\newtheorem{lemma}{Lemma}
\newtheorem{assumption}{Assumption}

\makeatletter

\numberwithin{equation}{section}
\numberwithin{figure}{section}
\numberwithin{table}{section}

\usepackage{graphicx}
\usepackage{fancyhdr}
\usepackage{geometry}

\usepackage{fancybox}
\usepackage{lastpage}
\usepackage{longtable}

\usepackage{subcaption}
\usepackage{mathptmx}
\DeclareMathOperator*{\R}{\mathbb{R}}
\DeclareMathOperator{\E}{\mathbb{E}}
\DeclareMathOperator{\F}{\mathcal{F}}
\DeclareMathOperator*{\Q}{\mathbb{Q}}
\DeclareMathOperator{\Var}{\mathrm{Var}}
\DeclareMathOperator{\Cov}{\mathrm{Cov}}
\usepackage{makecell}
\usepackage{placeins}
\usepackage{verbatim}

\numberwithin{equation}{section}

\numberwithin{figure}{section}

\numberwithin{table}{section}

\AtBeginDocument{
  
}

\makeatother

\begin{document}

\title{Finite-Difference Solution Ansatz approach in Least-Squares Monte Carlo}

\author{Jiawei Huo \thanks{\href{mailto:wayne.huo@citi.com}{wayne.huo@citi.com}. We thank Andrei N. Soklakov for useful discussion and review. The views expressed herein should not be considered as investment advice or promotion. They represent personal research of the author and do not necessarily reflect the view of his employers, or their associates or affiliates.}}

\date{June 17, 2025}
\maketitle

\listoffigures
\listoftables

\begin{abstract}
This article presents a simple but effective and efficient approach to improve the accuracy and stability of Least-Squares Monte Carlo. The key idea is to construct the ansatz of conditional expected continuation payoff using the finite-difference solution from one dimension, to be used in linear regression. This approach bridges between solving backward partial differential equations and Monte Carlo simulation, aiming at achieving the best of both worlds. In a general setting encompassing both local and stochastic volatility models, the ansatz is proven to act as a control variate, reducing the mean squared error, thereby leading to a reduction of the final pricing error. We illustrate the technique with realistic examples including Bermudan options, worst of issuer callable notes and expected positive exposure on European options under valuation adjustments.
\end{abstract}


\tableofcontents{}

\section{Introduction}

Solving partial differential equations (PDE) with finite-difference methods and Monte Carlo simulation are two major advanced numerical methods for exotic derivative pricing among practitioners. As the dimensionality of the problem increases, the PDE approach becomes less feasible due to implementation complexity and the numerical burden; specifically, there is no efficient and accurate implementation widely accepted in the industry with dimension greater than two. On the other hand, the Monte Carlo method is widely used in high-dimensional derivative pricing without suffering from the `curse of dimensionality' as PDE with grid approach does.

Derivatives with American-style early exercise features are especially difficult to price within Monte Carlo framework. The major challenge is the determination of the conditional expectation future payoff for an optimal exercise strategy, without further additional numerical procedures which cannot be obtained in a simple forward simulation. With this regard, a backward PDE solver seems outwit Monte Carlo method in low dimensions, as the former effectively stores conditional expectation payoff within the PDE grids when marching backward. 

Among pioneers to tackle this challenge within simulation framework(\cite{Barraquand1995,Broadie2004,tsitsiklis2001regression, Longstaff2001}), the least-squares Monte Carlo approach (\cite{Longstaff2001}), is the most popular algorithm due to its reliability and robustness(\cite{Stentoft2001}). The breakthrough is to allow efficient computation of conditional expectation for future payoff by a projection towards a given set of basis functions. One important detail to note is that the least-squares projection proxy is only used in the optimal stopping time determination, rather than in the value function evaluation itself, another variant of regression-based simulation approach proposed by \cite{tsitsiklis2001regression}. It has been showed by \cite{stentoft2014value} that the former algorithm achieves much less bias in absolute terms by comparison, under either two-period or multiple-period settings. Further theoretical analysis proved the almost sure convergence of the algorithm and the rate of convergence by combining simulation and least-squares method is asymptotically Gaussian normalized error (\cite{clement2002analysis}). 

After this milestone supported by solid theoretical foundation, there has been active research about further improvement on the least-squares method as main stream for American-style option pricing under arbitrary dimensions. For example, conventional variance reduction techniques applicable to European-style option pricing, e.g., antithetic variates, control variates and moment matching methods, were soon proved effective in least-squares Monte Carlo, by reducing the statistical error of finite samples in the payoff expectation estimator \cite{Tian2003}. After these achievements as low-hanging fruits, further focus has been skewed towards an enhanced estimator of conditional expectation until more recently. There have been diverse research activities with specific issues to tackle: \cite{fabozzi2017improved} suggested a weighted regression approach to manage heteroskedasticity in the regression; in order to eliminate the foresight bias introduced from an in-sample method, \cite{Boire2020} and \cite{woo2024leave} proposed different bias-adjusted corrections in the least-squares estimator, without doubling simulations. Worth to mention that the \textit{out-of-sample} implementation, i.e., additional independent simulation paths for regression only, is a much more robust alternative to address these statistical defects. 

Concurrently, there has been growing interest in developing efficient high-dimensional PDE solvers, particularly with the emergence of machine learning techniques, such as deep neural network (DNN) methods. The idea is to cast the high-dimensional PDE solving problem as a learning task, where a DNN is trained to fit the particular setting (\cite{han2018solving}). As applications to optimal stopping problems, effective forward DNN with reduced loss function and accelerated convergence speed to solve backward stochastic differential equations (BSDEs) have been proposed by \cite{Fujii2019} in American basket option pricing. Later, backward deep BSDEs method to better tackle the optimal stopping problem was proposed by \cite{Liang2020} and \cite{gao2023convergence}. However, none of these studies so far were able to provide convincing evidence that DNN methods outperform the least-squares method in terms of accuracy, stability and efficiency. For instance, when pricing Bermudan option with 5-10 underlyings, DNN approach is $\sim$5 time slower than least-squares Monte Carlo while achieving similar accuracy, and it has the edge over simulation methods in terms of memory efficiency only beyond 20 underlyings (\cite{Liang2020}). The advantage of the DNN approach in terms of memory efficiency has become less pronounced, as the continuous improvement of hardware has made larger memory capacities more readily available. While the DNN-based approaches hold promise, the least-squares method remains a more robust and competitive alternative.

In this paper, we propose an enhanced least-squares regression based method by \cite{Longstaff2001}, which aims at improving the projection of conditional expectations by constructing an ansatz from a finite-difference PDE solver, to be used in the linear regression. Previous studies in the literature have explored methodologies that aim to combine PDE-based and simulation-based approaches, to leverage the strengths of both. For instance, \cite{farahany2020mixing} introduced a method by mixing a Fourier transforms based PDE solver on discretized volatility space and regressing the value function onto these resulting conditional payoff profiles, under stochastic volatility model. However, our work is distinguished from their proposal in below aspects:
\begin{itemize}
	\item Our method is suitable under a wider range of model settings, not limited to stochastic volatility only.
	\item The regressed functional proxy used by \cite{farahany2020mixing} enters into value function calculation, in the same spirit of \cite{tsitsiklis2001regression}, rather than \cite{Longstaff2001}.
\end{itemize}

Therefore, our approach is the first attempt at integrating a PDE solver into least-squares Monte Carlo framework, to aid conditional expectation evaluation within optimal stopping time approximation, to the best of our knowledge.

Going beyond American-style derivatives pricing, the regression-based approach is also widely used in the context of Credit Valuation Adjustment (CVA), which has been imposed under the Basel III framework as risk capital charge. For CVA, as well as other related value adjustments such Debt Value Adjustment, Funding Value Adjustment and Capital Value Adjustment, all of which are now categorized as Valuation Adjustments (XVA) in a general term, the major challenge is the determination of uncertain future exposures of a given transaction. Instead of a brute-force approach to re-evaluate the entire portfolio in each simulated scenario, a practically prevalent approach in the industry is to approximate the financial product mark-to-market values using regression functions via the celebrated least-squares approach. As we can see later, the call for stability, on top of accuracy, on the regression method, plays an important role in this application within this regard, and our new method is also readily applicable to satisfy the requirement. 

The rest of this paper is organized as follows. Sec.\ref{sec:Theoretical-Framework} is the description of the framework from a theoretical standpoint, where the underlying model settings are presented in Sec.\ref{sec:ModelSettings}. Then we introduce background knowledge in Sec.\ref{sec:Theoretical-Background} in the context of American-style exercise derivatives pricing and CVA. After setting up the general problem in Sec.\ref{sec:Problem-Formulation}, we outline prevailing numerical methods in practice, including the least-squares Monte Carlo approach (Sec.\ref{sec:LeastSquaresMonteCarloMethod} and its limitation in Sec.\ref{sec:lsmlimitation}), and finite-difference PDE method (Sec.\ref{sec:fdpde}). With such foundation, and further motivated by the analysis of a limiting case in Sec.\ref{sec:FDSolution-LimitingCase}, we proceed to introduce our novel approach in Sec.\ref{sec:FDSolutionAnsatz}, by highlighting its enhancement over the classical approach in Sec.\ref{sec:FDSolution-ControlVariate} and subsequently performing error analysis in Sec.\ref{sec:Error-Analysis}. Then implementation details will be given in Sec.\ref{sec:ImplementationDetails}. The numerical results are delivered in Section \ref{sec:Numerical-Results}, which compares the new method against existing alternatives regarding accuracy, efficiency, and stability: Sec.\ref{sec:Numerical-Results:BermudaOption1d}, \ref{sec:Numerical-Results:BermudaOption2d4d}, and \ref{sec:Numerical-Results:BermudaOptionHeston} examine Bermudan option pricing under the Black-Scholes (up to 4D) and Heston models; Sec.\ref{sec:Numerical-Results:EPE4d} details CVA calculations for European options under 4D; last but not least, Sec.\ref{sec:Numerical-Results:WIC5d} presents examples of worst-of issuer callable notes under a multi-variate local volatility model. Finally, concluding remarks are given in Sec.\ref{sec:Conclusion}. 

\section{Theoretical Framework}\label{sec:Theoretical-Framework}

\subsection{Model Settings}\label{sec:ModelSettings}

For model settings on the underlying diffusion process, we consider multi-variate local volatility and single-asset stochastic volatility. In either case, an equivalent one-dimensional process (often a local volatility model) exists that matches the marginal distributions of the original process, as demonstrated by \cite{gyongy1986mimicking}.

\textbf{Multi-variate Local Volatility}\label{sec:ModelSettings-LocalVolatility}

Under a local volatility setting, we assume that the risk-neutral asset price $\vec{S}(t)=(S_1(t),\ldots,S_d(t))^T$ process follows the stochastic differential equation with the annualized risk-free rate $r$ to be deterministic and constant over time

\begin{equation}\label{eqn:localvoldynamics}
\frac{dS_i(t)}{S_i(t)} = (r-q_i)dt + \sigma_i(S_i(t), t) dB_i(t),
\end{equation}

for $i = 1\ldots d$, where $q_i$ and $\sigma_i(S_i(t), t)$ are the constant continuous dividend yield and local volatility of the $i$-th asset's motion, respectively; and $\{B_i(t)|i = 1\ldots d\}$ is a correlated $d$-dimensional Brownian motion, with $\E[dB_i(t)dB_j(t)] = \rho_{ij}dt := [\delta_{ij}+\rho(1-\delta_{ij})]dt$ using Kronecker delta notation $\delta_{ij}$ and $\rho$ denoting single correlation between assets. Since we only consider continuous (rather than discrete) dividends, we assume $\vec{S}(0)=(1,\ldots,1)^T$ for simplicity without loss of generality. In the limit of $\sigma_i(S_i(t), t)$ being constants, this setting degenerates to the celebrated Black-Scholes model.

\textbf{Stochastic Volatility: Heston Model}\label{sec:ModelSettings-Heston}

For stochastic volatility setting, let's consider below stochastic differential equation under the Heston model 
\begin{equation}\label{eqn:stochvoldynamics}
\frac{dS(t)}{S(t)} = (r-q)dt +\sqrt{\nu(t)} dB^S(t) \quad \text{and}\quad d\nu(t) =\kappa(\theta-\nu(t))dt + \xi\sqrt{\nu(t)}dB^\nu(t),
\end{equation}
with $\E[dB^S(t)dB^\nu(t)] = \rho^{S,\nu}dt$, where $q$ is the constant continuous dividend yield, $\theta$ is the long variance, $\kappa$ is the mean reverting rate, $\xi$ is the volatility of volatility and $\rho^{S,\nu}$ is the correlation between spot and variance. For the same reason as previous subsection, we assume $S(0)=1$ for simplicity. Given the initial variance $\nu(0)$, there is a well known result on the expectation value of $\nu(t)$ over time: $\E[\nu(t)]=\nu(0)e^{-\kappa t} +\theta(1-e^{-\kappa t})$.

\subsection{Theoretical Background}\label{sec:Theoretical-Background}

\subsubsection{Problem Formulation}\label{sec:Problem-Formulation}

As a general framework relevant for derivative valuation as well as regulatory CVA computations, we assume that throughout a given terminal expiry $T$ with the riskless interest rate $r$ defined previously, there exists a complete probability space ($\Omega,\F,(\F_t)_{0\leq t\leq T},\Q$), accommodating the model settings in Sec.\ref{sec:ModelSettings}, under risk-neutral measure $\Q$. Also, let $L^2(\Omega,\F,(\F_t)_{0\leq t\leq T},\Q)$, hereinafter simplified as $L^2$, denote the space of square integrable functions. As problem formulation, from $t=0$ and final expiry $T$ of an derivative contract, we define a set of discretized time points as $\pi(t):=\{t_k\ge t|t_k=k\Delta T, 1\leq k\leq M=\frac{T}{\Delta T}\}$ with a time interval $\Delta T$. $\pi(t)$ represents either the exercise times for American-style option, or discrete exposure monitoring times in CVA calculation. Within the complete probability space $\Omega$, there exists a discrete filtration $\{\F_{t_k}|1\leq k\leq M\}$ on $\pi(0)$. For notational clarity, we use $XYZ(\omega,t)$ to denote a stochastic variable under $\F_t$ with $\omega\in\Omega$. In this spirit, the state variables are denoted by $X(\omega,t_k)$, which is $\F_{t_k}$-measurable. 

To define the problem with maximum generality, we also have to define two important quantities: $Z(\omega,t_k)\equiv Z(X(\omega,t_k))$ as the exercise payoff at $t_k$, and $\tilde{V}(\omega,\tilde{\tau}(t_k))$ as the sum of discounted (back to $t_k$) future cash flows from $t_k$ and up to a stopping time $\tilde{\tau}(t_k)\in\pi(t_k)$. Different from $X(\omega,t_k)$ or $Z(\omega,t_k)$, $\tilde{V}(\omega,\tilde{\tau}(t_k))$ is only measurable with respect to the information in $\tilde{\tau}(t_k)$. For single cash flow derivatives, these two quantities are related to each other as $\tilde{V}(\omega,\tilde{\tau}(t_k))=e^{-r[\tilde{\tau}(t_k)-t_k]}Z(X(\omega,\tilde{\tau}(t_k)))$. Similar to previous literature, we limit the attention to square integrable payoff functions, i.e., $\tilde{V}(\omega,\cdot)\in L^2$.

With all the notations defined in placed, we can formally write down the final solution, representing either the option present value (PV) in the context of American-style option pricing or CVA in valuation adjustments, respectively, as

\begin{equation}\label{eqn:finalResultsForm}
PV(CVA)=\E[\Phi^{PV(CVA)}(\{Z(\omega, t_k),\tilde{V}(\omega,\tilde{\tau}(t_k)), \E[Y^{PV(CVA)}(\omega,t)|\F_{t_k}]|t_k\in\pi(0)\})|\F_{0}].
\end{equation}

The difficulty in prohibiting standard European-style Monte Carlo evaluation on $\Phi^{PV(CVA)}$ lies in its dependence on $\E[Y^{PV(CVA)}(\omega,t)|\F_{t_k}]$ where $t>t_k$.

\subsubsection{Least-Squares Monte Carlo}\label{sec:LeastSquaresMonteCarloMethod}
As can be seen in Eqn.\ref{eqn:finalResultsForm}, once the $\F_{t_k}$-measurable conditional expectations can be efficiently crystallized into a functional form depending on $X(\omega,t_k)$, the PV or CVA can be determined in a path-by-path manner independently and thus estimated by its corresponding Monte Carlo estimators. The core idea of regression based Least-Squares approach (LSM) is to approximate the parameterized form of $\E[Y(\omega,t)|\F_{t_k}]$ (omitting the superscript of $Y$) as a linear combination of a collection of $R$ basis functions $\vec{\phi}^P_R(x)=\{\phi^P_r(x)|r=1,\ldots,R\}^T$ in $\R^R$ space, where the coefficients are obtained via regression using the cross-sectional information computed in the simulation. 
For notational convenience, let's define the least-squares estimator of the conditional expectation, using the $L^2$ projection operator $P_k$ at $t_k$ under $\vec{\phi}^P_R(x)$, as
\begin{equation}\label{eqn:l2projection}
P_k[Y(\omega,t)] := \vec{\beta}^P_k \cdot \vec{\phi}^P_R(X(\omega,t_k)),
\end{equation}
where

\begin{equation}\label{eqn:objFunction}
\begin{split}
\vec{\beta}^P_k = \arg_{\vec{b}\in\mathbb{R}^R} \min\mathcal{L}(Y(\omega,t);\vec{b};\vec{\phi}^P_R(X(\omega,t_k))),\\
\mathcal{L}(Y(\omega,t);\vec{b};\vec{\phi}^P_R(X(\omega,t_k))):=\E\left[(Y(\omega,t) -\vec{b}\cdot\vec{\phi}^P_R(X(\omega,t_k)))^2\right|\F_{0}].
\end{split}
\end{equation}
$P_k$ is a projector onto a linear span of $L^2$, defined by a given basis functions $\vec{\phi}^P_R(X(\omega,t_k))$ as
\begin{equation}\label{eqn:spanDef}
\text{span}(\vec{\phi}^P_R(X(\omega,t_k))):=\{\vec{b}\cdot\vec{\phi}^P_R(X(\omega,t_k))|\vec{b}\in{\R}^R\}.
\end{equation}

In Eqn.\ref{eqn:objFunction}, $\mathcal{L}(Y(\omega,t);\vec{b};\vec{\phi}^P_R(X(\omega,t_k)))$ is the objective function to be minimized, which is the residual sum of squares (RSS). Its lower bound is given by, in the limit as $\vec{b}\cdot\vec{\phi}^P_R(X(\omega,t_k))\to\E[Y(\omega,t)|X(\omega,t_k)]$,
\begin{equation}\label{eqn:objFunctionInf}
\inf_{\vec{b}\in\mathbb{R}^R}\mathcal{L}(Y(\omega,t);\vec{b};\vec{\phi}^P_R(X(\omega,t_k)))=\E[\Var(Y(\omega,t)|\F_{t_k})|\F_{0}],
\end{equation}
where $\Var(\cdot|\F_{t_k})$ denotes the conditional variance.

The formal solution of Eqn.\ref{eqn:objFunction} as a function of $Y(\omega,t)$ is given by
\begin{equation}\label{eqn:objFunctionLSMBeta}
\vec{\beta}^P_k(Y(\omega,t)):=(\E[\vec{\phi}^P_R(X(\omega,t_k))\vec{\phi}^P_R(X(\omega,t_k))^T|\F_{0}])^{-1}\E[\vec{\phi}^P_R(X(\omega,t_k))Y(\omega,t)|\F_{0}].
\end{equation}

In practice, the Monte Carlo estimator of $\vec{\beta}^P_k$ can be obtained via regression as described in Eqn.\ref{eqn:solvebeta} of Appendix~\ref{sec:regressionleastsquare}.

The choice of stochastic variable $Y$ to be regressed on will be illustrated in a case-by-case basis below. 

\textbf{Option Pricing with American-style exercise}

Here we assume that the American-style option pricing is exercisable at $\pi(0)$ either on the option holder side or on the issuer side.

In the spirit of Eqn.\ref{eqn:finalResultsForm}, we have
\begin{equation}\label{eqn:optionpv}
\Phi^{PV} := 
\begin{cases}
\sup_{\tilde{\tau}(0)\in\pi(0)}\tilde{V}(\omega,\tilde{\tau}(0))\text{,  for holder exercise;} \\
\inf_{\tilde{\tau}(0)\in\pi(0)}\tilde{V}(\omega,\tilde{\tau}(0))\text{,  for issuer exercise.}
\end{cases}
\end{equation}

In this article, we study Bermudan options on constant-weighted basket $S_B(t)=\sum_{i=1}^d w_i S_i(t)$ with $w_i=\frac{1}{d}$ (see Appendix~\ref{sec:BermudaOption}), and worst of issuer callable notes (WIC) on worst-of basket $S_W(t) = \min_{i\in\{1,\ldots,d\}}S_i(t)$ (see Appendix~\ref{sec:WorstOfIssuerCallableNote}), respectively.

Under discrete-time formulation, a preferable way to solve the Eqn.\ref{eqn:optionpv} is to find out the optimal stopping time strategy, denoted by $\tau(t_k):=\tau(\omega, t_k)$ (removing the tilde compared to $\tilde{\tau}(t_k)$ for optimality and omitting the dependence on $\omega$ for notational convenience). For further notational clarity, we also introduce $V(\omega,\tau(t_{k+1}))$ to represent the option path value under optimal stopping time $\tau(t_k)$, differentiated from $\tilde{V}(\omega,\tilde{\tau}(t_{k+1}))$ by removing the tilde similarly to emphasize that it is calculated under an optimal stopping policy.

With the help of $V(\omega,\tau(t_{k+1}))$, we further define the so-called continuation value representing the conditional expected option value (discounted back to $t_k$), for ease of discussion, as $F(\omega,t_k):=\E[e^{-r\Delta T}V(\omega,\tau(t_{k+1}))|X(\omega,t_k)]$. 

Thanks to the $\F_{t_k}$-measurability of $F(\omega,t_k)$ on an equal footing with $Z(\omega,t_k)$, the formal solution of the optimal stopping times can be written, via a dynamic programming fashion (together with terminal condition $\tau(t_M)=T$), as 
\begin{equation}\label{eqn:tau_k}
\tau(t_k)=t_k\cdot I(\omega,t_k)+\tau(t_{k+1})\cdot\left[1-I(\omega,t_k)\right],
\end{equation}
where the exercise indicator function is defined as $I(\omega,t_k):=\Theta\left(\pm[Z(\omega,t_k)-F(\omega,t_k)]\right)$, with Heaviside step function denoted by $\Theta(x)$ and `$+$'(`$-$') for holder (issuer) exercise. The formal dynamic programming solution of $V(\omega,\tau(t_{k}))$ is also given by
\begin{equation}\label{eqn:v_k}
V(\omega,\tau(t_k)) = Z(\omega,t_k)\cdot I(\omega,t_k) + e^{-r\Delta T}V(\omega,\tau(t_{k+1}))\cdot[1-I(\omega,t_k)].
\end{equation}

Given the solution of Eqn.\ref{eqn:v_k}, we arrive at the desired result  

\begin{equation}\label{eqn:phiOptionpv}
\Phi^{PV}=V(\omega,\tau(0)),
\end{equation}
which will be useful in the theoretical analysis of pricing error by translating a global optimization problem in Eqn.\ref{eqn:optionpv} into a dynamic programming problem.

To proceed under LSM following Eqn.\ref{eqn:l2projection}, we introduce the regressed continuation value 
\[
F_k^P(X(\omega,t_k)):=P_k[e^{-r\Delta T}V(\omega,\tau(t_{k+1}))]
\]
as a proxy of $F(\omega,t_k)$.

In practice, we run two Monte Carlo simulations, where in the first regression stage $F_k^P(X(\omega,t_k))$ is determined, and then PV is evaluated in the subsequent pricing stage. It is worth mentioning that in the regression stage, when solving $\vec{\beta}^P_k$ iteratively using Eqn.\ref{eqn:v_k} and Eqn.\ref{eqn:l2projection}, \ref{eqn:objFunction}, with $e^{-r\Delta T}V(\omega,\tau(t_{k+1}))$ substituted for $Y$ ($Y^{PV}$) as the regression dependent variable, the least-squares estimator only enters into the calculation inside the exercise indicator $I(\omega,t_k)$, whereas the continued value function on the right-hand side of Eqn.\ref{eqn:v_k} is still calculated as pathwise cash flows from the simulation. \cite{stentoft2014value} showed that the approach by \cite{tsitsiklis2001regression}, which further replaces the value function outside the exercise indicator \textit{outright} with its estimator, results in a much higher absolute bias.

In the pricing stage after regression is done, one can then compute PV as

\begin{equation}\label{eqn:pvMC}
\widehat{PV} =\frac{1}{N_p}\sum_{i=1}^{N_p}V^{(i)}(\{X^{(i)}(t)|t\leq\tau^{(i)}(0)\},\tau^{(i)}(0)),
\end{equation}
where the superscript $(i)$ indicates that the variable is under $i^{th}$ of the total $N_P$ pricing paths; and the optimal stopping rule is to exercise the option only at

\begin{equation}\label{eqn:tauBermuda}
\tau^{(i)}(0)  = \inf\{t_k\in\pi(0)|I^{(i)}(\omega,t_k)>0\}.
\end{equation}

To strike a balance between convergence and calculation efficiency, one usually draws a different set of Monte Carlo paths in this pricing stage more than the previous regression stage, i.e., $N_P>N_R$. This is so-called \textit{out-of-sample} implementation to eliminate foresight bias without the need of sophisticated regression correctors as mentioned before. Also, escalating $N_P$ only is enough to effectively reduce Monte Carlo variance in the discounted pathwise cash flows calculation for complex derivative structures (e.g., WIC).

\textbf{Credit Valuation Adjustment and Expected Positive Exposure}

Now we turn to the example of CVA computation, as the expected loss resulting from the potential future default of the counterparty. For the ease of illustration without loss of generality, we assume that the product doesn't have early termination features, e.g., an European Option on $S_B(T)$. By choosing $Y^{CVA}(\omega,T):=\tilde{V}(\omega,\tilde{\tau}(t_k)=T)=e^{-r(T-t_k)}Z(S_B(T))$\footnote{One can consider the European option as a limiting case of Bermudan option by forcing all $\tilde{\tau}(t_k)=T$.} whose conditional expectation $\tilde{F}(\omega,t_k):=\E[Y^{CVA}(\omega,T)|{\F}_{t_k}]$ is the mark-to-market value of the option as of $t_k$, the CVA on this financial transaction can be derived, in the form of Eqn.\ref{eqn:finalResultsForm}, as:

\begin{equation}\label{eqn:cva}
\begin{split}
CVA & = (1-R_{ctpy})\E\left[\int_{0}^{T}e^{-r\cdot t}\cdot\max\left(0, \tilde{F}(\omega,t)\right) dPD(t) | {\F}_{0}\right]  \\
&= (1-R_{ctpy}) \E\left[\int_{0}^{T}e^{-r\cdot t}\cdot\max\left(0, \tilde{F}(\omega,t)\right) h_{cpty}(\omega,t) e^{-\int_{0}^{t}h_{cpty}(\omega,s)ds} dt| {\F}_0\right] \\
&\approx (1-R_{ctpy})\sum_{k=1}^{M}\E[\Phi^{CVA}_{t_k}|{\F}_0],
\end{split}
\end{equation}
with
\begin{equation}\label{eqn:phiCva}
\Phi^{CVA}_{t_k}:=e^{-r\cdot t_k}\cdot \max\left(0, \tilde{F}(\omega,t_k) \right) \cdot h_{cpty}(\omega,t_k)\cdot\Delta T \cdot e^{-\sum_{j=1}^{k}h_{cpty}(\omega,t_j)\Delta T},
\end{equation}
where $R_{ctpy}$ is the recovery rate and $PD(t)$ denotes the default probability of the counterparty at $t$, which depends on the counterparty hazard rate denoted by $h_{cpty}(\omega,t)\equiv h_{cpty}(\tilde{F}(\omega,t),t)$.

We define the Expected Positive Exposure (EPE), as well as its discounted version EPE*, as:
\begin{equation}\label{eqn:epe}
\begin{split}
EPE(t_k)     &:= \E\left[\max\left( 0, \tilde{F}(\omega,t_k)\right)| {\F}_{0}\right],\\
EPE^{*}(t_k) &:= e^{-r\cdot t_k}EPE(t_k).
\end{split}
\end{equation}

Introducing $\tilde{F}_k^{P}(X(\omega,t_k)):=P_k[e^{-r(T-t_k)}Z(S_B(T))]$ as the estimator of $\tilde{F}(\omega,t_k)$, the EPE and CVA can be calculated as:

\begin{equation}\label{eqn:epeMC}
\widehat{EPE}(t) = \frac{1}{N_p}\sum_{i=1}^{N_p} \max(0, \tilde{F}_k^{P}(X^{(i)}(t_k))),
\end{equation}
and
\begin{equation}\label{eqn:cvaMC}
\begin{split}
\widehat{CVA} = \frac{1-R_{ctpy}}{N_p}\sum_{i=1}^{N_p}\sum_{k=1}^{M}\Phi^{CVA(i)}_{t_k},
\end{split}
\end{equation}
with 
\[
\Phi^{CVA(i)}_{t_k} := e^{-r\cdot t_k}\cdot \max(0, \tilde{F}_k^{P}(X^{(i)}(t_k))) \cdot h_{cpty}(\tilde{F}_k^{P}(X^{(i)}(t_k)),t_k)\cdot\Delta T \cdot e^{-\sum_{j=1}^{k}h_{cpty}(\tilde{F}_j^{P}(X^{(i)}(t_j)),t_j)\Delta T},
\]
respectively. Here the superscript $(i)$ indicates that the variable is under $i^{th}$ of the total $N_P$ pricing paths, similar to PV problem in Eqn.\ref{eqn:pvMC}.

It is worth mentioning that in Eqn.\ref{eqn:epeMC} and \ref{eqn:cvaMC}, the value function are completely approximated with its regressed estimator. Admittedly, this approach is more akin to the procedure advocated by \cite{tsitsiklis2001regression}, which could result in biases that are difficult to control without using pathwise cash flows calculation in pricing stage; however, it is still the standard market practice for the indispensable sake of efficient trade aggregation at the netting set level. \footnote{Note that we are aware of alternative hybrid approach to calculate CVA/EPE, by making use of both the simulated cash flows and regressed function (demonstrated by \cite{Joshi2016} without considering Wrong Way Risk). The drawback of this proposal is that one has to generate all the cash flows at given individual product level, without being decoupled from the path generation at the upper portfolio level. It would result in an aggregation of contractual timelines, rather than simple algorithmic operation upon regressed functions at individual product level when it comes to portfolio aggregation.}

\subsubsection{Basis function and Limitations}\label{sec:lsmlimitation}

The basis functions $\{\phi^P_r(x)|r=1,\ldots,R\}$, also known as regressors, are the main source of approximation within LSM. Based on the original study, the LSM is remarkably robust to the choice of basis functions (\cite{Longstaff2001}). However, in a further analysis, it is concluded that for complex options, the robustness does not seem to be guaranteed. Nevertheless, there is no golden rule provided by the author as an optimal choice for any given product, and different choices of popular regressors can merely affect option prices slightly (\cite{Moreno2003}).

In practice, it is not worth using special functions involving exponential calculation, such as Laguerre, or Hermite polynomials, as they introduce unnecessary computational burden without actual benefit. Thus, simple monomials are widely used in LSM as regressors, i.e., $\phi^P_r(x)=x^{r-1}$ in the case of single state variable\footnote{It can be easily generalized to the case of multiple state variables, e.g., in the case of stochastic volatility, by specifying the total monomial degree across variables.}. There are several limitations as follows:

\begin{itemize}
	\item With increasing the cutoff $R$ for the sake of regression accuracy, it becomes more susceptible to overfitting.
	\item The fitting accuracy is poor away from at-the-money (ATM), as those regions are effectively out of scope in the objective function to be minimized. 
	\item As a solution of backward recursive optimization in the context of American-style option pricing, $F_k^P(X(\omega,t_k))$ would depend on the solution of the counterpart at $t_{k+1}$, which was obtained by another regression. That said, the numerical error propagates and accumulates from iteration to iteration. 
\end{itemize}

\subsubsection{Finite-difference approach in PDE} \label{sec:fdpde}

As opposed to LSM, alternative numerical approach to solve the American-style option pricing is the finite-difference (FD) method. The elegance of this method lies in the fact that it marches from maturity to present, which naturally fits into the requirement of backward recursive optimization in stopping time problems. We restrict the discussion within 1D for maximum efficiency. The local volatility setting as describe in Sec.\ref{sec:ModelSettings-LocalVolatility} with $d=1$ is solvable under 1D FD PDE, under which the option price $V^{FD}(S,t)$ as function of underlying spot $S$ (omitting the subscript $i$ for the mono underlying case) and time $t$ satisfies:
\begin{equation}\label{eqn:pde1d}
\frac{\partial V^{FD}(S,t)}{\partial t} = - \frac{1}{2}\sigma_{FD}^2(S,t) S^2\frac{\partial^2V^{FD}(S,t)}{\partial S^2} - [r-q_{FD}(S,t)]S\frac{\partial V^{FD}(S,t)}{\partial S} + rV^{FD}(S,t),
\end{equation}
within the time domain $(t_k,t_{k+1}]$, and subjected to the initial condition $V^{FD}(S,T)=Z(S,T)$ and transition conditions around $t_k$ due to obstacle problem as
\[
V^{FD}(S,t^-_k) = \max(Z(S,t_k),V^{FD}(S,t^+_k))
\]
for holder's exercise; and
\[
V^{FD}(S,t^-_k) = \min(Z(S,t_k),V^{FD}(S,t^+_k))
\]
for issuer's exercise, respectively. As we will see later, using the approach of \cite{gyongy1986mimicking}, for a more general higher-dimensional setting, the marginal distribution of original process after dimensional reduction is mimicked by a 1D PDE in Eqn.\ref{eqn:pde1d} with the coefficients $q_{FD}(S,t)$ and $\sigma_{FD}(S,t)$ (re-labeled from $q_1(S,t)$ and $\sigma_1(S,t)$ in Eqn.\ref{eqn:localvoldynamics}), corresponding to the effective single-asset continuous dividend yield and local volatility, respectively. The details of such dimensional reduction will  specified case by case, e.g., for multi-variate local volatility and single-asset stochastic volatility, respectively, in Sec.\ref{sec:ImplementationDetails}.

After solving the 1D backward propagating PDE via Crank-Nicolson method, from now on, we will denote by, $f^{FD}_k(S)$ the FD solution of continuation value at $t_k$, which is related to backward PDE solution as
\begin{equation}\label{eqn:FDContinuedValue}
f^{FD}_k(S)=V^{FD}(S,t^{+}_k).
\end{equation}

\subsection{FD Solution Ansatz based Least-Squares Monte Carlo} \label{sec:FDSolutionAnsatz}

\subsubsection{Limiting Case}\label{sec:FDSolution-LimitingCase}

Before proceeding, we introduce below lemma as a limiting case:
\begin{lemma}\label{lemma:UniqueExactRegressor}
	In a generalized LSM formulism, the continuation value estimator is exact if the exact continuation value is used as one of the regressors.
\end{lemma}

The proof of Lemma~\ref{lemma:UniqueExactRegressor} is omitted as it is an elementary consequence of conditional expectation as $L^2$ projection that minimizes the objective function of LSM. This conclusion is free from model setting assumption. 

For a general problem of interest, one could find an auxiliary variable with existing exact (or quasi-exact) solution as an ansatz to start with as one of the basis functions in the $L^2$ projection, with the expectation of dominating regression weight on this ansatz in the spirit of Lemma~\ref{lemma:UniqueExactRegressor}. Inspired by the simplicity of FD solution for 1D PDE in Sec.\ref{sec:fdpde} with efficiency and quasi-exact accuracy, we would intuitively take the option price under 1D process as auxiliary variable, whose FD solution of continuation value $f_k^{\text{FD}}(x)$ is given by Eqn.\ref{eqn:FDContinuedValue}. From now on, we call this new method by incorporating $f_k^{\text{FD}}(x)$ into one of the LSM basis functions as \textit{FD Solution Ansatz based Least-Squares Monte Carlo} (FD-LSM). 

Before further elaborating on FD-LSM, we introduce simplified notation for random variables to enhance clarity and conciseness from now on. Specifically, we suppress the spatial dependence and condense the notation by replacing arguments depending on time by subscripts, and then further replace $t_k$ with simply $k$ whenever applicable. For example, $X_{k}\equiv X(\omega,t_{k})$, $\vec{\phi}^{P}_{R,k}\equiv\vec{\phi}^{P}_R(X_k)$, $\vec{\phi}^{P^\text{FD}}_{k}\equiv\vec{\phi}^{P^\text{FD}}(X_k)$ (to be defined in next section), $V_{k+1}\equiv V(\omega,\tau(t_{k+1}))$, $Z_k\equiv Z(\omega,t_k)$, $F_k\equiv F(\omega,t_k)$ and $F^P_k\equiv F^P(\omega,t_k)$.

\subsubsection{FD Solution: Control Variate}\label{sec:FDSolution-ControlVariate}

Mathematically, FD-LSM is defined by a modified least-squares continuation estimator with a new $L^2$ projection operator $P_k^{\text{FD-LSM}}$, as opposed to the classic one $P_k^{\text{LSM}}$(previously denoted by $P$ defined in Eqn.\ref{eqn:l2projection}) operating on $\vec{\phi}_R^{P^\text{LSM}}(x)$(previously denoted by $\vec{\phi}^P_R(x)$). The new basis functions of $P_k^{\text{FD-LSM}}$ is given by $\vec{\phi}_R^{P^\text{FD-LSM}}(x) := \vec{\phi}^{P^\text{FD}}(x) \cup \vec{\phi}_{R-1}^{P^\text{LSM}}(x)$ ($R$ is decremented by one to cater for the added new FD ansatz), where $\vec{\phi}^{P^\text{FD}}(x):=\{1, f_k^{\text{FD}}(x)\}^T$\footnote{We allow a constant term in the regressors to cater for mismatch of center of mass between independent and dependent variables.} under which the associated projection operator $P_k^{\text{FD}}$ is also introduced for illustrative purpose later. 

To formally define $P_k^{\text{FD}}$, let's denote by $\hat{V}_{k+1}^{1D}$ the FD solvable auxiliary variable with the interpretation as the path option value under 1D process satisfying $e^{-r\Delta T}\E[\hat{V}_{k+1}^{1D}|X=x]=f_k^{\text{FD}}(x)$. It is the equivalent counterpart of $V_{k+1}$ but under a hypothetical 1D simulation process instead.

At $t_k$, by projecting a given stochastic variable $Y_{t\geq t_k}$ onto $\text{span}(\vec{\phi}^{P^\text{FD}}_k)$, defined similarly as Eqn.\ref{eqn:spanDef}, we have
\begin{equation}\label{eqn:fdProjection}
P_k^{\text{FD}}[Y_t] := \beta_k^{\text{FD}}(Y_t)f_k^{\text{FD}}(X_k)+\alpha_k^{\text{FD}}(Y_t),
\end{equation}
where from simple linear regression, the optimal coefficients as functions of $Y_t$ are given by 
\begin{equation}\label{eqn:optimalFDbeta}
\begin{split}
\beta_k^{\text{FD}}(Y_t)& :=\frac{\E[\Cov(e^{r\Delta T}Y_t,\hat{V}_{k+1}^{1D}|\F_{t_k})|\F_{0}]}{\E[\Var(\hat{V}_{k+1}^{1D}|\F_{t_k})|\F_{0}]},\\
\alpha_k^{\text{FD}}(Y_t) &:= \E(Y_t|\F_{0}) - \beta_k^{\text{FD}}(Y_t)e^{-r\Delta T}\E(\hat{V}_{k+1}^{1D}|\F_{0}),
\end{split}
\end{equation}
with $\Var(\cdot|\F_{t_k})$ and $\Cov(\cdot,\cdot\cdot|\F_{t_k})$ denoting the conditional variance and covariance operators, respectively. 
From Eqn.\ref{eqn:optimalFDbeta}, we can also see that
\begin{equation}\label{eqn:PFDVEqualToPFDF}
P_k^{\text{FD}}[F_k]=\beta_k^{\text{FD}}(F_k)f_k^{\text{FD}}(X_k)+\alpha_k^{\text{FD}}(F_k)=P_k^{\text{FD}}[e^{-r\Delta T}V_{k+1}]	
\end{equation}
because $\beta_k^{\text{FD}}(F_k)=\beta_k^{\text{FD}}(e^{-r\Delta T}V_{k+1})$ and $\alpha_k^{\text{FD}}(F_k)=\alpha_k^{\text{FD}}(e^{-r\Delta T}V_{k+1})$ given the identity $F_k=\E(e^{-r\Delta T}V_{k+1}|\F_{t_k})$.

We also define the annihilator under $\vec{\phi}^{P^\text{FD}}(x)$ as $M_k^{\text{FD}}:=1-P_k^{\text{FD}}$. Both of $P_k^{\text{FD}}$ and $M_k^{\text{FD}}$ are idempotent, i.e., $P_k^{\text{FD}}[P_k^{\text{FD}}[\cdot]]=P_k^{\text{FD}}[\cdot]$ and $M_k^{\text{FD}}[M_k^{\text{FD}}[\cdot]]=M_k^{\text{FD}}[\cdot]$.

With above definitions in place, it is ready to state and prove below lemma.

\begin{lemma}\label{lemma:pDecomposition}
$P_k^{\text{FD-LSM}}$ on the option price can be decomposed exactly into $P_k^{\text{FD},\text{FD}\perp\text{LSM}}$ and $P_k^{\text{LSM},\text{LSM}\perp\text{FD}}$: $P_k^{\text{LSM},\text{LSM}\perp\text{FD}}$ employs the same basis functions as $P_k^{\text{LSM}}$, but the regression coefficients are obtained as the ``net'' effect of $\vec{\phi}_{R-1}^{P^\text{LSM}}(x)$ on $V_{k+1}$ with the effect of $\vec{\phi}^{P^\text{FD}}(x)$ annihilated from the remaining system; $P_k^{\text{FD},\text{FD}\perp\text{LSM}}$ is defined analogously, but with the roles of $\vec{\phi}^{P^\text{FD}}(x)$ and $\vec{\phi}_{R-1}^{P^\text{LSM}}(x)$ interchanged.
\end{lemma}

\renewcommand{\qedsymbol}{$\blacksquare$}
\begin{proof}

Here, we decompose an $L^2$ projection using two sets of basis functions, $\vec{\phi}^{P^\text{FD}}(x)$ and $\vec{\phi}_{R-1}^{P^\text{LSM}}(x)$, effectively implementing a Frisch-Waugh-Lovell (FWL) type decomposition into $P_k^{\text{FD},\text{FD}\perp\text{LSM}}$ and $P_k^{\text{LSM},\text{LSM}\perp\text{FD}}$. We will focus on a detailed examination of $P_k^{\text{LSM},\text{LSM}\perp\text{FD}}$, noting that $P_k^{\text{FD},\text{FD}\perp\text{LSM}}$ can be analyzed symmetrically by swapping the roles of the FD and LSM basis functions.

Based on the FWL theorem\footnote{This theorem is well-recognized in the field of Econometrics, as a powerful tool for understanding and simplifying regression analysis. Its proof is provided in Section 2.4 of \cite{davidson2004econometric}, or more recently in \cite{lovell2008simple}.}, the mathematical definition of  $P_k^{\text{LSM},\text{LSM}\perp\text{FD}}$ is
\begin{equation}\label{eqn:l2perpProjection}
P_k^{\text{LSM},\text{LSM}\perp\text{FD}}[e^{-r\Delta T}V_{k+1}] := \vec{\beta}^{P^{\text{LSM}\perp\text{FD}}}_k \cdot \vec{\phi}^{P^\text{LSM}}_{R-1,k}
\end{equation}

with the coefficients obtained by 
\begin{equation}\label{eqn:objFunctionWeightedProjection}
\begin{split}
\vec{\beta}^{P^{\text{LSM}\perp\text{FD}}}_k &= \arg_{\vec{b}\in\mathbb{R}^{R-1}} \min\mathcal{L}(e^{-r\Delta T}V_{k+1}^{\perp\text{FD}};\vec{b};\vec{\phi}^{P^{\text{LSM}\perp\text{FD}}}_{R-1,k}).
\end{split}
\end{equation}
Here $\mathcal{L}$ adopts the same functional form as Eqn.\ref{eqn:objFunction} but operates on below arguments by partialling out the effect of $\vec{\phi}^{P^\text{FD}}(X_k)$
\begin{equation*}
\begin{split}
e^{-r\Delta T}V_{k+1}^{\perp\text{FD}}:&= M_k^{\text{FD}}[e^{-r\Delta T}V_{k+1}],\\
\vec{\phi}^{P^{\text{LSM}\perp\text{FD}}}_{R-1,k}:&=\{M_k^{\text{FD}}[\phi^{P^\text{LSM}}_{r,k}]|r\in[1,R-1]\}^T=M_k^{\text{FD}}[\vec{\phi}^{P^\text{LSM}}_{R-1,k}].
\end{split}
\end{equation*}

Similar to Eqn.\ref{eqn:objFunctionLSMBeta}, the formal solution of $\vec{\beta}^{P^{\text{LSM}\perp\text{FD}}}_k$ as a function of $e^{-r\Delta T}V_{k+1}$ is 
\begin{equation}\label{eqn:objFunctionLSMperpFDBeta}
\begin{split}
\vec{\beta}^{P^{\text{LSM}\perp\text{FD}}}_k(e^{-r\Delta T}V_{k+1}):=(\E[\vec{\phi}^{P^{\text{LSM}\perp\text{FD}}}_{R-1,k}\vec{\phi}^{P^{\text{LSM}\perp\text{FD}}T}_{R-1,k}|\F_{0}])^{-1}\E[\vec{\phi}^{P^{\text{LSM}\perp\text{FD}}}_{R-1,k}e^{-r\Delta T}V_{k+1}^{\perp\text{FD}}|\F_{0}].
\end{split}
\end{equation}

Given the symmetrical nature of $\vec{\phi}^{P^\text{FD}}(x)$ and $\vec{\phi}_{R-1}^{P^\text{LSM}}(x)$ with respect to the applicability of the FWL theorem, $P_k^{\text{FD},\text{FD}\perp\text{LSM}}[e^{-r\Delta T}V_{k+1}]$ is defined as a modified $P_k^{\text{FD}}[e^{-r\Delta T}V_{k+1}]$, controlling for $\vec{\phi}_{R-1}^{P^\text{LSM}}(X_k)$:

\begin{equation}\label{eqn:fdPertLSMProjection}
\begin{split}
P_k^{\text{FD},\text{FD}\perp\text{LSM}}[e^{-r\Delta T}V_{k+1}]& := \beta_k^{\text{FD}\perp\text{LSM}}(e^{-r\Delta T}V_{k+1})f_k^{\text{FD}}(X_k)+\alpha_k^{\text{FD}\perp\text{LSM}}(e^{-r\Delta T}V_{k+1}),\\
\beta_k^{\text{FD}\perp\text{LSM}}(e^{-r\Delta T}V_{k+1})& :=\frac{\E[\Cov(V_{k+1}^{\perp\text{LSM}},\hat{V}_{k+1}^{1D\perp\text{LSM}}|\F_{t_k})|\F_{0}]}{\E[\Var(\hat{V}_{k+1}^{1D\perp\text{LSM}}|\F_{t_k})|\F_{0}]},\\
\alpha_k^{\text{FD}\perp\text{LSM}}(e^{-r\Delta T}V_{k+1}) &:= e^{-r\Delta T}(\E(V_{k+1}^{\perp\text{LSM}}|\F_{0}) - \beta_k^{\text{FD}\perp\text{LSM}}(e^{-r\Delta T}V_{k+1})\E(\hat{V}_{k+1}^{1D\perp\text{LSM}}|\F_{0})),
\end{split}
\end{equation}

where $e^{-r\Delta T}Y_{k+1}^{\perp\text{LSM}}$, with $Y_{k+1}$ being either $V_{k+1}$ or $\hat{V}_{k+1}^{1D}$, is defined as:
\begin{equation*}
\begin{split}
e^{-r\Delta T}Y_{k+1}^{\perp\text{LSM}}:&=e^{-r\Delta T}Y_{k+1}-\arg_{\vec{b}\in\mathbb{R}^{R-1}} \min\mathcal{L}(e^{-r\Delta T}Y_{k+1};\vec{b};\vec{\phi}_{R-1}^{P^\text{LSM}}(X_k)) \cdot \vec{\phi}_{R-1}^{P^\text{LSM}}(X_k)\\
&=e^{-r\Delta T}Y_{k+1}-(\E[\vec{\phi}^{P^{\text{LSM}}}_{R-1,k}\vec{\phi}^{P^{\text{LSM}}T}_{R-1,k}|\F_{0}])^{-1}\E[\vec{\phi}^{P^{\text{LSM}}}_{R-1,k}e^{-r\Delta T}Y_{k+1}|\F_{0}]\cdot \vec{\phi}_{R-1}^{P^\text{LSM}}(X_k).
\end{split}
\end{equation*}

Combining Eqn.\ref{eqn:l2perpProjection} and \ref{eqn:fdPertLSMProjection}, we manage to arrive at
\begin{equation}\label{eqn:FDLSM_decomposition}
\begin{split}
P_k^{\text{FD-LSM}}[e^{-r\Delta T}V_{k+1}]&=P_k^{\text{FD},\text{FD}\perp\text{LSM}}[e^{-r\Delta T}V_{k+1}]+P_k^{\text{LSM},\text{LSM}\perp\text{FD}}[e^{-r\Delta T}V_{k+1}] \\
&=P_k^{\text{FD},\text{FD}\perp\text{LSM}}[e^{-r\Delta T}V_{k+1}]+P_k^{\text{LSM},\text{LSM}\perp\text{FD}}[e^{-r\Delta T}V_{k+1}^{\perp\text{FD}}],
\end{split}
\end{equation}
where in the last line we have applied $P_k^{\text{LSM},\text{LSM}\perp\text{FD}}[e^{-r\Delta T}V_{k+1}]=P_k^{\text{LSM},\text{LSM}\perp\text{FD}}[e^{-r\Delta T}V^{\perp\text{FD}}_{k+1}]$, induced by the idempotent property of $M_k^{\text{FD}}$, to facilitate further discussion.

\end{proof}

With the help of above lemma, below proposition can be established for the illustration of fundamental improvement of $P_k^{\text{FD-LSM}}$ over $P_k^{\text{LSM}}$ from a theoretical point of view.

\begin{prop}\label{prop:FD_LSM_CV_LSM}
	FD-LSM is equivalent to LSM with a modified projection ``kernel'' using the FD-solvable option value as a control variate. 
\end{prop}

\renewcommand{\qedsymbol}{$\blacksquare$}
\begin{proof}

For a given time step $t_k$, we would like to derive the relation between $P_k^{\text{FD-LSM}}[V_{k+1}]$ and $P_k^{\text{LSM}}[V_{k+1}]$ in the example of PV problem. Note that the final conclusion is equally applicable to the CVA problem. 

First of all, based on Lemma~\ref{lemma:pDecomposition}, let's apply FD-LSM in Eqn.\ref{eqn:FDLSM_decomposition} to the path option value under 1D process, which gives,
\begin{equation}
\begin{split}
	P_k^{\text{FD-LSM}}[e^{-r\Delta T}\hat{V}_{k+1}^{1D}]&=P_k^{\text{FD},\text{FD}\perp\text{LSM}}[e^{-r\Delta T}\hat{V}_{k+1}^{1D}]+P_k^{\text{LSM},\text{LSM}\perp\text{FD}}[e^{-r\Delta T}\hat{V}_{k+1}^{1D}-P_k^{\text{FD}}[e^{-r\Delta T}\hat{V}_{k+1}^{1D}]]\\
	&=f_k^{\text{FD}}(X_k) + P_k^{\text{LSM},\text{LSM}\perp\text{FD}}[e^{-r\Delta T}\hat{V}_{k+1}^{1D}-f_k^{\text{FD}}(X_k)],
\end{split}
\end{equation}
where we have utilized the identities $\beta_k^{\text{FD}\perp\text{LSM}}(e^{-r\Delta T}\hat{V}_{k+1}^{1D})=\beta_k^{\text{FD}}(e^{-r\Delta T}\hat{V}_{k+1}^{1D})=1$ and $\alpha_k^{\text{FD}\perp\text{LSM}}(e^{-r\Delta T}\hat{V}_{k+1}^{1D})=\alpha_k^{\text{FD}}(e^{-r\Delta T}\hat{V}_{k+1}^{1D})=0$.

Based on Lemma~\ref{lemma:UniqueExactRegressor}, the left-hand side in above is equal to $f_k^{\text{FD}}(X_k)$, which implies
\begin{equation}\label{eqn:v1dEqualsTofx}
P_k^{\text{LSM},\text{LSM}\perp\text{FD}}[e^{-r\Delta T}\hat{V}_{k+1}^{1D}-f_k^{\text{FD}}(X_k)]=0.
\end{equation}

Next, applying the projection $P_k^{\text{FD-LSM}}$ to the original option price $e^{-r\Delta T}V_{k+1}$ gives 

\begin{equation}\label{eqn:P_decomposition}
\begin{split}
P_k^{\text{FD-LSM}}[e^{-r\Delta T}V_{k+1}] &= P_k^{\text{FD},\text{FD}\perp\text{LSM}}[e^{-r\Delta T}V_{k+1}] + P_k^{\text{LSM},\text{LSM}\perp\text{FD}}[e^{-r\Delta T}V_{k+1} - P_k^{\text{FD}}[e^{-r\Delta T}V_{k+1}]] \\
&= \beta_k^{\text{FD}\perp\text{LSM}}(e^{-r\Delta T}V_{k+1})f_k^{\text{FD}}(X_k) +\alpha_k^{\text{FD}\perp\text{LSM}}(e^{-r\Delta T}V_{k+1}) \\
&\quad+ P_k^{\text{LSM},\text{LSM}\perp\text{FD}}[e^{-r\Delta T}V_{k+1} - \beta_k^{\text{FD}}(e^{-r\Delta T}V_{k+1})f_k^{\text{FD}}(X_k)-\alpha_k^{\text{FD}}(e^{-r\Delta T}V_{k+1})]\\
&= \beta_k^{\text{FD}\perp\text{LSM}}(e^{-r\Delta T}V_{k+1})f_k^{\text{FD}}(X_k)+ P_k^{\text{LSM},\text{LSM}\perp\text{FD}}[e^{-r\Delta T}V_{k+1} - \beta_k^{\text{FD}}(F_k)f_k^{\text{FD}}(X_k)],
\end{split}
\end{equation}
where in the last line, we have employed the identity $\beta_k^{\text{FD}}(e^{-r\Delta T}V_{k+1})=\beta_k^{\text{FD}}(F_k)$ from Eqn.\ref{eqn:PFDVEqualToPFDF}, and the constant term $\alpha_k^{\text{FD}\perp\text{LSM}}(\cdot)-\alpha_k^{\text{FD}}(\cdot)$ has been absorbed into $P_k^{\text{LSM},\text{LSM}\perp\text{FD}}$.

Given Eqn.\ref{eqn:v1dEqualsTofx}, we have $P_k^{\text{LSM},\text{LSM}\perp\text{FD}}[e^{-r\Delta T}\hat{V}_{k+1}^{1D}]=P_k^{\text{LSM},\text{LSM}\perp\text{FD}}[f_k^{\text{FD}}(X_k)]$. Thus we arrive at the desired result as a consequence of linearity property of $L^2$ projection:
\begin{equation}\label{eqn:projectionLSMandFDLSD}
P_k^{\text{FD-LSM}}[e^{-r\Delta T}V_{k+1}] =\beta_k^{\text{FD}\perp\text{LSM}}(e^{-r\Delta T}V_{k+1})f_k^{\text{FD}}(X_k) + P_k^{\text{LSM},\text{LSM}\perp\text{FD}}[e^{-r\Delta T}(V_{k+1} - \beta_k^{\text{FD}}(F_k)\hat{V}_{k+1}^{1D})].
\end{equation}
The following Sec.\ref{sec:Error-Analysis} will demonstrate that the primary source of regression error in this decomposition arises from the second term, equivalent to a customized projection on a control-variated ``kernel'' with the help of the FD ansatz.
\end{proof}

Based on Prop.\ref{prop:FD_LSM_CV_LSM}, proof of convergence on LSM from previous studies (\cite{Longstaff2001, Stentoft2001}) is equally applicable to FD-LSM. Both $P_k^{\text{FD-LSM}}$ and $P_k^{\text{LSM}}$ are unbiased estimators. However, the variance of projection ``kernel'' in the former is reduced tremendously in a control variated manner by observing (with the help of Eqn.\ref{eqn:optimalFDbeta})
\begin{equation}\label{eqn:optimizedVar}
\E[\Var(V_{k+1} - \beta_k^{\text{FD}}(F_k)\hat{V}_{k+1}^{1D}|\F_{t_k})|\F_{0}] = [1-\tilde{\rho}_k^2(V_{k+1},\hat{V}_{k+1}^{1D})]\cdot\E[\Var(V_{k+1}|\F_{t_k})|\F_{0}]
\end{equation}
with

\begin{equation}\label{eqn:correlationVV}
\tilde{\rho}_k(V_{k+1},\hat{V}_{k+1}^{1D}):=\frac{\E[\Cov(V_{k+1},\hat{V}_{k+1}^{1D}|\F_{t_k})|\F_{0}]}{\sqrt{\E[\Var(V_{k+1}|\F_{t_k})|\F_{0}]\cdot\E[\Var(\hat{V}_{k+1}^{1D}|\F_{t_k})|\F_{0}]}}
\end{equation}

defining the correlation between $V_{k+1}$ and $\hat{V}_{k+1}^{1D}$ at time step $t_k$.

Comparing the objective function in Eqn.\ref{eqn:objFunctionWeightedProjection} with Eqn.\ref{eqn:objFunctionInf} (re-labeling $P$ to $P^{\text{LSM}}$), and applying the aforementioned identity, one can see that the scaling factor that reduces the variance in the projection $P_k^{\text{LSM},\text{LSM}\perp\text{FD}}$ also proportionally reduces the RSS (lower bound):
\begin{equation*}
\begin{split}
\inf_{\vec{b}\in\mathbb{R}^{R-1}}\mathcal{L}(e^{-r\Delta T}(V_{k+1} - \beta_k^{\text{FD}}(F_k)\hat{V}_{k+1}^{1D});\vec{b};\vec{\phi}^{P^{\text{LSM}\perp\text{FD}}}_{R-1,k})&=\E[\Var(e^{-r\Delta T}V_{k+1} - e^{-r\Delta T}\beta_k^{\text{FD}}(F_k)\hat{V}_{k+1}^{1D}|\F_{t_k})|\F_{0}] \\
&=[1-\tilde{\rho}_k^2(V_{k+1},\hat{V}_{k+1}^{1D})]\cdot\E[\Var(e^{-r\Delta T}V_{k+1}|\F_{t_k})|\F_{0}]\\
&=[1-\tilde{\rho}_k^2(V_{k+1},\hat{V}_{k+1}^{1D})]\cdot\inf_{\vec{b}\in\mathbb{R}^R}\mathcal{L}(e^{-r\Delta T}V_{k+1};\vec{b};\vec{\phi}^{P^{\text{LSM}}}_{R,k}).
\end{split}
\end{equation*}
It means the stronger the correlation between $V_{k+1}$ and $\hat{V}_{k+1}^{1D}$, the more effective the variance reduction achieved in the new scheme; as always, such reduction directly translates into the decrease in $N_R$ required to fulfill suitable precision.

There are several further remarks on this result:
\begin{itemize}
	\item It is well known that control variates are very effective Monte Carlo variance reduction techniques, see e.g. Chapter 4 in \cite{glasserman2004}. In the context of American exercise option pricing, up til now, such toolkit has only been applied in the Monte Carlo pricing stage to reduce the final PV variance from literature (see \cite{sondergaard2002efficient,Tian2003}). On the contrary, FD-LSM is designed as equivalently applying the variance reduction technique during the regression stage, which is crucial for optimal stopping strategy determination subsequently in pricing stage. Such improvement cannot be achieved under the usual payoff control variates method.
	\item As mentioned before, $\hat{V}_{k+1}^{1D}$ can be seen as option value under a 1D simulation process. However, instead of employing the right-hand side in Eqn.\ref{eqn:projectionLSMandFDLSD} directly for variance reduction, Prop.\ref{prop:FD_LSM_CV_LSM} allows one to achieve exactly the same effect by just adding the FD ansatz into the projection basis functions. This helps mitigate computational cost as it saves additional tedious Monte Carlo simulation on $\hat{V}^{1D}$, as well as the variance and covariance calculation in Eqn.\ref{eqn:optimalFDbeta} and \ref{eqn:fdPertLSMProjection}.
	\item From the proof of Prop.\ref{prop:FD_LSM_CV_LSM}, up to first order of variance reduction, we don't see benefit of introducing cross term between $\vec{\phi}^{P^\text{FD}}(x)$ and $\vec{\phi}_{R-1}^{P^\text{LSM}}(x)$. This simplifies the construction of new basis and avoid overfitting due to further introduction of higher order terms.	
	\item The most effective FD auxiliary variable $\hat{V}^{1D}$ is obtained by achieving the highest possible correlation with $V$ based on Eqn.\ref{eqn:optimizedVar}. Heuristically, selecting a suitable ansatz can be seen as an effort to capture the dominant dynamics of the original high-dimensional problem within a computationally efficient and numerically tractable framework, e.g., 1D FD PDE, as an application of the Gy\"{o}ngy theorem. This is done even if the maximal correlation of the resulting option price after dimension reduction is not explicitly proven. For instance, under a multi-asset single-factor setting (e.g. Black Scholes or Local Volatility), we can choose the ansatz from single-factor 1D PDE after inter-asset dimension reduction; for single-asset multi-factor settings (e.g. Heston stochastic volatility), one can construct the equivalent 1D process (Local volatility in general) by matching same marginal distributions as mentioned before in Sec.\ref{sec:fdpde}; finally, for the most general case of multi-asset multi-factor setting, one could apply the dimension reduction on intra-asset level (reducing the factors other than spot, e.g. stochastic volatility on the same asset) to start with, and then on inter-asset level, eventually arriving at 1D problem which is solvable under Eqn.\ref{eqn:pde1d}.  
\end{itemize}

\subsubsection{Error Analysis}\label{sec:Error-Analysis}

After establishing the role of FD solution as a control variate in FD-LSM, we proceed with error analysis in the form of Eqn.\ref{eqn:finalResultsForm} as final solution of Least-Squares Monte Carlo. 

From this point on, we drop the $\F_0$ filtration for total expectation, variance and covariance operators by default unless explicitly specified, i.e, $\E(\cdot)\equiv\E(\cdot|\F_0)$, $\Var(\cdot)\equiv\Var(\cdot|\F_0)$ and $\Cov(\cdot,\cdot\cdot)\equiv\Cov(\cdot,\cdot\cdot|\F_0)$. 

To initiate the final pricing error analysis, we introduce below lemma to pin down the local regression error, i.e., the mean squared error (MSE) $\sqrt{\E[(F^P_k-F_k)^2]}$ of $F^P_k$ (introduced to conditional expectation in Eqn.\ref{eqn:finalResultsForm}) under a given projection method $P$ (either $P^{\text{LSM}}$ or $P^{\text{FD-LSM}}$) at $t_k$. 

\begin{lemma}\label{lemma:localRegressionError}
	Under the ceteris paribus assumption with $R\geq 3$, the local regression error (upper bound) of FD-LSM is reduced, compared to LSM, by a scaling factor\\ $\sqrt{\frac{R-1}{R}(1-\tilde{\rho}_k^2(V_{k+1},\hat{V}_{k+1}^{1D}))}$, where $\tilde{\rho}_k(V_{k+1},\hat{V}_{k+1}^{1D})$ is defined in Eqn.\ref{eqn:correlationVV}.
\end{lemma}

\renewcommand{\qedsymbol}{$\blacksquare$}
\begin{proof}

We denote by $\Xi^{P^{\text{LSM}}}_k[e^{-r\Delta T}V_{k+1}]$ the upper bound of $\sqrt{\E[(F^{P^{\text{LSM}}}_k-F_k)^2]}$. A detailed derivation of this bound is provided in Appendix \ref{sec:localRegressionErrorLSM}. One of the key points is to express $F_k$ in terms of exact regression coefficients $\vec{\beta}^{P^{\text{LSM}}0}$ satisfying
\begin{equation}\label{eqn:exactBeta0LSM}
F_k=\vec{\beta}^{P^{\text{LSM}}0}_k\cdot\vec{\phi}^{P^{\text{LSM}}}_{R,k}.
\end{equation}
From Eqn.\ref{eqn:lsmXiDerivation} in Appendix, it concludes that
\begin{equation}\label{eqn:xiPLSM}
\Xi^{P^{\text{LSM}}}_k[e^{-r\Delta T}V_{k+1}]=R\E[\Var(e^{-r\Delta T} V_{k+1}|\F_{t_k})].
\end{equation}
This result is consistent with previous studies on the asymptotic convergence rate of $\Xi^{P^{\text{LSM}}}_k$ going to zero under big $O_p$ notation by \cite{Stentoft2001} in their Theorem 1: the dominant term $R/N_R$ in the asymptotic from is related to the option variance and thus inversely proportional to $N_R$; whereas the other term $R^{-\frac{2s}{d}}$ (with $s$ denoting the number of continuous derivatives of $F_k$ that exist) is negligible, i.e., 
\begin{equation}\label{eqn:rangeOfR}
R^{-\frac{2s}{d}}\ll 1,
\end{equation}
in the case of $R\geq 2$ and large $s\gg\frac{d}{2}$, i.e., a smooth $F_k$\footnote{In the case of WIC in Appendix~\ref{sec:WorstOfIssuerCallableNote}, there are discontinuities on the digital coupons and the knocked-in short put option, respectively. We assume that such discontinuities can be mitigated to ensure smoothness by introducing overhedges, e.g., call spreads, which aligns with common practice.}, see \cite{newey1997convergence}. In fact, this condition is a justification of  the exact decomposition in Eqn.\ref{eqn:exactBeta0LSM}.

Apparently, the upper bound estimate using Eqn.\ref{eqn:xiPLSM} is also valid for FD-LSM considering all the basis functions as a whole, but it is a loose one which ignores the correlation between $\vec{\phi}^{P^{\text{FD-LSM}}}_{R,k}$ and $V_{k+1}$. Thanks to the decomposition in Eqn.\ref{eqn:projectionLSMandFDLSD} from Prop.\ref{prop:FD_LSM_CV_LSM}, we can achieve a tighter estimate as follows.

By construction, the MSE of $F^{P^{\text{FD-LSM}}}_k$ can be decomposed into individual $L^2$-norm square of orthogonal components $P_k^{\text{FD}}[F^{P^{\text{FD-LSM}}}_k-F_k]$ and $M^{\text{FD}}_k[F^{P^{\text{FD-LSM}}}_k-F_k]$, respectively. 

By definition, $P_k^{\text{FD}}$ projects a variable onto $\text{span}(\vec{\phi}^{P^\text{FD}}_{k})$ as a subset of $\text{span}(\vec{\phi}^{P^\text{FD-LSM}}_{k})$. Therefore, Law of Iterated Projection gives $P_k^{\text{FD}}[F^{P^{\text{FD-LSM}}}_k]=P_k^{\text{FD}}[P^{\text{FD-LSM}}_k[e^{-r\Delta T}V_{k+1}]]=P_k^{\text{FD}}[e^{-r\Delta T}V_{k+1}]$. Additionally, taking into account Eqn.\ref{eqn:PFDVEqualToPFDF}, one arrives at $P_k^{\text{FD}}[F^{P^{\text{FD-LSM}}}_k]=P_k^{\text{FD}}[F_k]$; therefore 
\begin{equation}\label{eqn:FDLSMMESSimplification}
\begin{split}
\E[(F^{P^{\text{FD-LSM}}}_k-F_k)^2]&=\E[(P_k^{\text{FD}}[F^{P^{\text{FD-LSM}}}_k-F_k])^2]+\E[(M^\text{FD}_k[F^{P^{\text{FD-LSM}}}_k-F_k])^2]\\
&=\E[(M^\text{FD}_k[F^{P^{\text{FD-LSM}}}_k]-M^\text{FD}_k[F_k])^2].
\end{split}
\end{equation}

Applying $M^\text{FD}_k$ to the right-hand side of Eqn.\ref{eqn:projectionLSMandFDLSD} and exploiting the orthogonality condition $M^\text{FD}_k[f^\text{FD}_k(X_k)]=f^\text{FD}_k(X_k)-P^\text{FD}_k[f^\text{FD}_k(X_k)]=0$, and then expressing $P_k^{\text{LSM},\text{LSM}\perp\text{FD}}$ via Eqn.\ref{eqn:l2perpProjection} and Eqn.\ref{eqn:objFunctionLSMperpFDBeta}, we have

\begin{equation}\label{eqn:mfdfpfdlsm}
\begin{split}
M^\text{FD}_k[F^{P^{\text{FD-LSM}}}_k] &=M^\text{FD}_k[P_k^{\text{LSM},\text{LSM}\perp\text{FD}}[e^{-r\Delta T}(V_{k+1} - \beta_k^{\text{FD}}(F_k)\hat{V}_{k+1}^{1D})]]\\
&=\vec{\beta}^{P^{\text{LSM}\perp{\text{FD}}}}_k(e^{-r\Delta T}(V_{k+1} - \beta_k^{\text{FD}}(F_k)\hat{V}_{k+1}^{1D}))\cdot M^\text{FD}_k[\vec{\phi}^{P^\text{LSM}}_{R-1,k}]\\
&=(\E[\vec{\phi}^{P^{\text{LSM}\perp\text{FD}}}_{R-1,k}\vec{\phi}^{P^{\text{LSM}\perp\text{FD}}T}_{R-1,k}])^{-1}
\E[\vec{\phi}^{P^{\text{LSM}\perp\text{FD}}}_{R-1,k}e^{-r\Delta T}(V_{k+1} - \beta_k^{\text{FD}}(F_k)\hat{V}_{k+1}^{1D})]
\cdot \vec{\phi}^{P^{\text{LSM}\perp\text{FD}}}_{R-1,k}.\\
\end{split}
\end{equation}

Similar to Eqn.\ref{eqn:exactBeta0LSM}, exact coefficients $\vec{\beta}^{P^{\text{LSM}\perp{\text{FD}}}0}_k$ satisfying $M^{\text{FD}}[F_k]=\vec{\beta}^{P^{\text{LSM}\perp{\text{FD}}}0}_k\cdot\vec{\phi}^{P^{\text{LSM}\perp\text{FD}}}_{R-1,k}$ are introduced. According to Eqn.\ref{eqn:objFunctionLSMperpFDBeta}, we have
\begin{equation}\label{eqn:mfdfpfdlsmExact}
\begin{split}
\vec{\beta}^{P^{\text{LSM}\perp{\text{FD}}}0}_k&=\vec{\beta}^{P^{\text{LSM}\perp{\text{FD}}}}_k(P_k^{\text{FD}}[F_k])\\
&=(\E[\vec{\phi}^{P^{\text{LSM}\perp\text{FD}}}_{R-1,k}\vec{\phi}^{P^{\text{LSM}\perp\text{FD}}T}_{R-1,k}])^{-1}
\E[\vec{\phi}^{P^{\text{LSM}\perp\text{FD}}}_{R-1,k}(F_k-\beta_k^{\text{FD}}(F_k)f^{\text{FD}}_k(X_k))].
\end{split}
\end{equation}

In Eqn.\ref{eqn:mfdfpfdlsm} and \ref{eqn:mfdfpfdlsmExact}, we have ignored the constant term $\alpha_k^{\text{FD}}(F_k)$ that can be absorbed in the projection $P_k^{\text{LSM},\text{LSM}\perp\text{FD}}$ as usual. 

As discussed before, $R \geq 2$ is required due to the constraint in Eqn.\ref{eqn:rangeOfR}. In fact, to account for the decrement in the number of basis functions when using $\vec{\phi}^{P^{\text{LSM}\perp\text{FD}}}_{R-1,k}$, we would enforce $R \geq 3$. Fortunately, the control variate makes the smoothness condition on $F_k-\beta_k^{\text{FD}}(F_k)f^{\text{FD}}_k(X_k)$ easier to achieve than on plain $F_k$.

In what follows, the calculation of $\E[(M^\text{FD}_k[F^{P^{\text{FD-LSM}}}_k]-M^\text{FD}_k[F_k])^2]$ mirrors the procedure used for $\E[(F^{P^{\text{LSM}}}_k-F_k)^2]$ after performing below transformation:
\begin{equation}\label{eqn:lsmperpfdmappingtolsm}
\begin{split}
F_k &\rightarrow F_k-\beta_k^{\text{FD}}(F_k)f^{\text{FD}}_k(X_k)\\
\vec{\phi}^{P^\text{LSM}}_{R,k} &\rightarrow\vec{\phi}^{P^{\text{LSM}\perp\text{LSM}}}_{R-1,k}\\
\vec{\beta}^{P^{\text{LSM}}}_k(\cdot) &\rightarrow \vec{\beta}^{P^{\text{LSM}\perp{\text{FD}}}}_k(\cdot)\\
e^{-r\Delta T}V_{k+1} &\rightarrow e^{-r\Delta T}(V_{k+1} - \beta_k^{\text{FD}}(F_k)\hat{V}_{k+1}^{1D})
\end{split}.
\end{equation}

Along with Eqn.\ref{eqn:FDLSMMESSimplification}, it yields
\begin{equation}
\begin{split}
&\E[(M^\text{FD}_k[F^{P^{\text{FD-LSM}}}_k]-M^\text{FD}_k[F_k])^2]\\
\leq&(R-1)\cdot\E[(e^{-r\Delta T}(V_{k+1} - \beta_k^{\text{FD}}(F_k)\hat{V}_{k+1}^{1D})-(F_k-\beta_k^{\text{FD}}(F_k)f^{\text{FD}}_k(X_k)))^2]\\
=&(R-1)\cdot\E[\Var( e^{-r\Delta T}(V_{k+1} - \beta_k^{\text{FD}}(F_k)\hat{V}_{k+1}^{1D})|\F_{t_k})]\\
=&(R-1)[1-\tilde{\rho}_k^2(V_{k+1},\hat{V}^{1D}_{k+1})]\E[\Var(e^{-r\Delta T}V_{k+1}|\F_{t_k})],
\end{split}
\end{equation}
where the identity $\E[e^{-r\Delta T}(V_{k+1} - \beta_k^{\text{FD}}(F_k)\hat{V}_{k+1}^{1D})|\F_{t_k}]=F_k-\beta_k^{\text{FD}}(F_k)f^{\text{FD}}_k(X_k)$ and Eqn.\ref{eqn:optimizedVar} have been used in the penultimate and last line, respectively, apart from following the derivation described in Appendix~\ref{sec:localRegressionErrorLSM} with the mapping in Eqn.\ref{eqn:lsmperpfdmappingtolsm}.

Denoting $\Xi_k^{P^{\text{FD-LSM}}}[e^{-r\Delta T}V_{k+1}]$ as the upper bound of $\E[(F^{P^{\text{FD-LSM}}}_k-F_k)^2]$, and comparing above with Eqn.\ref{eqn:xiPLSM}, we have

\begin{equation}\label{eqn:varReductionXi}
\begin{split}
\Xi_k^{P^{\text{FD-LSM}}}[e^{-r\Delta T}V_{k+1}] =\frac{R-1}{R}[1-\tilde{\rho}_k^2(V_{k+1},\hat{V}^{1D}_{k+1})]\cdot\Xi_k^{P^{\text{LSM}}}[e^{-r\Delta T}V_{k+1}],
\end{split}
\end{equation}
which claims the lemma after taking a square root on both hand sides.

\end{proof}

Let's delve deeper into Lemma~\ref{lemma:localRegressionError} before proceeding:

\begin{itemize}
	\item In the limit as $\tilde{\rho}_k^2(V_{k+1},\hat{V}^{1D}_{k+1})\to 1$, the result degenerates to Lemma~\ref{lemma:UniqueExactRegressor} as expected.	
	\item In the limit as $\tilde{\rho}_k^2(V_{k+1},\hat{V}^{1D}_{k+1})\to 0$, the control variate has null effect as $f^{\text{FD}}_k(X_k)$ is orthogonal to $F_k$ in $L^2$. In this regime, the remaining mild error reduction, going as $\sqrt{\frac{R-1}{R}}$ is purely from reducing susceptibility of overfitting by removing unnecessary high order monomial component under the constraint of Eqn.\ref{eqn:rangeOfR}; further taking the limit of $R\to\infty$, $P_k^{\text{LSM},\text{LSM}\perp\text{FD}}$ converges to $P_k^{\text{LSM}}$ with MSE being equal to each other.
\end{itemize}

In the light of Lemma~\ref{lemma:localRegressionError} as an excellent implication of Prop.\ref{prop:FD_LSM_CV_LSM}, we state the main theoretical result, and its assumption(s), in this section as follows.

\begin{assumption}\label{assumption:ignoreVarCondition}	
	In the option PV problem, below inequality (with $e^{-r\Delta T}$ dropped out) holds for all $k$:
	\begin{equation*}
	\Var(B^P_k)\geq\Var(V^P_{k+1} - V_{k+1})\cdot\mathcal{X}^P_k,
	\end{equation*}
where
\begin{equation}\label{eqn:definitionsMathcalBX}
\begin{split}
B^P_k&:=(Z_k-F_k)(\Theta(Z_k-P_k[V^P_{k+1}])-\Theta(Z_k-F_k)),\\
\mathcal{X}^P_k&:=\E(\Theta(|P_k[V^P_{k+1}]-F_k|-|F_k-Z_k|)),
\end{split}
\end{equation}
with $V^P_{k+1}$ denoting the approximated option price under projection $P$, as opposed to the exact $V_{k+1}$.
\end{assumption}

\begin{assumption}\label{assumption:k0limit}
Further defining below quantity in the option PV problem
\begin{equation}\label{eqn:definitionsMathcalY}
\begin{split}
\mathcal{Y}^P_k&:=\E(\Theta(P_k[V^P_{k+1}]-Z_k)),
\end{split}
\end{equation}
below inequality is true
\begin{equation*}
	\mathcal{X}^P_0\leq(1-\mathcal{Y}^P_0)^2,
\end{equation*}
in the limit as $M\to\infty$, where $\mathcal{X}^P_0:=\lim_{\substack{k\to 1 \\ M\to\infty}}\mathcal{X}^P_k$ and $\mathcal{Y}^P_0:=\lim_{\substack{k\to 1 \\ M\to\infty}}\mathcal{Y}^P_k$. For avoidance of doubt, we use ``0'' as the subscript to represent this limiting case where $k$ is indexed from 1 as $k\in[1,M]$.
\end{assumption}

\begin{prop}\label{prop:Err_FD_LSM}
The pricing error of FD-LSM converges to zero as $\tilde{\rho}_{m^*}(V_{m^*+1},\hat{V}^{1D}_{m^*+1})\to 1$, where $m^*:=\arg_{m\in[1,M]} \min \tilde{\rho}_m(V_{m+1},\hat{V}^{1D}_{m+1})$ represents the time step at which the minimum correlation between $V$ and $\hat{V}^{1D}$ is achieved. This convergence holds in a generic multi-period (or multi-bucket) setting, while the error of the LSM counterpart remains finite. For the option PV problem, Assumption \ref{assumption:ignoreVarCondition} is required; furthermore, if Assumption \ref{assumption:k0limit} is also satisfied, the convergence of the FD-LSM error remains valid even as $M\to\infty$.
\end{prop}

\renewcommand{\qedsymbol}{$\blacksquare$}
\begin{proof}
The proof will be conducted for American-style option PV in Eqn.\ref{eqn:phiOptionpv} and CVA in Eqn.\ref{eqn:phiCva}, separately. As we can see, the derivations for the two categories (with or without the need of dynamic programming) are generic and representative enough to cover practical use cases of both categories. 

\textbf{American-style option PV: Error accumulation in a multi-period setting}

Here we tackle the error analysis on Eqn.\ref{eqn:phiOptionpv} by considering a most generic multi-period setting, which give rises to the error accumulation under dynamic programming on LSM and FD-LSM. This piece of analysis is largely an extension of previous studies, with a focus on relating the accumulated option pricing errors to local regression error during the backward evolution. We restrict the discussion in the context of holder exercisability, but the conclusion can be applied to issuer exercisability equivalently.

With the discounting factor from period to period $e^{-r\Delta T}$ dropped out for ease of illustration from now on (see \cite{clement2002analysis}), the dynamic programming solution of option value for both exact $V_k$ and approximated $V_k^P$ can be re-written here from Eqn.\ref{eqn:v_k} as
\begin{equation}
\begin{aligned}
V_k &= Z_k \Theta(Z_k-F_k) + V_{k+1}(1-\Theta(Z_k-F_k)),\\
&\quad\quad\quad\quad\quad\quad\quad\quad\text{and}\\
V_k^P &= Z_k \Theta(Z_k- P_k[V_{k+1}^P]) + V_{k+1}^P(1-\Theta(Z_k-P_k[V_{k+1}^P])),
\end{aligned}
\end{equation}
respectively, where $P_k$ can be either $P_k^{\text{LSM}}$ or $P_k^{\text{FD-LSM}}$ as usual, and $P_k[V^P_{k+1}]$ is the projection on the option value with error accumulated up to the next step due to approximation from backward. The option PV at $t=0$ is $V_{k=1}$( and its proxy $V^P_{k=1}$) due to omitting $e^{-r\Delta T}$.

We are interested in the total expectation value of the difference between $V_k^P$ and $V_k$, and it can be calculated as:

\begin{equation}\label{eqn:E_V_P_k}
\begin{split}
\E(V_k^P-V_k) &=\E(\E([Z_k-V_{k+1}][\Theta(Z_k-P_k[V^P_{k+1}])-\Theta(Z_k-F_k)]|\F_{t_k})) + \\
&\qquad\E[(V_{k+1}^P-V_{k+1})\Theta(P_k[V^P_{k+1}]-Z_k)] \\
&=\E([Z_k-\E(V_{k+1}|\F_{t_k})][\Theta(Z_k-P_k[V^P_{k+1}])-\Theta(Z_k-F_k)]) + \\
&\qquad\E[(V_{k+1}^P-V_{k+1})\Theta(P_k[V^P_{k+1}]-Z_k)] \\
&=\E((Z_k-F_k)[\Theta(Z_k-P_k[V^P_{k+1}])-\Theta(Z_k-F_k)]) + \\
&\qquad\E(V_{k+1}^P-V_{k+1})\E(\Theta(P_k[V^P_{k+1}]-Z_k))+\Cov[(V_{k+1}^P-V_{k+1}),\Theta(P_k[V^P_{k+1}]-Z_k)], \\
\end{split}
\end{equation}

where $\E(\E(V_{k+1}^P|\F_{t_k}))\equiv\E(\E(V_{k+1}^P|\F_{t_{k+1}}))$ and $\E(\E(V_{k+1}|\F_{t_k}))\equiv\E(\E(V_{k+1}|\F_{t_{k+1}}))$ have been applied starting with law of total expectation, i.e., $\E(\cdot)\equiv\E(\E(\cdot|\F_{t_k}))$.

We can see easily that $\E(V_k^P-V_k)$ in Eqn.\ref{eqn:E_V_P_k} is non-positive by induction starting from $\E(V_{M}^P)=\E(V_{M})$. Intuitively, the reason is that the least-squares approximation introduces a sub-optimal exercise strategy giving rise to biased low error for holder exercisable option price. The last covariance term is arguably small which is neglected by previous analysis on the same matter in \cite{clement2002analysis}; here we would quantify it as non-negative in a more precise measure, given the fact that as $V^P_{k+1}\to V_{k+1}$, both sides tend to increase under the covariance operator, indicating a non-negative correlation between the two. 

Denoting $e^{PV,P}_k=|\E(V_k^P-V_k)|$, we have 
\begin{equation}\label{eqn:e_P_k}
e^{PV,P}_k \leq |\E(B^P_k)| + \mathcal{Y}^P_k e^{PV,P}_{k+1},
\end{equation}
where the identities of $B^P_k$ from Eqn.\ref{eqn:definitionsMathcalBX} and $\mathcal{Y}^P_k$ from Eqn.\ref{eqn:definitionsMathcalY} have been applied, respectively; and `$\leq$' stems from a different sign of the covariance term compared to  $\E(V_k^P-V_k)$ in Eqn.\ref{eqn:E_V_P_k}, as analysed previously. 

For $|\E(B^P_k)|^2$, we have, 
\begin{eqnarray*}
|\E(B^P_k)|^2 &=&\E((B^P_k)^2)-\Var(B^P_k)     \\
&=& \E(|Z_k - F_k|^2|\Theta(F_k-Z_k)\Theta(Z_k-P_k[V^P_{k+1}])-\Theta(P_k[V^P_{k+1}]-Z_k)\Theta(Z_k-F_k)|^2)-\Var(B^P_k) \\
                &\leq& \E(|Z_k - F_k|^2\Theta(|P_k[V^P_{k+1}]-F_k|-|F_k-Z_k|)) -\Var(B^P_k)\\ 
                &\leq& \E(|P_k[V^P_{k+1}] - F_k|^2)\E(\Theta(|P_k[V^P_{k+1}]-F_k|-|F_k-Z_k|))-\Var(B^P_k)\\
                &=& \E(|P_k[V^P_{k+1}] - F_k|^2)\mathcal{X}^P_k-\Var(B^P_k),
\end{eqnarray*}
where we have made use of $\mathcal{X}^P_k$ (which in practice is small as $P_k[V^P_{k+1}]$ is close to $F_k$) from Eqn.\ref{eqn:definitionsMathcalBX} in the penultimate line.

Applying basic properties of projection to the first term of the last line in above (see similar derivation in the proof of Theorem 3.1 in \cite{clement2002analysis}), we have

\begin{eqnarray*}
\E(|P_k[V^P_{k+1}] - F_k|^2) &\leq& \E(|P_k[V^P_{k+1}] - P_k[V_{k+1}]|^2) + \E(|P_k[V_{k+1}] - F_k|^2) \\ 
&\leq& \E(|V^P_{k+1} - V_{k+1}|^2) + \E(|P_k[V_{k+1}] - F_k|^2) \\
&=& (e^{PV,P}_{k+1})^2 + \Var(V^P_{k+1} - V_{k+1}) + \E(|P_k[V_{k+1}] - F_k|^2),\\
\end{eqnarray*}
where in the last line, the last term is exactly the local regression error with upper bound $\Xi_k^P[V_{k+1}]$ due to the projection on $P_k[V_{k+1}]$, which doesn't consider accumulated error yet (as opposed to $P_k[V^P_{k+1}]$). To get rid of the two offsetting variance terms within $|\E(B^P_k)|^2$, we need to make use of Assumption \ref{assumption:ignoreVarCondition}, i.e., $\Var(B^P_k)\geq\Var(V^P_{k+1} - V_{k+1})\cdot\mathcal{X}^P_k$. Then we arrive at
\begin{equation}\label{eqn:B_P_k_2}
\begin{split}
|\E(B^P_k)|^2 &\leq \mathcal{X}^P_k\cdot((e^{PV,P}_{k+1})^2 + \Xi_k^P[V_{k+1}]+\Var(V^P_{k+1} - V_{k+1})) -\Var(B^P_k)\\
&\leq \mathcal{X}^P_k\cdot((e^{PV,P}_{k+1})^2 + \Xi_k^P[V_{k+1}]).
\end{split}
\end{equation} 

Note that as $V^P_{k+1}$ approaches the exact solution, both sides of the inequality in Assumption \ref{assumption:ignoreVarCondition} tend to zero, it conjectures that the right-hand side, with $V^P_k-V_k\to 0$ and $\mathcal{X}^P_k\to 0$, converges faster than the left-hand side with finite $Z_k-F_k$. In practical terms, this assumption can be relaxed as
\[
\mathcal{X}^P_k\cdot(e^{PV,P}_{k+1})^2+\Var(B^P_k)\gg\mathcal{X}^P_k\cdot\Var(V^P_{k+1} - V_{k+1}),
\]
in such a way that $\mathcal{X}^P_k\cdot(e^{PV,P}_{k+1})^2$ outweighs the difference between the two variance terms, even if that difference is positive.

Next, $\Xi_k^P[V_{k+1}]$ can be bounded by $\Xi_{Max}^P$ (for both LSM and FD-LSM, making use of Eqn.\ref{eqn:varReductionXi} from Lemma~\ref{lemma:localRegressionError}) as
\begin{equation}
\begin{split}
\Xi_k^{P^{\text{LSM}}}[V_{k+1}]&\leq\Xi_{n^*}^{P^{\text{LSM}}}[V_{n^*+1}]\\
&=\Xi_{Max}^{P^{\text{LSM}}},\\
\Xi_k^{P^{\text{FD-LSM}}}[V_{k+1}] &=\frac{R-1}{R}[1-\tilde{\rho}_k^2(V_{k+1},\hat{V}^{1D}_{k+1})]\cdot\Xi_k^{P^{\text{LSM}}}[V_{k+1}]\\
&\leq\frac{R-1}{R}[1-\tilde{\rho}_{m^*}^2(V_{{m^*}+1},\hat{V}^{1D}_{{m^*}+1})]\cdot\Xi_k^{P^{\text{LSM}}}[V_{k+1}]\\
&\leq\frac{R-1}{R}[1-\tilde{\rho}_{m^*}^2(V_{{m^*}+1},\hat{V}^{1D}_{{m^*}+1})]\cdot\Xi_{n^*}^{P^{\text{LSM}}}[V_{n^*+1}]\\
&=\Xi_{Max}^{P^{\text{FD-LSM}}},
\end{split}
\end{equation}
where $n^*=\arg_{n\in[1,M]} \max \Xi_n^{P^{\text{LSM}}}[V_{n+1}]$.
 
Combining above with Eqn.\ref{eqn:e_P_k} and \ref{eqn:B_P_k_2}, the dynamical programming solution on the upper bound of $e^{PV,P}_{k}$, denoted by $\epsilon^{PV,P}_{k}$ for a given projection $P$, is 
\begin{equation}\label{eqn:epsilon_PV_k}
\epsilon^{PV,P}_{k} =\sqrt{\mathcal{X}^P_k\cdot((\epsilon^{PV,P}_{k+1})^2 + \Xi_{Max}^P)} + \mathcal{Y}^P_k \epsilon^{PV,P}_{k+1},
\end{equation}
with initial condition $\epsilon^{PV,P}_{M}=0$. It defines the relation between accumulated errors and local regression errors on option prices under least-squares Monte Carlo at each step. 

The problem now lies in analyzing the characteristics of an iterative backward sequence, governed by the equation $x_{k}=\sqrt{a_{k}(x_{k+1}^2 + b)} + c_{k} x_{k+1}$. The (non-negative) parameters are defined as follows: $a_k=\mathcal{X}^P_k$, $b=\Xi_{Max}^P$, and $c_k=\mathcal{Y}^P_k$. This sequence is initialized with the condition $x_M = 0$. A detailed analysis of this sequence's properties is presented in Appendix~\ref{sec:sequenceProperties}.

For the general multi-period setting under finite $M$, following Eqn.\ref{eqn:sequenceResultsInduction} as a proof by induction in Appendix~\ref{sec:sequenceFiniteInduction}, we have 
\begin{equation}\label{eqn:epsilonFDLSMUB}
\epsilon^{PV,P^{\text{FD-LSM}}}_{k}\sim O(\sqrt{\Xi_{Max}^{P^{\text{FD-LSM}}}})\sim O(\sqrt{1-\tilde{\rho}_{m^*}(V_{m^*+1},\hat{V}^{1D}_{m^*+1})})
\end{equation}
as $\tilde{\rho}_{m^*}(V_{m^*+1},\hat{V}^{1D}_{m^*+1})\to 1$, whereas $\epsilon^{PV,P^{\text{LSM}}}_{k}\sim O(\sqrt{\Xi_{Max}^{P^{\text{LSM}}}})$ remains finite, for any $k\in[1,M]$.

Next, we consider the asymptotic behavior of the error in the limit as $M\to\infty$ and $k\to 1$. As we can see from the fixed point analysis (Appendix~\ref{sec:sequenceFixedPointLimit}) of the backward sequence, if Assumption~\ref{assumption:k0limit} also holds, the fixed point condition of Eqn.\ref{eqn:fixedPointCondition} is satisfied with $a_0=\mathcal{X}^P_0$ and $c_0=\mathcal{Y}^P_0$. Consequently, using Eqn.\ref{eqn:fixedPointSolution} and substituting $b=\Xi_{Max}^P$, one can see that the error saturates at 
\begin{equation}\label{eqn:limit_epsilon_PV_k}
\epsilon^{PV,P}_0:=\lim_{\substack{k\to 1 \\ M\to\infty}}\epsilon^{PV,P}_k = \sqrt{\frac{\Xi_{Max}^P}{\frac{(1-\mathcal{Y}^P_0)^2}{\mathcal{X}^P_0}-1}},
\end{equation}
as going backward. From above, we arrive at $\epsilon^{PV,P^{\text{FD-LSM}}}_0\sim O(\sqrt{1-\tilde{\rho}_{m^*}(V_{m^*+1},\hat{V}^{1D}_{m^*+1})})$, similar to the case in Eqn.\ref{eqn:epsilonFDLSMUB} with finite $M$.

For a more tractable analysis of Eqn.\ref{eqn:epsilon_PV_k}, let's assume $\mathcal{X}^P_k=\mathcal{X}$ and $\mathcal{Y}^P_k=\mathcal{Y}$ for all $k$ in both LSM and FD-LSM.

Using $\tilde{\rho}_{m^*}(V_{m^*+1},\hat{V}^{1D}_{m^*+1})=90\%$ and $R=4$, $\Xi_{Max}^{P^{\text{LSM}}}=5\%^2$ and thus\\ $\Xi_{Max}^{P^{\text{FD-LSM}}} = \left(5\%\times\sqrt{\frac{4-1}{4}(1-90\%^2)}\right)^2=1.89\%^2$ as a numerical example along with $\mathcal{X}=0.01$ and $\mathcal{Y}=0.8$, we have the plots for $\epsilon^{PV,P}_{k}$ as depicted in Fig.\ref{fig:accumulatedError}. The asymptotic limits obtained through backward iteration are consistent with the theoretical predictions of Eqn.\ref{eqn:limit_epsilon_PV_k}: specifically, $\epsilon^{PV,P^{\text{LSM}}}_0=2.89\%$ for LSM and $\epsilon^{PV,P^{\text{FD-LSM}}}_0=1.09\%$ for FD-LSM.

\begin{figure}[ht]
	\begin{varwidth}{\linewidth}
		\centerline{\includegraphics[width=0.6\textwidth]{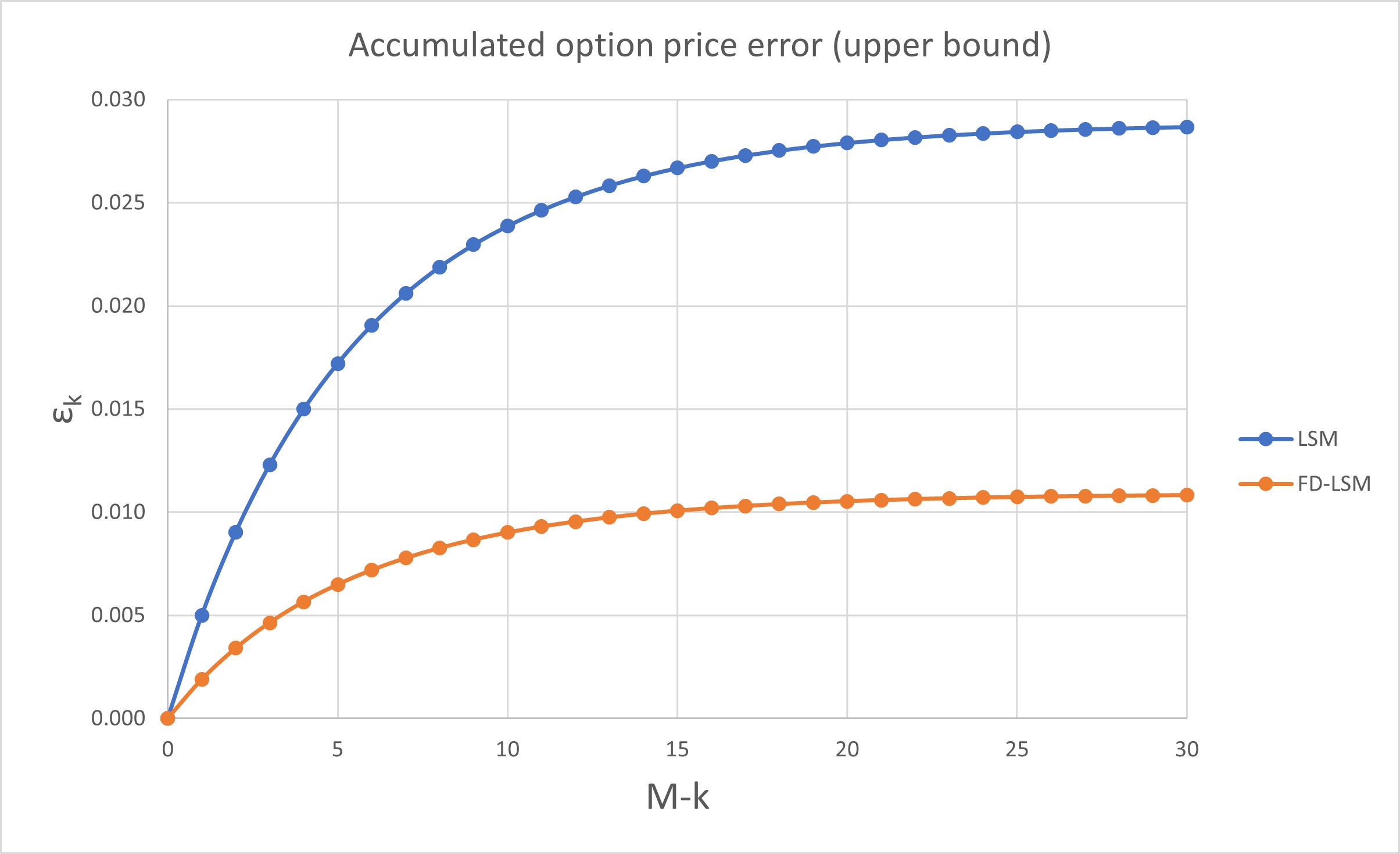}}
		\caption{The upper bound of accumulated option price error for LSM and FD-LSM at each step going backward, using the parameters mentioned in above. The $x$-axis is $M-k$ where M=60. }
		\label{fig:accumulatedError}
	\end{varwidth}
\end{figure}

Our quantitative result in Eqn.\ref{eqn:limit_epsilon_PV_k} is consistent with previous qualitative study on the same matter by \cite{stentoft2014value}: due to the fact that the projection approximation in LSM is only utilized to estimate stopping time rather than value function, the magnitude of accumulated error is relatively small. In fact, as stated by \cite{tsitsiklis2001regression} in their Theorem 1, the accumulated error (upper bound) on option value scales as $\sqrt{M}$ when $k\to 1$ by applying the projection approximation to the value function itself. Now with a customized ansatz strongly correlated to the original problem, we show that theoretically in Eqn.\ref{eqn:epsilonFDLSMUB}, FD-LSM could achieve even much lower level of accumulated errors.

\textbf{Error analysis in CVA: Multi-bucket setting}

Now we turn to the error analysis of the CVA problem in a similar manner. As one could expect, the derivation of the proof for CVA problem in Eqn.\ref{eqn:cva} is much less involved, due to the absence of a dynamic programming control coupled with least-squares projection, i.e., the local regression error due to $\tilde{F}_k\to\tilde{F}^P_k$ at each time step is decoupled and independent from other steps without being accumulated. 

In a generic multi-bucket setting, the CVA error under projection $P$ at the $t_k$ bucket is defined as
\begin{equation}
e^{CVA,P}_k:=|\E[(\Phi^{CVA,P}_k-\Phi^{CVA}_k)]|,
\end{equation}
by differentiating between $\Phi^{CVA}_k$ in Eqn.\ref{eqn:phiCva} and its proxy defined as
\begin{equation}\label{eqn:phiCVAk}
\Phi^{CVA,P}_k:=e^{-r\cdot t_k}\cdot \max\left(0, \tilde{F}^{P}_k \right) \cdot h_{cpty}(\tilde{F}^{P}_k,t_k)\cdot\Delta T \cdot e^{-\sum_{j=1}^{k}h_{cpty}(\tilde{F}^{P}_j,t_j)\Delta T},
\end{equation}
where $h_{cpty}(\tilde{F}^{P}_k,t_k)$ is approximated due to its dependence on $\tilde{F}^{P}_k$ subjected to the least-squares approximation. 

Given the explicit functional dependence on $\tilde{F}_i$ ($i\in[1,k]$) in Eqn.\ref{eqn:phiCVAk}, the upper bound of $e^{CVA,P}_k$, denoted by $\epsilon^{CVA,P}_k$, can be derived as (defining $\Xi^P_{i}[\tilde{V}]:=\E[(\tilde{F}^{P}_i-\tilde{F}_i)^2]$)
\begin{equation}
\begin{split}
e^{CVA,P}_k &\leq \E(|\Phi^{CVA}_{k}-\Phi^{CVA,P}_{k}|)\\
&\approx\E[|\sum^{k}_{i=1}\frac{\partial \Phi^{CVA}_k}{\partial \tilde{F}_i}\cdot(\tilde{F}^{P}_i-\tilde{F}_i)|]\\
&\leq\sum^{k}_{i=1}\E[|\frac{\partial \Phi^{CVA}_k}{\partial \tilde{F}_i}\cdot(\tilde{F}^{P}_i-\tilde{F}_i)|]\\
&\leq\sum^{k}_{i=1}\sqrt{\E[\left(\frac{\partial \Phi^{CVA}_k}{\partial \tilde{F}_i}\right)^2]}\cdot\sqrt{\E[(\tilde{F}^{P}_i-\tilde{F}_i)^2]},\\
\epsilon^{CVA,P}_k&:=\sum^{k}_{i=1}\sqrt{\E[\left(\frac{\partial \Phi^{CVA}_k}{\partial \tilde{F}_i}\right)^2]}\cdot \sqrt{\Xi^P_i[\tilde{V}]}
\end{split}
\end{equation}
where we apply the first-order Taylor series approximation and the Cauchy-Schwarz inequality in the second and forth lines, respectively. 

To proceed, let's denote $\tilde{m}^*=\arg_{m\in[1,M]} \min \tilde{\rho}_m(\tilde{V},\hat{\tilde{V}}^{1D})$, where $\tilde{\rho}_k(\tilde{V},\hat{\tilde{V}}^{1D})$ is the correlation between European options $\tilde{V}$ and $\hat{\tilde{V}}^{1D}$, a direct analogue of $\tilde{\rho}_k(V_{k+1},\hat{V}_{k+1}^{1D})$ in the previous problem. After some simple algebra and making use of Eqn.\ref{eqn:varReductionXi}, we have

\begin{equation}\label{eqn:CVAerrorAnalysisRes}
\begin{split}
\epsilon^{CVA,P^{\text{FD-LSM}}}_k &=\sum^{k}_{i=1}\sqrt{\E[\left(\frac{\partial \Phi^{CVA}_k}{\partial \tilde{F}_i}\right)^2]}\cdot \sqrt{\frac{R-1}{R}[1-\tilde{\rho}^2_i(\tilde{V},\hat{\tilde{V}}^{1D})]}\cdot\sqrt{\Xi^{P^{\text{LSM}}}_i[\tilde{V}]}\\
&\leq\sum^{k}_{i=1}\sqrt{\E[\left(\frac{\partial \Phi^{CVA}_k}{\partial \tilde{F}_i}\right)^2]}\cdot \sqrt{\frac{R-1}{R}[1-\tilde{\rho}^2_{\tilde{m}^*}(\tilde{V},\hat{\tilde{V}}^{1D})]}\cdot\sqrt{\Xi^{P^{\text{LSM}}}_i[\tilde{V}]}\\
&=\sqrt{\frac{R-1}{R}[1-\tilde{\rho}^2_{\tilde{m}^*}(\tilde{V},\hat{\tilde{V}}^{1D})]}\epsilon^{CVA,P^{\text{LSM}}}_k.
\end{split}
\end{equation}

It is worth to note that the CVA error under LSM could be difficult to control, since value function is approximated directly; in the new method, by virtual of the 1D ansatz as a control variate, it provides an invaluable error reduction $\epsilon^{CVA,P^{\text{FD-LSM}}}_k\to0$ since $\tilde{\rho}_{\tilde{m}^*}(\tilde{V},\hat{\tilde{V}}^{1D})\to 1$, based on above Eqn.\ref{eqn:CVAerrorAnalysisRes}. 

In summary, Eqn.\ref{eqn:epsilonFDLSMUB} and \ref{eqn:CVAerrorAnalysisRes} are conclusive as a proof of Prop.\ref{prop:Err_FD_LSM} in the context of option PV and CVA, respectively.

\end{proof}

\subsubsection{Implementation Details}\label{sec:ImplementationDetails}

Dimension reduction into a 1D problem lies into the heart of the additional FD ansatz construction within FD-LSM.

At single-asset multi-factor level, the basic idea is to construct a 1D Markov process matching the marginal distribution of the original process, i.e., we take the conditional expectation of the instantaneous variance to construct the effective 1D local volatility:

\begin{equation}
\sigma^2_{FD}(S,t) = \E[\nu(t)|S(t)=S].
\end{equation} 
For Heston model in Eqn.\ref{eqn:stochvoldynamics} without spot dependence on variance, we have 
\[
\sigma_{FD}(t) = \sqrt{\E[\nu(t)]}=\sqrt{\nu(0)e^{-\kappa t} +\theta(1-e^{-\kappa t})}.
\]

Now we discuss the dimension reduction at multi-asset level. First of all, we can formally write down the effective 1D continuous dividend yield and local volatility as (see \cite{avellaneda2002reconstruction,gyongy1986mimicking})
\begin{equation}
\begin{split}
q_{FD}(X,t)&: =\E\left[\frac{1}{X}\sum_{i}q_i\cdot\frac{\partial S_{B(W)}}{\partial S_i}\cdot S_i\bigg|\sum_{i}\frac{\partial S_{B(W)}}{\partial S_i}\cdot S_i=X\right], \\
\sigma^2_{FD}(X,t)&:=\E\left[\frac{1}{X^2}\sum_{ij}\rho_{ij}\sigma_i(S_i, t)\sigma_j(S_j, t)\frac{\partial S_{B(W)}}{\partial S_i}\frac{\partial S_{B(W)}}{\partial S_j}S_iS_j\bigg|\sum_{i}\frac{\partial S_{B(W)}}{\partial S_i}\cdot S_i=X\right],
\end{split}
\end{equation}
 where $S_{B(W)}$ represents the spot of constant weighted basket (worst of basket) with subscript $B$($W$) and we omit the time dependence on $S_i$ and $S_j$ for notational convenience. 

For worst of basket where $\frac{\partial S_{W}}{\partial S_i}=\delta_{i,\arg_{j\in\{1,\ldots,d\}} \min S_j}$, observing that
\[
(q_{FD}(X,t), \sigma_{FD}(X,t))=(q_i,\sigma_i(S_i,t))\delta_{i,\arg_{j\in\{1,\ldots,d\}} \min S_j},
\]
we would first solve all set of uncorrelated 1D PDE on original underlyings individually to obtain corresponding FD solutions 
\[
\{f^{FD,i}_{k}(x)\text{ as solution of Eqn.\ref{eqn:pde1d} with } q_{FD}=q_i \text{ and } \sigma_{FD}(S,t)=\sigma_i(S,t) |1\le i \le d\}
\]
for given time step $t_k$, and then construct the ansatz as
\[
f^{FD,W}_{k}(\{S_i(t_k)\}) = \sum_i \delta_{i,i^*(\{S_i(t_k)\})} f^{FD,i}_{k}(\min S_i(t_k)),
\]
where $i^*(\{S_i(t_k)\})=\arg_{i\in\{1,\ldots,d\}} \min S_i(t_k)$ \footnote{In practice when the basis function is used, $i^*$ is determined pathwise based on simulated spots, and the multi-asset correlation effect is wired-in via statistical average over different trajectories.}.

For constant weighted basket with $X=S_B(t)$, there has been proposals to derive the effective local volatility approximately under the most general local volatility setting, see for example \cite{avellaneda2002reconstruction}. In the limit case of constant volatilities, i.e., Black-Scholes settings, we have below analytically tractable solution up to $2^{nd}$ order moment matching 

\begin{eqnarray*}
	\bar{q}&:=&-\frac{1}{T}\ln(\frac{1}{d}\sum_{i=1}^{d}e^{-q_iT}),\\
	\bar{\sigma}^2 &:=& \frac{1}{T} \ln \left(
	\frac{\frac{1}{d^2}\sum_{i,j}e^{(-q_i-q_j+\rho_{i,j}\sigma_i\sigma_i)T}}{(\frac{1}{d}\sum_{i=1}^{d}e^{-q_iT})^2}\right).
\end{eqnarray*}
Plugging these into Eqn.\ref{eqn:pde1d} as $q_{FD}=\bar{q}$ and $\sigma_{FD}=\bar{\sigma}$, we obtain the FD solution of continuation value $f^{FD,B}_{k}(x)$ after solving the PDE. 

In practice, to be used in the simulation, $f^{FD,B(W)}_{k}(x)$ is approximated as an interpolation function using a cubic spline on $V^{FD}(S, t^+_k)$ along the spatial direction in the PDE grid subjected to natural boundary condition; outside the PDE grid, a linear extrapolation is used instead. In the case of worst of basket, one needs to build all individual underlying cubic splines on $\{f^{FD,i}_{k}(x)|1\le i \le d\}$ for a given $t_k$; and then for a given pathwise input $\{S_i(t_k)\}$, we use the calculated $i^*$-th spline to do interpolation on $\min S_i(t_k)$ as spatial argument.

Finally, we can summarize the procedures of FD-LSM:
\begin{enumerate}
	\item Perform dimension reduction, from intra-asset (reducing factors within the same asset) to inter-asset (reducing factors across multi-asset), to construct the auxiliary 1D PDE up to $d$ assets. With a backward solver on each PDE, for each early exercise date $t_k$ within one sweep, we retrieve either $f^{FD,B}_{k}(x)$ for $X=S_B(t_k)$, or $\{f^{FD,i}_{k}(x)|1\le i \le d\}$ for $X=S_W(t_k)$.
	\item Construct the interpolation/extrapolation function proxy $\hat{f}^{FD,B(W)}_{k}(x)$ for $f^{FD,B(W)}_{k}(x)$ to be used in the regression and pricing stages.
	\item Follow the usual regression based procedures in LSM in Sec.\ref{sec:LeastSquaresMonteCarloMethod} to calculate the product PV using the basis functions (for each $t_k$) $\vec{\phi}_R^{P^\text{FD-LSM}}(x)$ concatenating $\{\hat{f}^{FD,B(W)}_{k}(x)\}$ and $\vec{\phi}_{R-1}^{P^\text{LSM}}(x)$.
\end{enumerate}

\section{Numerical Results}\label{sec:Numerical-Results} 
We performed our numerical experiments on a desktop, equipped with Intel(R) Xeon(R) E-2276G 3.80 GHz CPU, 6 cores/12 threads and 64GB RAM. As a starting point, we consider Bermudan option pricing under Black-Scholes in Sec.\ref{sec:Numerical-Results:BermudaOption1d} (\ref{sec:Numerical-Results:BermudaOption2d4d}) with $d=1(2,4)$ and under Heston Stochastic Volatility in Sec.\ref{sec:Numerical-Results:BermudaOptionHeston}, respectively. In Sec.\ref{sec:Numerical-Results:EPE4d}, we examine the stability of the new method in the example of EPE/CVA for European option with $d=4$ under Black-Scholes. Finally, we push the limit of multi-dimensional ($d$ up to 50) WIC pricing with a vast range of contractual and market parameters stressed under Black-Scholes and Local Volatility in Sec.\ref{sec:Numerical-Results:WIC5d}.

\subsection{Bermudan Option with $d=1$ under Black-Scholes}\label{sec:Numerical-Results:BermudaOption1d} 
In Eqn.\ref{eqn:localvoldynamics}, we use $r=3.96\%$ for numerical experiments in Bermudan options. Unless specified otherwise, the option maturity is set as $T=5$ with monthly exercise, i.e., $\Delta T=\frac{1}{12}$. In the simulation, we use $N_R=2^{13}$ and $N_P=2^{16}$ paths for regression and pricing stages, respectively, with weekly discretization steps. Sobol sequence is used in random number generation, which is shown as one of the most important variance reduction techniques (\cite{Areal2008}).

Now we present the results for the mono underlying case with $\sigma_1=30\%$ and $q_1=0$. From previous literature, this case has been discussed primarily on put options under LSM framework with relatively short tenor, e.g., $T\leq 2$ (\cite{Longstaff2001, Fabozzi2017}). In this study, we focus on both put and call options with longer maturity ($T=5$), such as to expose methodological limitations as much as possible. Especially for a call option on a non-dividend paying stock, it is not optimal to exercise the option before maturity due to its time value. Nevertheless, pricing American or Bermudan call options with exercise boundary at infinity, i.e., away from the ATM region, would be a very stringent test with respect to the accuracy and stability of the numerical algorithm.  

We will compare the option $PV$ under different approaches: PDE1D, LSM, $\text{LSM}^\text{CV}$, FD-LSM and $\text{FD-LSM}^\text{CV}$ as discussed in Sec.\ref{sec:Theoretical-Framework}. The superscript (CV) indicates the use of 1d analytic European option values as the control variates in pricing stage (\cite{Tian2003}).

In the light of the fact that PDE ansatz is approximated with cubic splines, we first compare FD-LSM using FD ansatz as main regressor only ($R=1$) against LSM using the same degree of highest polynomial power, i.e., $R=4$. The results are shown in Tab.\ref{tab:pv1d}, for the LSM ($R=4$) (with or without CV), the PV error against PDE1D benchmark is around 1\% and 2\%+ for put and call options, respectively; the CV toolset merely helps reduce the standard errors (s.e.), but having negligible impact on PV. On the other hand, under FD-LSM ($R=1$), PV error is reduced to no more than few basis points, much smaller than s.e.; also, the effect of variance reduction with CV is more prominent than LSM, giving rise to minimal s.e. among all simulation methods. Given the improvement due to CV is mainly on reducing s.e., we will not apply this technique anymore for $d>1$ where the analytic solution for European option is absent anyway. 

\begin{table}[htbp]
	\fontsize{8pt}{8pt}\selectfont	
	\centering
	\begin{tabular}{lrrrrrrrrrrrrrrr}
		\multicolumn{1}{c}{\multirow{2}[3]{*}{P/C}} & \multicolumn{1}{c}{\multirow{2}[3]{*}{$K$}} & \multicolumn{2}{c}{PDE1D} & \multicolumn{3}{c}{LSM ($R=4$)} & \multicolumn{3}{c}{FD-LSM ($R=1$)} & \multicolumn{3}{c}{$\text{LSM}^\text{CV}$ ($R=4$)} & \multicolumn{3}{c}{$\text{FD-LSM}^\text{CV}$ ($R=1$)} \\
		\cmidrule(rl){3-4}\cmidrule(rl){5-7}\cmidrule(rl){8-10}\cmidrule(rl){11-13}\cmidrule(rl){14-16}            &       & \multicolumn{1}{l}{PV} & \multicolumn{1}{l}{c.t.} & \multicolumn{1}{l}{Error} & \multicolumn{1}{l}{s.e.} & \multicolumn{1}{l}{c.t.} & \multicolumn{1}{l}{Error} & \multicolumn{1}{l}{s.e.} & \multicolumn{1}{l}{c.t.} & \multicolumn{1}{l}{Error} & \multicolumn{1}{l}{s.e.} & \multicolumn{1}{l}{c.t.} & \multicolumn{1}{l}{Error} & \multicolumn{1}{l}{s.e.} & \multicolumn{1}{l}{c.t.} \\
		\midrule
		P     & 100\% & 18.46\% & 0.03  & -1.08\% & 0.07\% & 1.7   & 0.05\% & 0.07\% & 2.0   & -1.08\% & 0.05\% & 1.8   & 0.05\% & 0.05\% & 2.0 \\
		P     & 80\%  & 9.58\% & 0.02  & -0.97\% & 0.05\% & 1.7   & 0.03\% & 0.05\% & 1.9   & -0.97\% & 0.03\% & 1.8   & 0.03\% & 0.03\% & 2.0 \\
		P     & 120\% & 30.19\% & 0.03  & -0.96\% & 0.09\% & 1.7   & 0.03\% & 0.09\% & 1.9   & -0.96\% & 0.07\% & 2.1   & 0.03\% & 0.06\% & 2.2 \\
		C     & 100\% & 33.85\% & 0.02  & -2.47\% & 0.19\% & 1.7   & -0.05\% & 0.25\% & 1.8   & -2.46\% & 0.12\% & 1.9   & -0.03\% & 0.06\% & 2.0 \\
		C     & 80\%  & 42.84\% & 0.03  & -2.29\% & 0.20\% & 1.7   & 0.02\% & 0.27\% & 1.8   & -2.28\% & 0.13\% & 1.8   & 0.04\% & 0.04\% & 2.0 \\
		C     & 120\% & 26.83\% & 0.03  & -2.13\% & 0.18\% & 1.6   & -0.04\% & 0.23\% & 1.8   & -2.12\% & 0.11\% & 1.8   & -0.02\% & 0.05\% & 1.8 \\
	\end{tabular}%
	\caption{For call (C) and put (P) options with different strikes ($K$), we calculate the $PV$ under various methods as comparison. The PV Error is calculated against the PDE1D benchmark. We also show the calculation time (c.t.) in seconds. }
\label{tab:pv1d}%
\end{table}%

\begin{figure}[ht]
	\begin{varwidth}{\linewidth}
		\centerline{\includegraphics[width=0.8\textwidth]{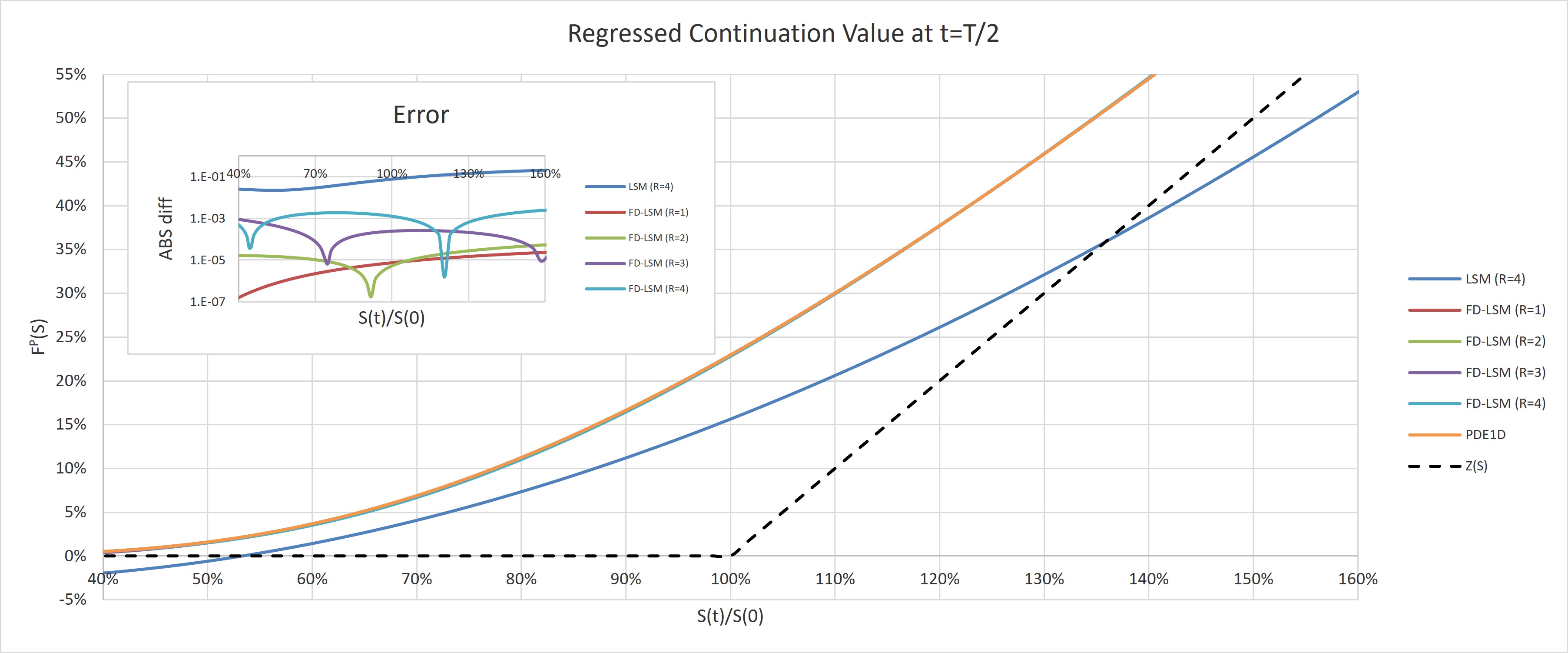}}
		\caption{The continuation value after regression as a function of underlying spot $S$ at $t=\frac{T}{2}$, i.e., $k=30$, for a call option striking at $K=100\%$. The dash line depicts the exercise payoff $Z(S)$ as reference. The inner figure on the upper left shows the absolute errors of various regression methods against $f^{FD}_{k=30}(S)$ in PDE1D.}
		\label{fig:continuedvalue1D}
	\end{varwidth}
\end{figure}

In retrospect of large pricing error in LSM for call options as seen in Tab.\ref{tab:pv1d}, we plot the regressed continuation value $F^P_{k=30}(S)$ for both LSM ($R=4$) and FD-LSM ($R=1\ldots 4$) in Fig.\ref{fig:continuedvalue1D} at $t_k=\frac{T}{2}$. Note that the FD result $f^{FD}_{k=30}(S)$ is also shown as benchmark, and the inner figure shows the absolute differences of various regression methods against this benchmark. First of all, for $R=1$, using the FD solution as the only regressor, the regressed value does converge to this basis function within numerical errors between $10^{-7}$ and $10^{-5}$. Secondly, FD-LSM shows excellent agreement against benchmark, with maximum absolute differences $\sim$ 10 basis points even with $R=4$ as monomials regressors are introduced as correction.

However, the traditional LSM is showing poor agreement with discrepancy up to $10\%+$ for high spot; for low spots below 55\% and high spots above 135\%, the fitted curve is even below the exercise payoff $Z(S)$ and thus arbitragable, which mistakenly instructs the option holder to exercise immediately. Apart from mis-fitting away from ATM, another cause of poor results is that at $t_{k=30}=\frac{T}{2}$ the numerical errors have been accumulated after 30 LSM iterations from maturity. Nevertheless, FD-LSM doesn't suffer from these two issues thanks to the use of the FD solution as the ansatz to be regressed on.

\begin{figure}[ht]
	\begin{varwidth}{\linewidth}
		\centerline{\includegraphics[width=0.7\textwidth]{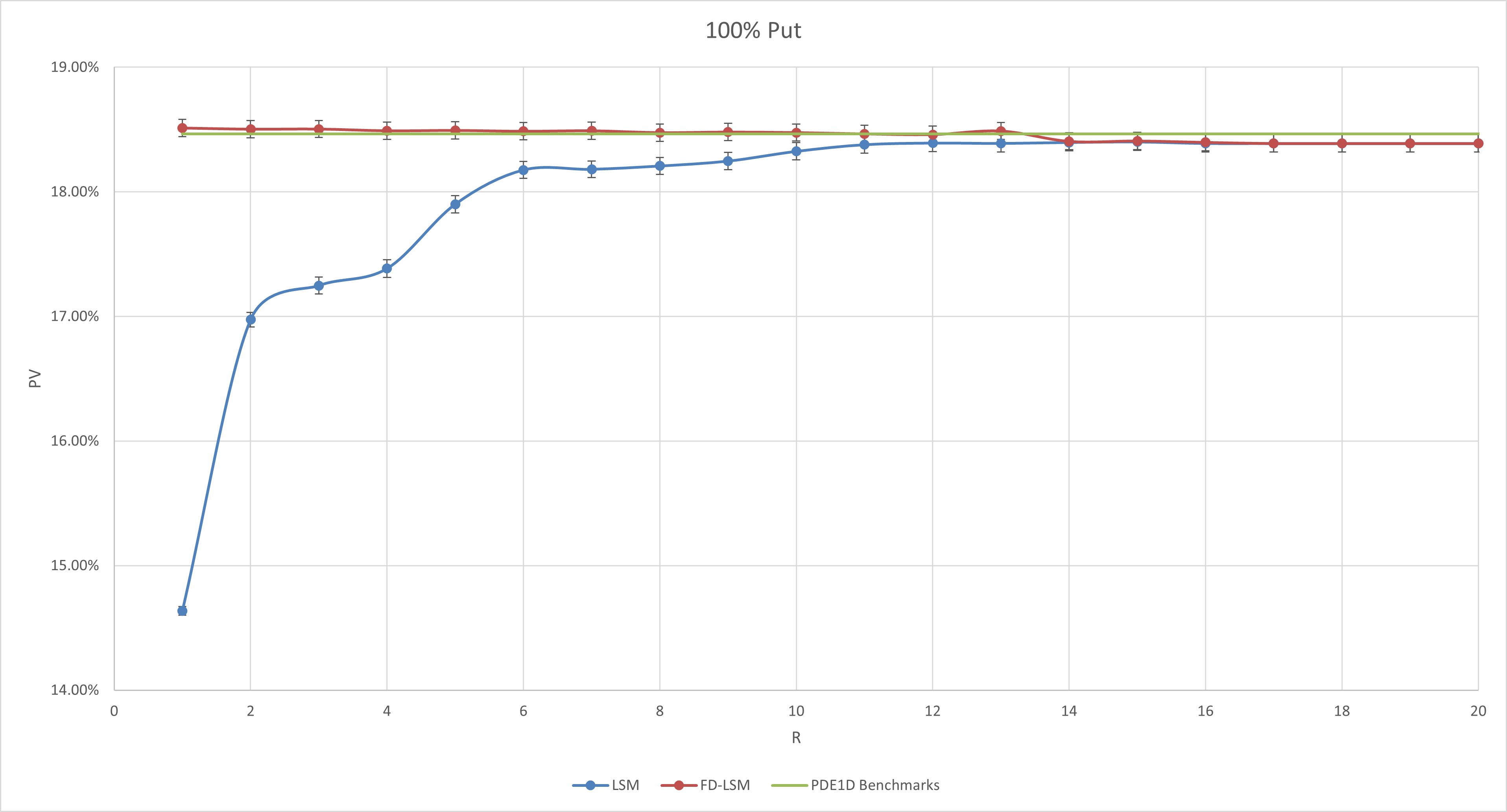}}
		\centerline{\includegraphics[width=0.7\textwidth]{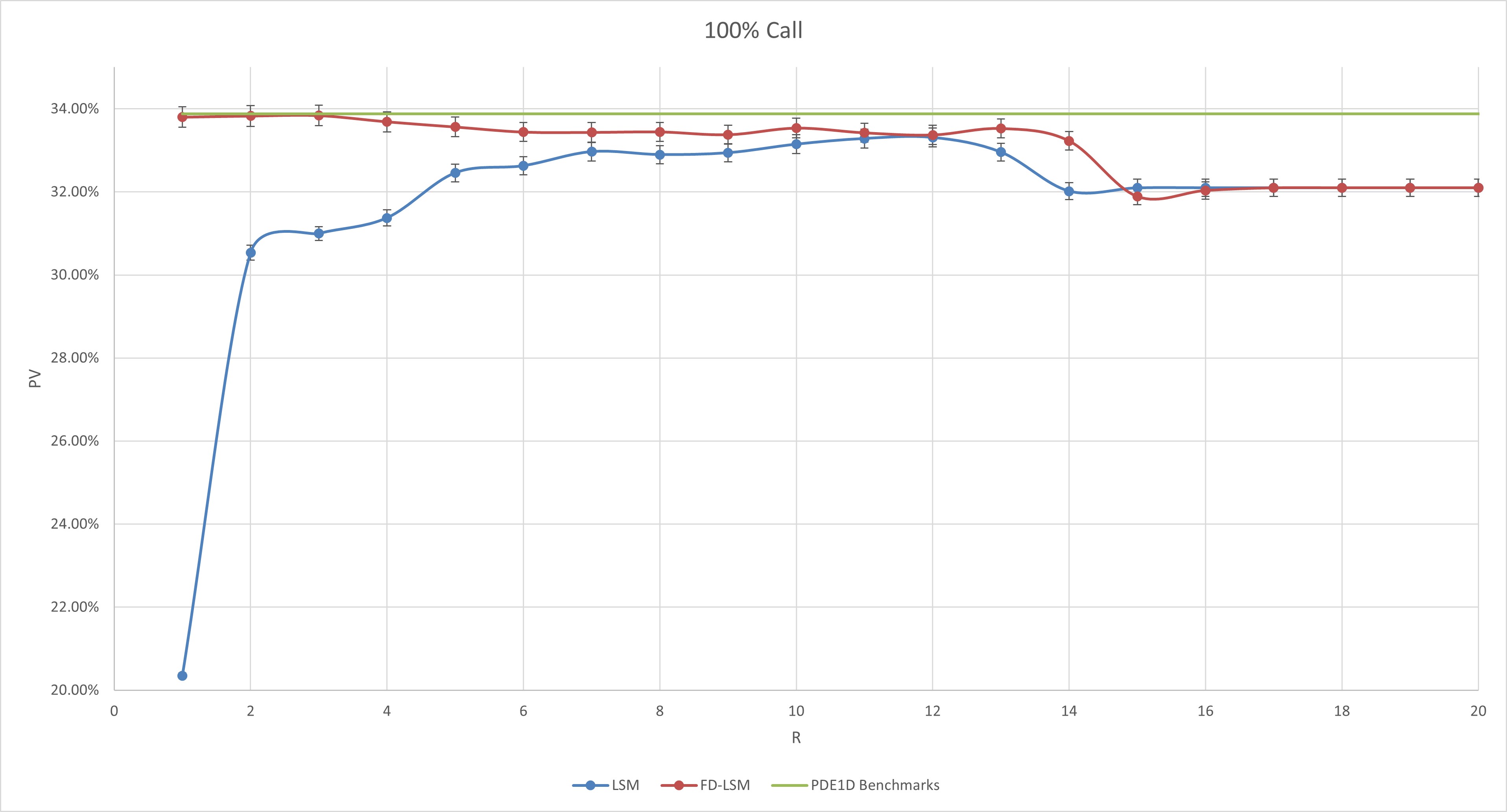}}
		\caption{Option PV with increasing cutoff $R$ for Bermudan put (upper panel) and call (lower panel) options, respectively. The s.e. is shown as error bar in the graph.}
		\label{fig:CallPutR}
	\end{varwidth}
\end{figure}

As a further examination on the stability of the regression method, we show the $PV$ with arbitrary cutoff $R$ in Fig.\ref{fig:CallPutR}. Moving from low number of $R$ to medium level, e.g., $R=10$, we can see a significant improvement in terms of accuracy in LSM, where the $PV$ is more or less monotonically converging towards benchmark, largely and qualitatively consistent with previous literature (\cite{Fabozzi2017}) focusing on put options, except the minor deterioration from $R=7$ to $R=8$ in the call option case; However, beyond $R=12$, the results deviate clearly away from benchmark in the call case, which is a clear sign of overfitting. On the other hand, for FD-LSM, the PV error against benchmark is bounded at 50 bps even up to $R=14$ for the call case. Although in this example the new regression method is exhibiting amazing robustness against overfitting, it is inadvisable to perform linear regression with high order of monomial terms in practice. From now on, we restrict and cap the cutoff to 4, unless specified otherwise.

\subsection{Bermudan Option with $d=2$ and $d=4$ under Black-Scholes}\label{sec:Numerical-Results:BermudaOption2d4d} 
Next, we further look at the case of $d=2$, where $\sigma_1=\sigma_2=30\%$ and $q_1=q_2=0$. For a fair comparison between FD-LSM and LSM, we use the same number of regressors with $R=4$ fixed: LSM makes use of full monomials up to the cubic term, whereas FD-LSM only uses perturbational monomials up to the quadratic term, apart form the main FD regressor. Looking at PV results as seen in Tab.\ref{tab:pv2d}, the maximum of difference in FD-LSM against FD PDE2D benchmark \footnote{We use alternating direction implicit method (\cite{McKee1970}) in PDE2D.} is merely 17 bps, whereas the counterpart in LSM is more than 2\%, well beyond the simulation standard errors. Also, the computational overhead (single-digit in percentage) to achieve such accuracy is moderate as shown in the c.t. comparison against LSM. The PDE2D solver's c.t, approximately 25\% of that required by LSM, is non-negligible compared to the 1D calculations. This also justifies our choice of using the 1D PDE as the ansatz. Last but not least, the Optimized Exercise Boundaries method (Opt-EB), as mentioned in Appendix~\ref{sec:OptimizedExerciseBoundaries}, can provide a valid independent check, although it tends to slightly underprice the option due to partial convergence in the external optimization.

Moving to $d=4$, where $\sigma_1=\sigma_2=30\%$, $\sigma_3=\sigma_4=20\%$, and $q_1=q_2=q_3=q_4=0$, The PV's and c.t.'s under each various methods are shown in Tab.\ref{tab:pv4d}. Different from low dimensions, full FD PDE result is no longer available; fortunately, we still have Opt-EB as an independent lower bound estimate.

Similar to lower dimensions, the new FD-LSM exhibits excellent agreement against benchmark consistently in terms of PV difference, outperforming traditional LSM for various put/call and $\rho$ settings: the maximum of differences between LSM and Opt-EB are 0.62\% and 1.56\% for put and call, respectively, and LSM underprices all test cases; whereas FD-LSM outperforms Opt-EB by up to 20 bps for put options and only underprices call options by 5 bps as maximum. For computational efficiency, as we can see from the c.t. columns between the two methods, the overhead due to additional PDE1D solver and cubic spline interpolation is insignificant. 

\begin{table}[htbp]
	\fontsize{9pt}{9pt}\selectfont	
	\centering
	\begin{tabular}{cclrlllrlllrll}
		\multirow{2}[3]{*}{P/C} & \multirow{2}[3]{*}{$\rho$} & \multicolumn{2}{c}{PDE2D} & \multicolumn{4}{c}{LSM ($R$=4)}  & \multicolumn{4}{c}{FD-LSM ($R$=4)} & \multicolumn{2}{c}{Opt-EB} \\
		\cmidrule(rl){3-4}   \cmidrule(rl){5-8} \cmidrule(rl){9-12} \cmidrule(rl){13-14}      &       & PV    & c.t. & PV    & s.e.  & Diff  & c.t. & PV    & s.e.  & Diff  & c.t. & PV    & Diff \\
		\midrule
		P     & 90\%  & 17.94\% & 0.5   & 17.02\% & 0.07\% & -0.91\% & 2.5   & 17.94\% & 0.07\% & 0.01\% & 2.8   & 17.85\% & -0.09\% \\
		P     & 50\%  & 15.59\% & 0.4   & 14.81\% & 0.06\% & -0.77\% & 2.6   & 15.56\% & 0.06\% & -0.03\% & 2.8   & 15.48\% & -0.11\% \\
		P     & 10\%  & 13.06\% & 0.4   & 12.42\% & 0.05\% & -0.64\% & 2.5   & 13.05\% & 0.05\% & -0.01\% & 2.7   & 12.92\% & -0.14\% \\
		C     & 90\%  & 33.35\% & 0.4   & 31.22\% & 0.19\% & -2.13\% & 2.4   & 33.17\% & 0.23\% & -0.17\% & 2.7   & 33.32\% & -0.03\% \\
		C     & 50\%  & 31.10\% & 0.5   & 29.10\% & 0.17\% & -2.00\% & 2.5   & 30.98\% & 0.21\% & -0.12\% & 2.9   & 31.08\% & -0.02\% \\
		C     & 10\%  & 28.72\% & 0.5   & 26.98\% & 0.15\% & -1.74\% & 2.6   & 28.56\% & 0.18\% & -0.16\% & 2.9   & 28.71\% & -0.01\% \\
	\end{tabular}%
		\caption{For call (C) and put (P) options striking at $K=100\%$ with different correlation $\rho$, we calculate the $PV$ under various methods as comparison. For LSM and FD-LSM, we also show the s.e. in the simulation as reference. The PV  difference (diff) is calculated between LSM/FD-LSM/Opt-EB and the PDE2D benchmark. For PDE2D/LSM/FD-LSM, we also show the calculation time (c.t.) in seconds. }
	\label{tab:pv2d}%
\end{table}%

\begin{table}[htbp]
	\fontsize{9pt}{9pt}\selectfont	
	\centering
	\begin{tabular}{lrrrrrrrrrr}
		\multicolumn{1}{c}{\multirow{2}[3]{*}{PC}} & \multicolumn{1}{c}{\multirow{2}[3]{*}{$\rho$}} & \multicolumn{1}{c}{\multirow{2}[3]{*}{Opt-EB}} & \multicolumn{4}{c}{LSM ($R$=4)}        & \multicolumn{4}{c}{FD-LSM ($R$=4)} \\
		\cmidrule(rl){4-7}    \cmidrule(rl){8-11}       &       &       & \multicolumn{1}{l}{PV} & \multicolumn{1}{l}{s.e.} & \multicolumn{1}{l}{Diff} & \multicolumn{1}{l}{c.t.} & \multicolumn{1}{l}{PV} & \multicolumn{1}{l}{s.e.} & \multicolumn{1}{l}{Diff} & \multicolumn{1}{l}{c.t.} \\
		\midrule
		P     & 90\%  & 13.80\% & 13.18\% & 0.06\% & -0.62\% & 4.0   & 13.84\% & 0.05\% & 0.04\% & 4.5 \\
		P     & 50\%  & 10.68\% & 10.23\% & 0.05\% & -0.45\% & 4.0   & 10.77\% & 0.04\% & 0.09\% & 4.4 \\
		P     & 10\%  & 6.86\% & 6.57\% & 0.03\% & -0.29\% & 4.2   & 7.06\% & 0.03\% & 0.20\% & 4.7 \\
		C     & 90\%  & 29.30\% & 27.74\% & 0.16\% & -1.56\% & 4.5   & 29.29\% & 0.19\% & -0.01\% & 4.6 \\
		C     & 50\%  & 26.33\% & 24.92\% & 0.14\% & -1.41\% & 4.1   & 26.28\% & 0.15\% & -0.05\% & 4.6 \\
		C     & 10\%  & 22.91\% & 21.56\% & 0.10\% & -1.35\% & 4.2   & 22.87\% & 0.12\% & -0.04\% & 4.5 \\
	\end{tabular}%
	\caption{For call (C) and put (P) options striking at $K=100\%$ with different correlation $\rho$, we calculate the $PV$ various methods as comparison. For LSM and FD-LSM, we also show the s.e. and c.t., respectively. The PV difference (diff) is calculated between LSM/FD-LSM and Opt-EB.}
\label{tab:pv4d}%
\end{table}%

\subsection{Bermudan Option under Heston Stochastic Volatility}\label{sec:Numerical-Results:BermudaOptionHeston} 
Here we consider Bermudan option pricing under Heston model as mentioned in Sec.~\ref{sec:ModelSettings-Heston}. For Heston parameters, we use $r=0.02$, $q=0$, $\nu(0)=0.15$, $\kappa=5$, $\theta=0.16$, $\xi=0.9$ and $\rho^{S,\nu}=0.1$. We consider a monthly exercisable put option with $T=1$ and $K=1$; $N_R=2^{14}$ and $N_P=2^{18}$ are used in the simulation. In LSM, we use monomials regressors up to cubic terms, i.e., $\{S^m\nu^n|m+n\leq3, n\in\mathbb{N}_0, m\in\mathbb{N}_0\}$ and in FD-LSM, up to quadratic terms are used. We compare LSM and FD-LSM with the method ``LSMC-PDE'' proposed by \cite{farahany2020mixing} using the same settings. As we can see in Tab.~\ref{tab:pv1dHeston}, both FD-LSM and LSMC-PDE are returning PVs closer to PDE2D benchmark than classical LSM. 

One interesting observation is that LSMC-PDE seems to be biased high compared to the benchmark, whereas LSM and FD-LSM are biased low. As mentioned before, LSMC-PDE is essentially applying the least-squares estimator on the value function as an extension of \cite{tsitsiklis2001regression}, whereas LSM and FD-LSM apply the approximation in the stopping time determination only following \cite{Longstaff2001}. Previous study by \cite{stentoft2014value} has concluded that the latter achieves a smaller absolute low bias with sub-optimality and less error accumulation compared to the former, based on a thorough comparison from both theoretical and numerical perspectives. 

\begin{table}[htbp]
	\fontsize{8pt}{8pt}\selectfont	
	\centering
	\begin{tabular}{rrrrrrrrrrrrr}
		\multicolumn{3}{c}{LSM (cubic)} & \multicolumn{3}{c}{FD-LSM (quadratic)} & \multicolumn{3}{c}{LSMC-PDE} &  \multicolumn{1}{c}{PDE2D} \\
		\cmidrule(rl){1-3}\cmidrule(rl){4-6}\cmidrule(rl){7-9}\cmidrule(rl){10-10} \multicolumn{1}{c}{PV} & \multicolumn{1}{l}{s.e.} & \multicolumn{1}{l}{c.t.} & \multicolumn{1}{l}{PV} & \multicolumn{1}{l}{s.e.} & \multicolumn{1}{l}{c.t.} & \multicolumn{1}{l}{PV} & \multicolumn{1}{l}{s.e.} & \multicolumn{1}{l}{c.t.} & \multicolumn{1}{c}{PV} \\
		\midrule
     14.41\% & 0.03\% & 2.9   & 14.49\% & 0.03\% & 3.0   & 14.53\% & 0.01\% & 7.0 & 14.51\% \\
	\end{tabular}%
	\caption{For ATM Bermudan put option, we calculate the $PV$ under various methods as comparison. We also show the standard error (s.e.) and calculation time (c.t.) in seconds. The results of ``LSMC-PDE'' and ``PDE2D'' are taken from Tab.9 and Tab.10 from \cite{farahany2020mixing}.}
	\label{tab:pv1dHeston}%
\end{table}%

\subsection{EPE/CVA for European option in $d=4$ under Black-Scholes}\label{sec:Numerical-Results:EPE4d} 

In this section, we evaluate the performance of FD-LSM in the context of EPE/CVA calculation for European options. Assuming we (the party) are long an uncollateralized European call option issued by the counterparty, we apply the same market parameters as Sec.\ref{sec:Numerical-Results:BermudaOption2d4d} with $d=4$. The option is struck at 100\% with $T=5$, and the EPE is evaluated with monthly discretized steps. 

Since the portfolio has only one positive payoff to the party, it follows that its discounted option value under $\Q$ measure is a martingale, hence its expected value, i.e. EPE*, is constant over time:
\begin{equation*}
EPE^*(t)=\E[e^{-rt}\tilde{F}(t)] = V_0,
\end{equation*}
where $V_0$ is the European option value at $t=0$ that can be computed exactly with usual simulation approach as benchmark.

\begin{figure}[ht]
	\begin{varwidth}{\linewidth}
		\centerline{\includegraphics[width=0.7\textwidth]{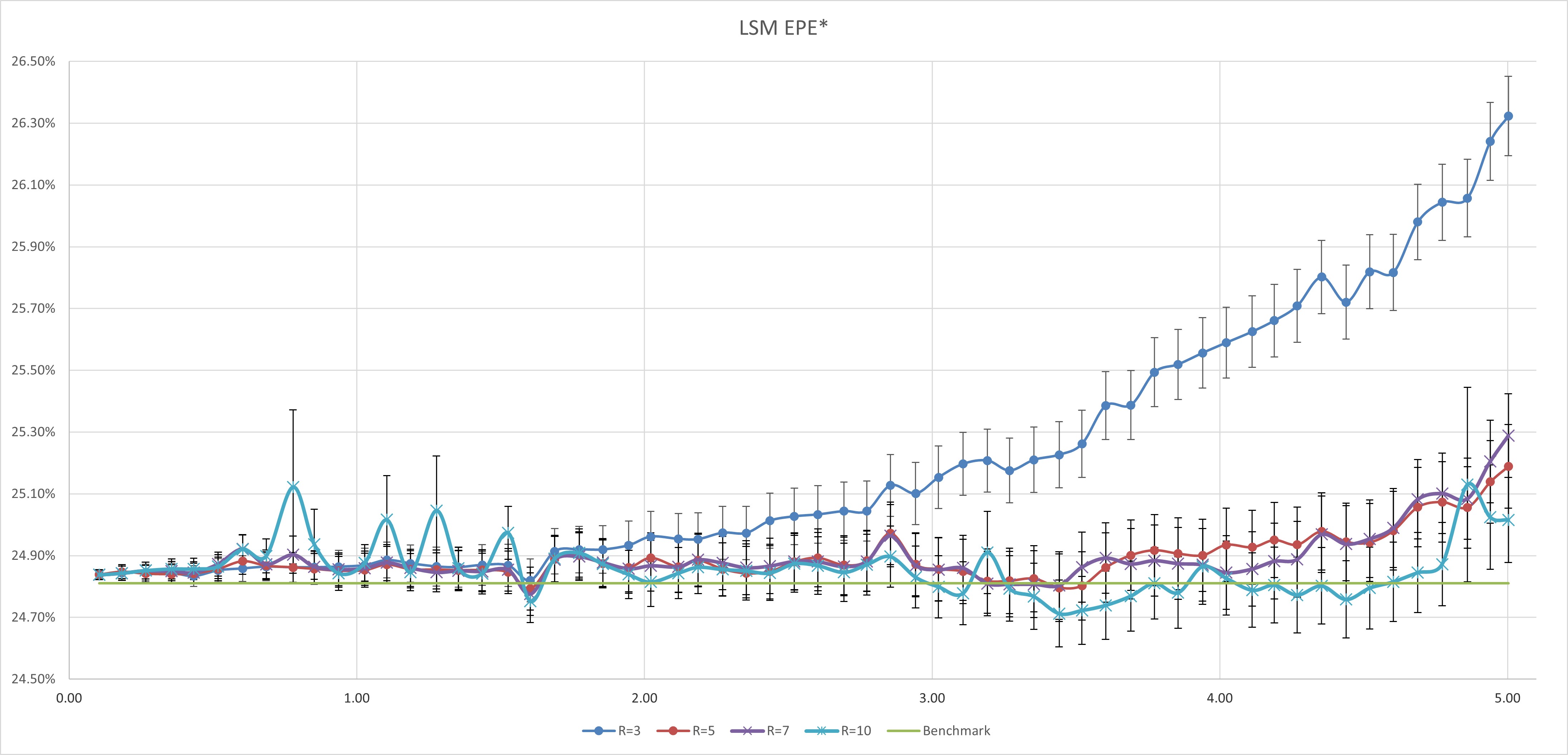}}
		\centerline{\includegraphics[width=0.7\textwidth]{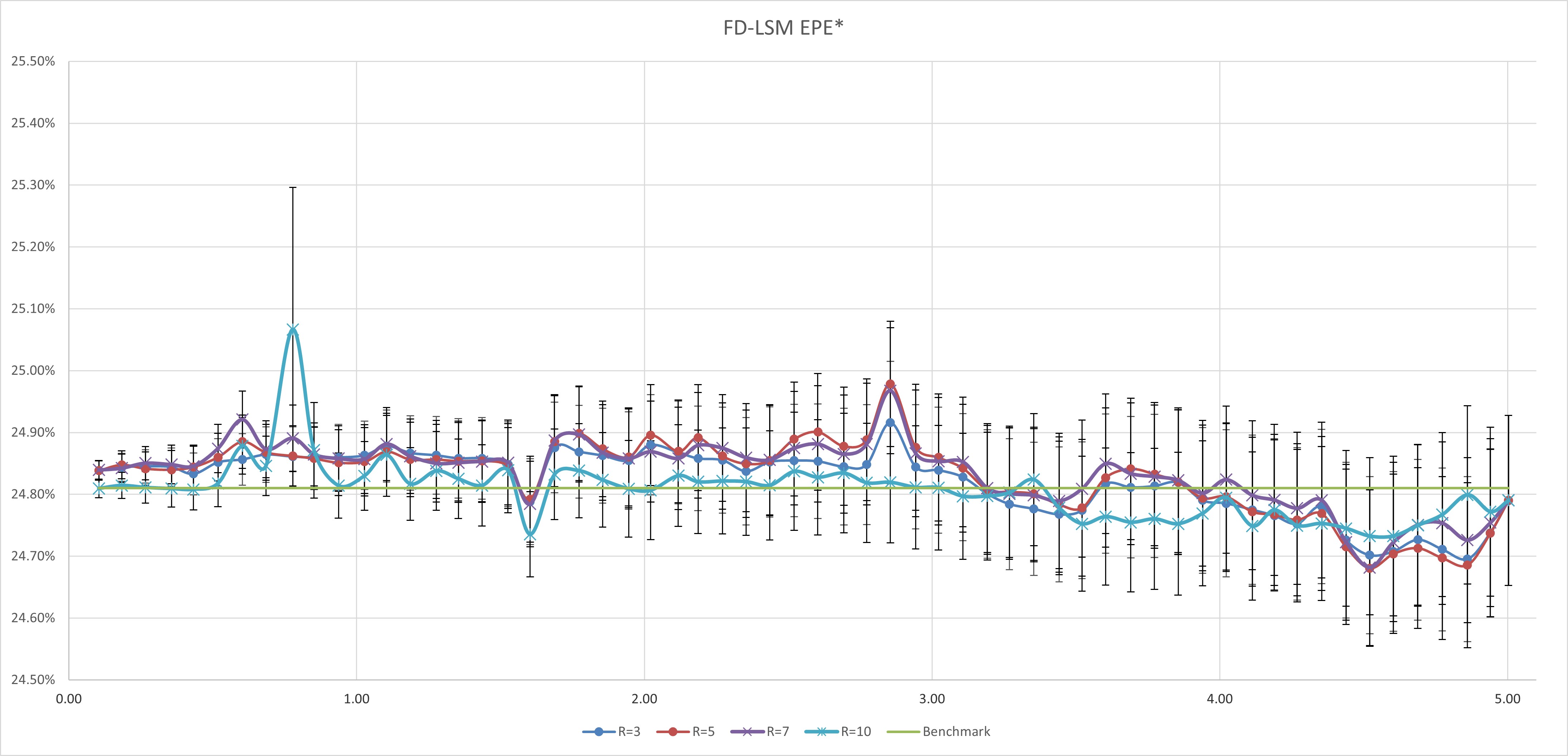}}
		\caption{EPE* profiles with various cutoff $R$ as a function of time, calculated using LSM (upper panel) and FD-LSM (lower panel), respectively. The s.e. is shown as error bar in the graph.}
		\label{fig:EPE4d}
	\end{varwidth}
\end{figure}

First, we inspect the EPE* profile, which is relevant to the CVA without considering Wrong Way Risk (WWR). In Fig.\ref{fig:EPE4d}, we plot the this profile using various cutoff $R$ under LSM and FD-LSM, where the s.e. is shown as error bar for indication. For LSM, EPE* deviates away from benchmark by up to $\sim$10 times of s.e. especially near the terminal end, with low level of $R=3$; by increasing $R$, the discrepancy is reduced in the longer end; however, at the same time, in the short end, there are spikes with abnormally sizable error bar arising around $t=1$ for $R=10$, indicting some kind of numerical instability. 

On the other hand, for FD-LSM, we see very accurate agreement between calculated EPE* against theoretical benchmark for various $R$ in general. It is worth noticing that even with $R=10$, the profile is much smoother with less angularities in terms of both frequency and intensity, than the LSM counterpart. Nevertheless, in both regression methods, the case with $R=10$ exhibits noticeable instability due to overfitting, as mentioned and shown in Bermuda option pricing previously; thus in practice, such high number of cutoff should be avoided for stability.

Now we continue to evaluate the total CVA using Eqn.\ref{eqn:cvaMC}. 

To further take into account the WWR, we consider a hazard rate with below functional dependence on the option mark-to-market value $\tilde{F}(\omega,t)$(\cite{hull2012cva}):
\[
h_{cpty}(\tilde{F}(\omega,t),t) = \ln[1+e^{a_{WWR}(t)+b_{WWR}\tilde{F}(\omega,t)}],
\]
where $a_{WWR}(t)$ and $b_{WWR}$ are WWR model parameters.

As numerical settings, we use $R_{cpty}=0$; for the hazard rate, $a_{WWR}(t)=-4$ and $b_{WWR}=0.1$ are used, which are assumed to have been calibrated to counterparty credit spread using exact mark-to-market option values, and thus are independent of least-squares approximation. As seen in Tab.~\ref{tab:CVA4d}, as $R$ increases, CVA under LSM is exhibiting unstable move especially for $R\geq 10$, and the s.e. is showing explosive values up to 78 bps ($R=11$). By contrast, FD-LSM generates much more stable CVA all the way up to very high cutoff with a maximum s.e. capped at 27 bps ($R=11$).

\begin{table}[htbp]
	\centering
	\begin{tabular}{rrrrr}
		\multicolumn{1}{c}{\multirow{2}[3]{*}{$R$}} & \multicolumn{2}{c}{LSM} & \multicolumn{2}{c}{FD-LSM} \\
		\cmidrule(rl){2-3}\cmidrule(rl){4-5}          & \multicolumn{1}{l}{CVA} & \multicolumn{1}{l}{s.e.} & \multicolumn{1}{l}{CVA} & \multicolumn{1}{l}{s.e.} \\
		\midrule
		3     & 2.30\% & 0.01\% & 2.27\% & 0.01\% \\
		4     & 2.28\% & 0.01\% & 2.27\% & 0.01\% \\
		5     & 2.27\% & 0.01\% & 2.27\% & 0.01\% \\
		6     & 2.27\% & 0.01\% & 2.27\% & 0.01\% \\
		7     & 2.27\% & 0.01\% & 2.27\% & 0.01\% \\
		8     & 2.29\% & 0.03\% & 2.28\% & 0.02\% \\
		9     & 2.36\% & 0.10\% & 2.36\% & 0.10\% \\
		10    & 2.52\% & 0.26\% & 2.37\% & 0.11\% \\
		11    & 3.71\% & 0.78\% & 2.54\% & 0.27\% \\
	\end{tabular}%
	\caption{The calculated CVA between LSM and FD-LSM for various cutoff $R$ as comparison. We also show the standard error (s.e.).}
	\label{tab:CVA4d}%
\end{table}%

In the context of EPE and CVA calculation, we not only further showcased the accuracy of FD-LSM over LSM, but also demonstrated the stability of the new method over time direction, i.e., time span between the regression variables and realized cash flows.

\subsection{WIC from $d=5$ to $d=50$ under Black-Scholes and Local Volatility}\label{sec:Numerical-Results:WIC5d} 
As a finial numerical example, we will look at a WIC with 5-50 underlyings in the basket. For contractual parameters, we use coupon barrier $B_c=70\%$, knock-in barrier $B_P=50\%$ and strike $K=100\%$. Unless specified otherwise, underlying market data used are shown in Tab.\ref{tab:icmktdata5d}, where we consider both Black-Scholes and Local Volatility settings, respectively. In the simulation, we use $N_R=2^{13}$ and $N_P=2^{17}$ paths for regression and pricing, respectively, with weekly discretization steps. 

\begin{table}[htbp]
	\fontsize{9pt}{9pt}\selectfont	
	\centering
	\begin{tabular}{|l|r|r|r|r|r|}
		\hline
		Stock tag $i$ with $j\in \mathbb{N}_0$ & $1 + 5j$     & $2 + 5j$     & $3 + 5j$     & $4 + 5j$     & $5 + 5j$ \\ \hline
		Dividend rate $q_i$ & 3\%   & 2\%   & 5\%   & 0\%   & 4\% \\ \hline
		Black-Scholes Volatility $\sigma_i^{BS}$ & 20\%  & 30\%  & 25\%  & 24\%  & 15\% \\ \hline
		Local Volatility $\sigma^{LV}_i(S_i(t),t)$ & \multicolumn{5}{c|}{$\sigma_i^{BS}e^{-0.05\sqrt{t}}\left(1.5-e^{-0.1t-5e^{-0.05t}\ln^2\frac{S_i(t)}{S(0)}}\right)$} \\ \hline
	\end{tabular}%
	\caption{Underlying market data.}
	\label{tab:icmktdata5d}%
\end{table}%
\textbf{Black-Scholes setting}: As presented in Tab.\ref{tab:icpv5d}, we show the PV and expected life $\E[\tau]$ between LSM and FD-LSM on WIC for various underlying market data and contractual parameters. In $(1+14n)$-th ($n=0,1,2,3$) rows of Tab.\ref{tab:icpv5d} the same parameters as \cite{Liang2020} in their Tab. 5.5 are used and our LSM results are perfectly in-line\footnote{We do not suffer from any memory issues up to 50 dimension under x64 system as experienced by Liang et al.}, although FD-LSM gives marginally lower PV. However, in these cases, the coupon rate $c=20\%$ is too high, in particular compared to $r=1\%$, and thus the issuer will almost surely call the product on 1-st exercise date as indicated by $\E[\tau]=0.29$ for $d=5$. If we stress $c$ to $1\%$ in $(2+14n)$-th rows of Tab.\ref{tab:icpv5d}, the FD-LSM gives much lower PV than LSM by up to 3.3 times of s.e.. Moreover, if we change to $\rho=90\%,c=1\%,r=5\%$, the difference ratio becomes 10+ as seen in $(3+14n)$-th rows, which is significant for a short dated option with few exercise dates. 
As a more comprehensive experiment to examine the accuracy of regression schemes, we use a much wider range of parameter sets by varying $\rho,c,r,T,\Delta T$ as well as the dimension $d$ (number of assets) in Tab.\ref{tab:icpv5d}. Since the optimal stopping strategy is a result of equilibrium due to competition among all these factors, with a wide range of parameter sets, it is expected to drive the change of exercise boundary, generating various testing scenarios. Going through all rows in Tab.\ref{tab:icpv5d}, FD-LSM always outperforms LSM by achieving a lower PV, and thus better results due to the nature of minimization problem using the same simulation paths, with difference ratio ranging from 0.2 to 28.8; and the outperformance is consistent from moderate ($d=5$) to high ($d=50$) dimensions. In practice, apart from achieving more accurate PV, we expect that the new method would naturally suppress the instability due to hopping between sub-optimal states in the scheme of bumping market factors and re-evaluation for greeks calculation. 

\clearpage
\begin{table}
	\fontsize{9pt}{9pt}\selectfont
	\centering
	\begin{tabular}{cccccccrrrrrrcc}
		\multirow{2}[3]{*}{\#} & \multirow{2}[3]{*}{$d$} & \multirow{2}[3]{*}{$\rho$} & \multirow{2}[3]{*}{$T$} & \multirow{2}[3]{*}{$\frac{1}{\Delta T}$} & \multirow{2}[3]{*}{$c$} & \multirow{2}[3]{*}{$r$} & \multicolumn{3}{c}{LSM ($R$=4)} & \multicolumn{3}{c}{FD-LSM ($R$=4)} & \multirow{2}[3]{*}{$\frac{\text{PV Diff}}{\text{s.e.}}$} & \multirow{2}[3]{*}{c.t. Diff\%} \\
		\cmidrule(rl){8-10}\cmidrule(rl){11-13}  &    &    &       &       &       &       & \multicolumn{1}{l}{PV(\%)} & \multicolumn{1}{l}{s.e.(\%)} & \multicolumn{1}{l}{$\E[\tau]$} & \multicolumn{1}{l}{PV(\%)} & \multicolumn{1}{l}{s.e.(\%)} & \multicolumn{1}{l}{$\E[\tau]$} & & \\
		\midrule
    1     & 5     & 30\%  & 1     & 4     & 20\%  & 1\%   & 104.43 & 0.01  & 0.29  & 104.43 & 0.01  & 0.29  & 0.4   & 8\% \\
2     & 5     & 30\%  & 1     & 4     & 1\%   & 1\%   & 98.57 & 0.02  & 0.80  & 98.49 & 0.02  & 0.99  & 3.3   & 5\% \\
3     & 5     & 90\%  & 1     & 4     & 1\%   & 5\%   & 95.56 & 0.02  & 0.96  & 95.39 & 0.02  & 1.01  & 10.3  & 6\% \\
4     & 5     & 30\%  & 5     & 4     & 1\%   & 5\%   & 63.59 & 0.07  & 4.79  & 62.88 & 0.07  & 5.00  & 10.2  & 2\% \\
5     & 5     & 90\%  & 5     & 4     & 1\%   & 5\%   & 71.33 & 0.06  & 4.64  & 70.16 & 0.06  & 5.00  & 18.8  & 5\% \\
6     & 5     & 30\%  & 10    & 4     & 1\%   & 5\%   & 41.30 & 0.07  & 9.76  & 40.65 & 0.06  & 10.01 & 9.8   & 3\% \\
7     & 5     & 90\%  & 10    & 4     & 1\%   & 5\%   & 51.80 & 0.07  & 9.43  & 50.30 & 0.06  & 10.01 & 22.7  & 4\% \\
8     & 5     & 30\%  & 1     & 12    & 20\%  & 1\%   & 101.54 & 0.01  & 0.13  & 101.52 & 0.01  & 0.14  & 1.7   & 5\% \\
9     & 5     & 30\%  & 1     & 12    & 1\%   & 1\%   & 98.17 & 0.02  & 0.82  & 98.01 & 0.02  & 1.01  & 6.9   & 6\% \\
10    & 5     & 90\%  & 1     & 12    & 1\%   & 5\%   & 95.11 & 0.02  & 0.95  & 94.89 & 0.02  & 1.01  & 13.7  & 6\% \\
11    & 5     & 30\%  & 5     & 12    & 1\%   & 5\%   & 62.06 & 0.07  & 4.81  & 61.35 & 0.07  & 5.00  & 10.4  & 4\% \\
12    & 5     & 90\%  & 5     & 12    & 1\%   & 5\%   & 69.39 & 0.06  & 4.67  & 68.16 & 0.06  & 5.00  & 20.3  & 4\% \\
13    & 5     & 30\%  & 10    & 12    & 1\%   & 5\%   & 39.16 & 0.06  & 9.78  & 38.46 & 0.06  & 10.00 & 11.0  & 6\% \\
14    & 5     & 90\%  & 10    & 12    & 1\%   & 5\%   & 48.74 & 0.06  & 9.47  & 47.15 & 0.06  & 10.01 & 25.3  & 4\% \\
\Xhline{0.01\arrayrulewidth}
    15    & 10    & 30\%  & 1     & 4     & 20\%  & 1\%   & 103.93 & 0.02  & 0.35  & 103.92 & 0.02  & 0.35  & 0.6   & 6\% \\
16    & 10    & 30\%  & 1     & 4     & 1\%   & 1\%   & 97.32 & 0.03  & 0.89  & 97.24 & 0.03  & 0.98  & 2.5   & 4\% \\
17    & 10    & 90\%  & 1     & 4     & 1\%   & 5\%   & 95.34 & 0.02  & 0.94  & 95.09 & 0.02  & 1.01  & 12.9  & 4\% \\
18    & 10    & 30\%  & 5     & 4     & 1\%   & 5\%   & 54.31 & 0.07  & 4.85  & 53.84 & 0.07  & 4.98  & 6.3   & 4\% \\
19    & 10    & 90\%  & 5     & 4     & 1\%   & 5\%   & 68.77 & 0.07  & 4.68  & 67.71 & 0.06  & 5.00  & 16.0  & 3\% \\
20    & 10    & 30\%  & 10    & 4     & 1\%   & 5\%   & 31.68 & 0.06  & 9.88  & 31.36 & 0.06  & 9.98  & 5.2   & 4\% \\
21    & 10    & 90\%  & 10    & 4     & 1\%   & 5\%   & 48.67 & 0.07  & 9.50  & 47.37 & 0.06  & 10.01 & 19.1  & 2\% \\
22    & 10    & 30\%  & 1     & 12    & 20\%  & 1\%   & 101.58 & 0.01  & 0.10  & 101.58 & 0.01  & 0.09  & 0.2   & 9\% \\
23    & 10    & 30\%  & 1     & 12    & 1\%   & 1\%   & 97.46 & 0.03  & 0.76  & 97.29 & 0.03  & 0.86  & 5.4   & 7\% \\
24    & 10    & 90\%  & 1     & 12    & 1\%   & 5\%   & 95.69 & 0.02  & 0.86  & 95.12 & 0.02  & 1.01  & 28.8  & 7\% \\
25    & 10    & 30\%  & 5     & 12    & 1\%   & 5\%   & 54.61 & 0.07  & 4.80  & 53.95 & 0.07  & 4.96  & 8.8   & 5\% \\
26    & 10    & 90\%  & 5     & 12    & 1\%   & 5\%   & 69.27 & 0.07  & 4.55  & 67.77 & 0.06  & 5.00  & 22.3  & 4\% \\
27    & 10    & 30\%  & 10    & 12    & 1\%   & 5\%   & 31.92 & 0.06  & 9.84  & 31.49 & 0.06  & 9.98  & 7.0   & 7\% \\
28    & 10    & 90\%  & 10    & 12    & 1\%   & 5\%   & 49.22 & 0.07  & 9.36  & 47.50 & 0.06  & 10.01 & 24.8  & 3\% \\
\Xhline{0.01\arrayrulewidth}
    29    & 20    & 30\%  & 1     & 4     & 20\%  & 1\%   & 102.43 & 0.04  & 0.50  & 102.42 & 0.04  & 0.50  & 0.2   & 5\% \\
30    & 20    & 30\%  & 1     & 4     & 1\%   & 1\%   & 95.13 & 0.04  & 0.92  & 95.07 & 0.04  & 0.94  & 1.3   & 4\% \\
31    & 20    & 90\%  & 1     & 4     & 1\%   & 5\%   & 95.00 & 0.02  & 0.94  & 94.73 & 0.02  & 1.01  & 11.8  & 2\% \\
32    & 20    & 30\%  & 5     & 4     & 1\%   & 5\%   & 44.53 & 0.07  & 4.93  & 44.35 & 0.07  & 4.98  & 2.5   & 5\% \\
33    & 20    & 90\%  & 5     & 4     & 1\%   & 5\%   & 66.17 & 0.07  & 4.73  & 65.30 & 0.07  & 5.00  & 12.5  & 3\% \\
34    & 20    & 30\%  & 10    & 4     & 1\%   & 5\%   & 23.45 & 0.05  & 9.96  & 23.39 & 0.05  & 9.98  & 1.3   & 1\% \\
35    & 20    & 90\%  & 10    & 4     & 1\%   & 5\%   & 45.66 & 0.07  & 9.62  & 44.66 & 0.07  & 10.01 & 14.4  & 4\% \\
36    & 20    & 30\%  & 1     & 12    & 20\%  & 1\%   & 101.31 & 0.02  & 0.17  & 101.29 & 0.02  & 0.17  & 0.9   & 7\% \\
37    & 20    & 30\%  & 1     & 12    & 1\%   & 1\%   & 95.29 & 0.04  & 0.84  & 95.20 & 0.04  & 0.87  & 2.1   & 3\% \\
38    & 20    & 90\%  & 1     & 12    & 1\%   & 5\%   & 95.39 & 0.02  & 0.85  & 94.77 & 0.02  & 1.01  & 27.2  & 4\% \\
39    & 20    & 30\%  & 5     & 12    & 1\%   & 5\%   & 44.79 & 0.07  & 4.90  & 44.48 & 0.07  & 4.96  & 4.3   & 4\% \\
40    & 20    & 90\%  & 5     & 12    & 1\%   & 5\%   & 66.64 & 0.07  & 4.62  & 65.38 & 0.07  & 5.00  & 17.9  & 6\% \\
41    & 20    & 30\%  & 10    & 12    & 1\%   & 5\%   & 23.63 & 0.05  & 9.95  & 23.55 & 0.05  & 9.97  & 1.6   & 2\% \\
42    & 20    & 90\%  & 10    & 12    & 1\%   & 5\%   & 45.99 & 0.07  & 9.54  & 44.76 & 0.07  & 10.01 & 17.6  & 7\% \\
\Xhline{0.01\arrayrulewidth}
    43    & 50    & 30\%  & 1     & 4     & 20\%  & 1\%   & 97.62 & 0.06  & 0.74  & 97.59 & 0.06  & 0.76  & 0.4   & 4\% \\
44    & 50    & 30\%  & 1     & 4     & 1\%   & 1\%   & 90.43 & 0.06  & 0.92  & 90.37 & 0.06  & 0.95  & 1.1   & 3\% \\
45    & 50    & 90\%  & 1     & 4     & 1\%   & 5\%   & 94.51 & 0.03  & 0.93  & 94.21 & 0.03  & 1.01  & 11.2  & 4\% \\
46    & 50    & 30\%  & 5     & 4     & 1\%   & 5\%   & 33.34 & 0.06  & 4.98  & 33.31 & 0.06  & 4.99  & 0.5   & 1\% \\
47    & 50    & 90\%  & 5     & 4     & 1\%   & 5\%   & 62.91 & 0.07  & 4.80  & 62.26 & 0.07  & 5.00  & 8.9   & 3\% \\
48    & 50    & 30\%  & 10    & 4     & 1\%   & 5\%   & 16.09 & 0.03  & 10.00 & 16.08 & 0.03  & 10.00 & 0.4   & 2\% \\
49    & 50    & 90\%  & 10    & 4     & 1\%   & 5\%   & 42.03 & 0.07  & 9.75  & 41.39 & 0.07  & 10.01 & 9.4   & 2\% \\
50    & 50    & 30\%  & 1     & 12    & 20\%  & 1\%   & 98.25 & 0.05  & 0.60  & 98.23 & 0.05  & 0.61  & 0.3   & 4\% \\
51    & 50    & 30\%  & 1     & 12    & 1\%   & 1\%   & 90.65 & 0.06  & 0.87  & 90.54 & 0.06  & 0.91  & 1.9   & 5\% \\
52    & 50    & 90\%  & 1     & 12    & 1\%   & 5\%   & 94.84 & 0.03  & 0.85  & 94.22 & 0.03  & 1.01  & 23.0  & 4\% \\
53    & 50    & 30\%  & 5     & 12    & 1\%   & 5\%   & 33.52 & 0.06  & 4.97  & 33.43 & 0.06  & 4.98  & 1.5   & 1\% \\
54    & 50    & 90\%  & 5     & 12    & 1\%   & 5\%   & 63.43 & 0.07  & 4.68  & 62.35 & 0.07  & 5.00  & 14.5  & 3\% \\
55    & 50    & 30\%  & 10    & 12    & 1\%   & 5\%   & 16.21 & 0.04  & 9.99  & 16.17 & 0.03  & 10.00 & 0.9   & 3\% \\
56    & 50    & 90\%  & 10    & 12    & 1\%   & 5\%   & 42.48 & 0.07  & 9.63  & 41.49 & 0.07  & 10.01 & 14.0  & 3\% \\
	\end{tabular}%
	\caption{PV and expected life $\E[\tau]$ of WIC under LSM and FD-LSM under various dimensions $d$, correlation $\rho$, tenor $T$ (in year), coupon rate $c$ and risk-free interest rate $r$. We also show the s.e. in the simulation as reference. In the last but one column, we show the ratio of PV difference (LSM less FD-LSM) against s.e. of LSM. And the last column shows the relative difference (FD-LSM less LSM in percentage of LSM) of computation time (c.t.). Black-Scholes volatilities are used in Tab.\ref{tab:icmktdata5d}. }
	\label{tab:icpv5d}%
\end{table}%

Finally, we also check the calculation time in last column of Tab.~\ref{tab:icpv5d}. In general, there is expected mild deterioration when switching to the new method, contributed by the additional PDE1D solvers and the cubic spline interpolation. However, the relative impacts are merely in single-digit range; thus we don't expect this amount of impact would become a ``showstopper'' when applying the new regression scheme for generic option pricing and risk management from practitioners' point of view.
\clearpage
\textbf{Local Volatility setting}: We repeat previous numerical experiment under a local volatility setting with
\[
\sigma^{LV}_i(S_i(t),t)=\sigma_i^{BS}e^{-0.05\sqrt{t}}\left(1.5-e^{-0.1t-5e^{-0.05t}\ln^2\frac{S_i(t)}{S(0)}}\right).
\]
This hypothetical functional form can generate a steep skew to mimic a stressed market condition. The results are presented in Tab.\ref{tab:icpv5dlocalvol}. We observe that in all test scenarios, FD-LSM outperforms LSM, with a maximum PV difference ratio reaching 32.6, compared to 28.8 under constant volatilities. In terms of calculation time, mild relative impacts are observed too. Therefore, we can conclude that FD-LSM is performing well even under a general multi-variate local volatility setting. 

\begin{table}[htbp]
	\fontsize{9pt}{9pt}\selectfont
	\centering
	\begin{tabular}{cccccccrrrrrrcc}
		\multirow{2}[3]{*}{\#} & \multirow{2}[3]{*}{$d$} & \multirow{2}[3]{*}{$\rho$} & \multirow{2}[3]{*}{$T$} & \multirow{2}[3]{*}{$\frac{1}{\Delta T}$} & \multirow{2}[3]{*}{$c$} & \multirow{2}[3]{*}{$r$} & \multicolumn{3}{c}{LSM ($R$=4)} & \multicolumn{3}{c}{FD-LSM ($R$=4)} & \multirow{2}[3]{*}{$\frac{\text{PV Diff}}{\text{s.e.}}$} & \multirow{2}[3]{*}{c.t. Diff\%} \\
		\cmidrule(rl){8-10}\cmidrule(rl){11-13}  &    &    &       &       &       &       & \multicolumn{1}{l}{PV(\%)} & \multicolumn{1}{l}{s.e.(\%)} & \multicolumn{1}{l}{$\E[\tau]$} & \multicolumn{1}{l}{PV(\%)} & \multicolumn{1}{l}{s.e.(\%)} & \multicolumn{1}{l}{$\E[\tau]$} & & \\
		\midrule
		1     & 5     & 30\%  & 1     & 4     & 20\%  & 1\%   & 104.77 & 0.00  & 0.26  & 104.77 & 0.00  & 0.29  & 0.0   & 6\% \\
		2     & 5     & 30\%  & 1     & 4     & 1\%   & 1\%   & 99.84 & 0.01  & 0.64  & 99.81 & 0.01  & 0.99  & 3.6   & 9\% \\
		3     & 5     & 90\%  & 1     & 4     & 1\%   & 5\%   & 96.06 & 0.01  & 1.01  & 96.05 & 0.01  & 1.01  & 0.3   & 6\% \\
		4     & 5     & 30\%  & 5     & 4     & 1\%   & 5\%   & 72.15 & 0.06  & 4.79  & 71.47 & 0.06  & 5.00  & 11.4  & 10\% \\
		5     & 5     & 90\%  & 5     & 4     & 1\%   & 5\%   & 76.55 & 0.05  & 4.68  & 75.52 & 0.05  & 5.00  & 20.4  & 10\% \\
		6     & 5     & 30\%  & 10    & 4     & 1\%   & 5\%   & 46.58 & 0.07  & 9.76  & 45.94 & 0.07  & 10.01 & 9.6   & 4\% \\
		7     & 5     & 90\%  & 10    & 4     & 1\%   & 5\%   & 55.37 & 0.06  & 9.46  & 53.97 & 0.06  & 10.01 & 22.1  & 10\% \\
		8     & 5     & 30\%  & 1     & 12    & 20\%  & 1\%   & 101.43 & 0.00  & 0.26  & 101.43 & 0.00  & 0.14  & 0.0   & 9\% \\
		9     & 5     & 30\%  & 1     & 12    & 1\%   & 1\%   & 99.33 & 0.01  & 0.81  & 99.24 & 0.01  & 1.01  & 11.2  & 0\% \\
		10    & 5     & 90\%  & 1     & 12    & 1\%   & 5\%   & 95.58 & 0.01  & 0.98  & 95.50 & 0.01  & 1.01  & 15.9  & 3\% \\
		11    & 5     & 30\%  & 5     & 12    & 1\%   & 5\%   & 70.49 & 0.06  & 4.67  & 69.27 & 0.06  & 5.00  & 20.7  & 11\% \\
		12    & 5     & 90\%  & 5     & 12    & 1\%   & 5\%   & 74.57 & 0.05  & 4.58  & 73.02 & 0.05  & 5.00  & 31.1  & 2\% \\
		13    & 5     & 30\%  & 10    & 12    & 1\%   & 5\%   & 43.76 & 0.07  & 9.68  & 42.78 & 0.06  & 10.00 & 15.0  & 8\% \\
		14    & 5     & 90\%  & 10    & 12    & 1\%   & 5\%   & 52.01 & 0.06  & 9.34  & 50.01 & 0.06  & 10.01 & 32.6  & 9\% \\
		\Xhline{0.01\arrayrulewidth}
		15    & 10    & 30\%  & 1     & 4     & 20\%  & 1\%   & 104.76 & 0.00  & 0.35  & 104.76 & 0.00  & 0.35  & 0.0   & 3\% \\
		16    & 10    & 30\%  & 1     & 4     & 1\%   & 1\%   & 99.61 & 0.01  & 0.89  & 99.59 & 0.01  & 0.98  & 1.6   & 2\% \\
		17    & 10    & 90\%  & 1     & 4     & 1\%   & 5\%   & 96.01 & 0.01  & 0.94  & 96.01 & 0.01  & 1.01  & 0.7   & 7\% \\
		18    & 10    & 30\%  & 5     & 4     & 1\%   & 5\%   & 65.27 & 0.07  & 4.85  & 64.72 & 0.07  & 4.98  & 7.7   & 1\% \\
		19    & 10    & 90\%  & 5     & 4     & 1\%   & 5\%   & 74.83 & 0.06  & 4.68  & 73.82 & 0.05  & 5.00  & 18.2  & 2\% \\
		20    & 10    & 30\%  & 10    & 4     & 1\%   & 5\%   & 36.82 & 0.07  & 9.88  & 36.53 & 0.07  & 9.98  & 4.3   & 3\% \\
		21    & 10    & 90\%  & 10    & 4     & 1\%   & 5\%   & 52.44 & 0.07  & 9.50  & 51.15 & 0.06  & 10.01 & 19.3  & 2\% \\
		22    & 10    & 30\%  & 1     & 12    & 20\%  & 1\%   & 101.61 & 0.00  & 0.10  & 101.61 & 0.00  & 0.09  & 0.0   & 8\% \\
		23    & 10    & 30\%  & 1     & 12    & 1\%   & 1\%   & 99.65 & 0.01  & 0.76  & 99.63 & 0.01  & 0.86  & 2.0   & 6\% \\
		24    & 10    & 90\%  & 1     & 12    & 1\%   & 5\%   & 96.12 & 0.01  & 0.86  & 96.02 & 0.01  & 1.01  & 14.5  & 8\% \\
		25    & 10    & 30\%  & 5     & 12    & 1\%   & 5\%   & 65.77 & 0.07  & 4.80  & 64.87 & 0.07  & 4.96  & 12.8  & 5\% \\
		26    & 10    & 90\%  & 5     & 12    & 1\%   & 5\%   & 75.33 & 0.06  & 4.55  & 73.85 & 0.05  & 5.00  & 26.3  & 4\% \\
		27    & 10    & 30\%  & 10    & 12    & 1\%   & 5\%   & 37.13 & 0.07  & 9.84  & 36.68 & 0.07  & 9.98  & 6.7   & 12\% \\
		28    & 10    & 90\%  & 10    & 12    & 1\%   & 5\%   & 52.93 & 0.07  & 9.36  & 51.25 & 0.06  & 10.01 & 24.9  & 1\% \\
		\Xhline{0.01\arrayrulewidth}
		29    & 20    & 30\%  & 1     & 4     & 20\%  & 1\%   & 104.75 & 0.00  & 0.26  & 104.75 & 0.00  & 0.26  & 0.0   & 5\% \\
		30    & 20    & 30\%  & 1     & 4     & 1\%   & 1\%   & 99.24 & 0.02  & 0.82  & 99.23 & 0.02  & 0.82  & 0.9   & 5\% \\
		31    & 20    & 90\%  & 1     & 4     & 1\%   & 5\%   & 95.93 & 0.01  & 1.00  & 95.93 & 0.01  & 1.01  & 1.0   & 2\% \\
		32    & 20    & 30\%  & 5     & 4     & 1\%   & 5\%   & 56.70 & 0.08  & 4.88  & 56.38 & 0.08  & 4.96  & 4.2   & 2\% \\
		33    & 20    & 90\%  & 5     & 4     & 1\%   & 5\%   & 72.77 & 0.06  & 4.76  & 72.00 & 0.06  & 5.00  & 12.8  & 2\% \\
		34    & 20    & 30\%  & 10    & 4     & 1\%   & 5\%   & 27.62 & 0.06  & 9.95  & 27.50 & 0.06  & 9.99  & 2.0   & 4\% \\
		35    & 20    & 90\%  & 10    & 4     & 1\%   & 5\%   & 49.60 & 0.07  & 9.59  & 48.51 & 0.07  & 10.01 & 15.8  & 1\% \\
		36    & 20    & 30\%  & 1     & 12    & 20\%  & 1\%   & 101.61 & 0.00  & 0.08  & 101.61 & 0.01  & 0.08  & 0.0   & 7\% \\
		37    & 20    & 30\%  & 1     & 12    & 1\%   & 1\%   & 99.34 & 0.02  & 0.56  & 99.30 & 0.02  & 0.68  & 2.5   & 5\% \\
		38    & 20    & 90\%  & 1     & 12    & 1\%   & 5\%   & 96.14 & 0.01  & 0.95  & 95.94 & 0.01  & 1.01  & 22.6  & 4\% \\
		39    & 20    & 30\%  & 5     & 12    & 1\%   & 5\%   & 57.06 & 0.08  & 4.81  & 56.53 & 0.08  & 4.93  & 6.9   & 2\% \\
		40    & 20    & 90\%  & 5     & 12    & 1\%   & 5\%   & 73.48 & 0.06  & 4.56  & 72.06 & 0.06  & 5.00  & 23.3  & 3\% \\
		41    & 20    & 30\%  & 10    & 12    & 1\%   & 5\%   & 27.86 & 0.06  & 9.92  & 27.65 & 0.06  & 9.98  & 3.6   & 6\% \\
		42    & 20    & 90\%  & 10    & 12    & 1\%   & 5\%   & 49.99 & 0.07  & 9.49  & 48.62 & 0.07  & 10.01 & 19.7  & 6\% \\
		\Xhline{0.01\arrayrulewidth}
		43    & 50    & 30\%  & 1     & 4     & 20\%  & 1\%   & 104.66 & 0.01  & 0.28  & 104.66 & 0.01  & 0.28  & 0.0   & 2\% \\
		44    & 50    & 30\%  & 1     & 4     & 1\%   & 1\%   & 98.18 & 0.03  & 0.88  & 98.16 & 0.03  & 0.89  & 0.8   & 4\% \\
		45    & 50    & 90\%  & 1     & 4     & 1\%   & 5\%   & 95.84 & 0.01  & 1.00  & 95.83 & 0.01  & 1.01  & 0.6   & 1\% \\
		46    & 50    & 30\%  & 5     & 4     & 1\%   & 5\%   & 43.95 & 0.08  & 4.96  & 43.88 & 0.08  & 4.98  & 0.9   & 0\% \\
		47    & 50    & 90\%  & 5     & 4     & 1\%   & 5\%   & 70.30 & 0.07  & 4.77  & 69.58 & 0.06  & 5.00  & 11.1  & 1\% \\
		48    & 50    & 30\%  & 10    & 4     & 1\%   & 5\%   & 18.30 & 0.04  & 9.99  & 18.27 & 0.04  & 10.00 & 0.6   & 4\% \\
		49    & 50    & 90\%  & 10    & 4     & 1\%   & 5\%   & 46.08 & 0.07  & 9.72  & 45.37 & 0.07  & 10.01 & 10.2  & 1\% \\
		50    & 50    & 30\%  & 1     & 12    & 20\%  & 1\%   & 101.61 & 0.00  & 0.08  & 101.61 & 0.00  & 0.08  & 0.0   & 3\% \\
		51    & 50    & 30\%  & 1     & 12    & 1\%   & 1\%   & 98.31 & 0.03  & 0.69  & 98.22 & 0.03  & 0.77  & 3.4   & 3\% \\
		52    & 50    & 90\%  & 1     & 12    & 1\%   & 5\%   & 96.06 & 0.01  & 0.95  & 95.85 & 0.01  & 1.01  & 19.9  & 3\% \\
		53    & 50    & 30\%  & 5     & 12    & 1\%   & 5\%   & 44.29 & 0.08  & 4.91  & 44.07 & 0.08  & 4.95  & 2.9   & 2\% \\
		54    & 50    & 90\%  & 5     & 12    & 1\%   & 5\%   & 70.94 & 0.07  & 4.60  & 69.66 & 0.06  & 5.00  & 19.4  & 4\% \\
		55    & 50    & 30\%  & 10    & 12    & 1\%   & 5\%   & 18.44 & 0.04  & 9.98  & 18.40 & 0.04  & 9.99  & 1.0   & 8\% \\
		56    & 50    & 90\%  & 10    & 12    & 1\%   & 5\%   & 46.53 & 0.07  & 9.59  & 45.46 & 0.07  & 10.01 & 15.1  & 2\% \\
	\end{tabular}%
	\caption{Counterpart of Tab.~\ref{tab:icpv5d} under local volatility settings.}
	\label{tab:icpv5dlocalvol}%
\end{table}%

\clearpage
\section{Conclusion}\label{sec:Conclusion}

In this manuscript, we presented a novel regression based approach for American-style option pricing. The formulism is to construct the ansatz to be regressed on using the 1D FD solution obtained from a backward PDE solver. Theoretical analysis indicates that the most effective error reduction can be obtained by choosing a 1D payoff process with highest correlation against the original one. Effective dimension reduction methodologies to arrive at the 1D auxiliary process have also been discussed throughout various model settings, ranging from local volatility to stochastic volatility (Heston model). Under this wide range of processes, numerical tests on Bermudan options and WIC indicate that FD-LSM produces accurate option prices benchmarking against FD-based full PDE approach in low dimensions ($d\leq 2$), and consistently overshadows classical LSM under high dimensions (up to 50). By stressing the market and contractual parameters, we showed that the combination of the ad-hoc FD-based ansatz and perturbational monomials as regressors, gives rise to stable results especially with exercise frontiers far away from the ATM region. At the same time, there has not been noticeable degradation in terms of the computational efficiency with the additional calculation in PDE solver and spline interpolation; the increase in runtime is primarily single-digit percentage, and only reaches low double digits in exceptional cases, see e.g. the last column of Tab.\ref{tab:icpv5dlocalvol}. This effective and efficient method can be implemented as a generic framework for pricing and risk management on structure derivative products with American-style exercise features. Going beyond option pricing, since LSM is also widely used to approximate the future exposure for products that cannot be valued analytically in the context of XVA, the new method is applicable there for the sake of accuracy and numerical stability. 

\section*{Declarations of Interest}
The author reports no conflicts of interest. The author alone is responsible for the content and writing of the paper. The views expressed in this article are those of the author alone and do not necessarily represent those of Citigroup. All errors are the author's responsibility.

\section*{Acknowledgments}
The author would like to thank his team members and fellow quants for stimulating discussions on the subject and is particularly grateful to Andrei N. Soklakov. 

\begin{center}
	\rule[1ex]{.25\textwidth}{.5pt}
\end{center}
\vspace*{2cm}

\renewcommand{\thesubsection}{A.\arabic{subsection}}
\section*{Appendix}
\addcontentsline{toc}{section}{\protect\numberline{}Appendix}
\subsection{Bermudan option}\label{sec:BermudaOption}
A Bermudan option is an American-style option that is written on the spot price of a basket $S_B(t)$, expressed as a weighted average of $d$ individual assets. Note that for mono-underlying structure, $S_B(t)$ is just simply $S(t)$, where we ignore the subscript. It is exercisable by the holder on $\pi(0)$.

The exercise payoff $Z(\omega,t_k)$ is given by 
\[
Z(\omega,t_k)=Z(S_B(t_k))=
\begin{cases}
\max(K-S_B(t_k),0)\text{, for put option;}\\
\max(S_B(t_k)-K,0)\text{, for call option.}
\end{cases}
\]

The state variables $X(\omega,t_k)$ can be reduced to $S_B(t_k)$ as the only explanatory independent variable, considering the functional dependence of the payoff $Z(\omega,t_k)$. Here $K$ is the strike price of the option.

Finally, $\tilde{V}(\omega,\tilde{\tau}(t))$ is given by
\[
\tilde{V}(\omega,\tilde{\tau}(t))=e^{-r(\tilde{\tau}(t)-t)}Z(S_B(\tilde{\tau}(t))).
\]

\subsection{Worst Of Issuer Callable Note}\label{sec:WorstOfIssuerCallableNote}

The worst of issuer callable note (WIC) is one of the most popular yield enhancement products in retail structured note market. On one hand, the investment performance of this product is capped by a equity dependent coupon rate; one the other hand, the issuer has the right to call the product with a return of principal, at his or her discretion, on exercise dates $\pi(0)$ on or before maturity $T$, i.e., the exercisability is on the issuer side. The performance is liked to a worst-of function on $d$ underlying assets $S_W(t)$.

There are a series of equity-dependent coupons announced on record dates in-line exactly with early exercise dates $\pi(0)$, and the coupon payments are in digital form with amount $c\cdot\Delta T$ subjected to a coupon barrier $B_c$:
\[
\mathcal{C}(S_W(t_k),t_k) = c\cdot\Delta T\cdot\Theta(S_W(t_k)-B_c),
\]
where $c$ is annualized coupon rate per unit notional and $\frac{1}{\Delta T}$ is the annual coupon frequency, as well as the exercise frequency; the Heaviside step function is denoted by $\Theta(x)$.

At maturity $T$ only, which is also the last coupon announcement date, there is a return of principal, as well as a short put option striking at $K$ with knocked-in barrier $B_P$ embedded: 
\[
\mathcal{P}(S_W(T),T) = -\Theta(B_P-S_W(T))\max(K-S_W(T), 0),
\]
where we have assumed the notional of the WIC is 1 for simplicity.

Different from a simple Bermudan option discussed previously, WIC is a much more complex product. There are several competing driving factors for the optimal stopping strategy to achieve a minimization of PV: 
\begin{itemize}
	\item On the upside, there are digital coupons whose amounts are linked to underlying asset performance. 
	\item On the downside, the chance of being knocked in as a put option is against early termination by issuer. 
	\item The interest rate is also playing a vital role as it impacts the PV of the returned principal.
\end{itemize}

It is expected to pose much more challenge on the numerical approach in terms of accuracy, due to the presence of discontinuity (e.g. the knocked-in and coupon digital barrier) in the payoff. 

In this problem, the worst of performance $S_W(t_k)$ is the explanatory variable $X(\omega,t_k)$, which is the only driver for both the coupon performance and value of the knock in put option. Mathematically, the exercise payoff and option value are given by 
\[
Z(\omega,t_k)=\mathcal{C}(S_W(t_k),t_k) + 1 +\delta_{t_k=T}\mathcal{P}(S_W(T),T)
\]
and 
\[
\tilde{V}(\omega,\tilde{\tau}(t)) = \sum_{t_i=\inf\pi(t)}^{\tilde{\tau}(t)}e^{-r (t_i-t)} \mathcal{C}(S_W(t_i),t_i)+1\cdot e^{-r (\tilde{\tau}(t)-t)} + \delta_{\tilde{\tau}(t)=T}\cdot e^{-r (T-t)}\mathcal{P}(S_W(T),T),
\]
respectively.

\subsection{Regression Approach in Monte Carlo framework} \label{sec:regressionleastsquare}

Translating objective function to be minimized in Eqn.\ref{eqn:objFunction} into the language within Monte Carlo framework, we have
\[
\frac{1}{N_R}\sum_{j=1}^{N_R}\left[Y^{(j)} - \sum_{r=1}^{R}\beta_{k,r}\phi^P_r(X^{(j)}(t_k))\right]^2,
\]
where the superscript $(j)$ indicates that the variable is under $j^{th}$ of total $N_R$ regression Monte Carlo paths. 

After some algebra, the Monte Carlo estimator for the solution of the minimization problem can be written as: 
\begin{equation}\label{eqn:solvebeta}
\hat{\vec{\beta}}^P_k=(\mathbb{X^P}^T_k\mathbb{X^P}_k)^{-1}\mathbb{X^P}^T_k\vec{Y},
\end{equation}

where the matrix $\mathbb{X^P}_k$ and vector $\vec{Y}$ are defined as:
\[
(\mathbb{X^P}_k)_{i,j}=\frac{1}{\sqrt{N_R}}\phi^P_i(X^{(j)}(t_k))
\]
and 
\[
(\vec{Y})_{j}=\frac{1}{\sqrt{N_R}}Y^{(j)},
\]
respectively.

\subsection{Local Regression Error in LSM}\label{sec:localRegressionErrorLSM}

To begin with, we analyze the MSE for LSM by writing down the exact projection coefficients as
\begin{equation}\label{eqn:objFunctionLSMBeta0}
\vec{\beta}^{P^{\text{LSM}}0}_k=\vec{\beta}^{P^{\text{LSM}}}_k(F_k)=(\E[\vec{\phi}^{P^\text{LSM}}_{R,k}\vec{\phi}^{P^\text{LSM}T}_{R,k}])^{-1}\E[\vec{\phi}^{P^\text{LSM}}_{R,k}F_k].
\end{equation}

Introducing $\Delta\vec{\beta}^{P^{\text{LSM}}}_k=\vec{\beta}^{P^{\text{LSM}}}_k(e^{-r\Delta T}V_{k+1})-\vec{\beta}^{P^{\text{LSM}}0}_k=\vec{\beta}^{P^{\text{LSM}}}_k(e^{-r\Delta T}V_{k+1}-F_k)$,
where $\vec{\beta}^{P^{\text{LSM}}}_k(\cdot)$ is obtained from Eqn.\ref{eqn:objFunctionLSMBeta} after re-labeling $P$ to $P^{\text{LSM}}$, the local error is derived as
\begin{equation}\label{eqn:lsmXiDerivation}
\begin{split}
&\E[(e^{-r\Delta T}V_{k+1}-F_k)^2]\\
=&\E[(\Delta\vec{\beta}^{P^{\text{LSM}}}_k\cdot\vec{\phi}^{P^{\text{LSM}}}_{R,k})^2]\\
=&\E[\text{tr}(\Delta\vec{\beta}^{P^{\text{LSM}}T}_k\vec{\phi}^{P^{\text{LSM}}}_{R,k}\Delta\vec{\beta}^{P^{\text{LSM}}T}_k\vec{\phi}^{P^{\text{LSM}}}_{R,k})]\\
=&\E[\text{tr}(\vec{\phi}^{P^{\text{LSM}}T}_{R,k}(e^{-r\Delta T}V_{k+1}-F_k)^2(\E[\vec{\phi}^{P^{\text{LSM}}}_{R,k}\vec{\phi}^{P^{\text{LSM}}T}_{R,k}])^{-1}\vec{\phi}^{P^{\text{LSM}}}_{R,k}\vec{\phi}^{P^{\text{LSM}}T}_{R,k}\\ \quad&(\E[\vec{\phi}^{P^{\text{LSM}}}_{R,k}\vec{\phi}^{P^{\text{LSM}}T}_{R,k}])^{-1}\vec{\phi}^{P^{\text{LSM}}}_{R,k})] \qquad\qquad\qquad\qquad\quad\text{  (Law of Iterated Expectation)}\\
=&\E[(e^{-r\Delta T}V_{k+1}-F_k)^2\text{tr}(\vec{\phi}^{P^{\text{LSM}}}_{R,k}\vec{\phi}^{P^{\text{LSM}}T}_{R,k}(\E[\vec{\phi}^{P^{\text{LSM}}}_{R,k}\vec{\phi}^{P^{\text{LSM}}T}_{R,k}])^{-1}\vec{\phi}^{P^{\text{LSM}}}_{R,k}\vec{\phi}^{P^{\text{LSM}}T}_{R,k}\\ 
\quad&(\E[\vec{\phi}^{P^{\text{LSM}}}_{R,k}\vec{\phi}^{P^{\text{LSM}}T}_{R,k}])^{-1})]\qquad\qquad\qquad\qquad\qquad\qquad\text{(Cyclic property of matrix trace)}\\
\leq&\E[(e^{-r\Delta T}V_{k+1}-F_k)^2]\cdot\E[\text{tr}(\vec{\phi}^{P^{\text{LSM}}}_{R,k}\vec{\phi}^{P^{\text{LSM}}T}_{R,k}(\E[\vec{\phi}^{P^{\text{LSM}}}_{R,k}\vec{\phi}^{P^{\text{LSM}}T}_{R,k}])^{-1}\vec{\phi}^{P^{\text{LSM}}}_{R,k}\vec{\phi}^{P^{\text{LSM}}T}_{R,k}\\ 
\quad&(\E[\vec{\phi}^{P^{\text{LSM}}}_{R,k}\vec{\phi}^{P^{\text{LSM}}T}_{R,k}])^{-1})]\qquad\qquad\qquad\qquad\qquad\qquad\text{(Cauchy-Schwarz inequality)}\\
\leq&\E[(e^{-r\Delta T}V_{k+1}-F_k)^2]\cdot\text{tr}(\E[\vec{\phi}^{P^{\text{LSM}}}_{R,k}\vec{\phi}^{P^{\text{LSM}}T}_{R,k}](\E[\vec{\phi}^{P^{\text{LSM}}}_{R,k}\vec{\phi}^{P^{\text{LSM}}T}_{R,k}])^{-1}\E[\vec{\phi}^{P^{\text{LSM}}}_{R,k}\vec{\phi}^{P^{\text{LSM}}T}_{R,k}]\\ 
\quad&(\E[\vec{\phi}^{P^{\text{LSM}}}_{R,k}\vec{\phi}^{P^{\text{LSM}}T}_{R,k}])^{-1})\qquad\qquad\qquad\qquad\qquad\qquad\text{ (Cauchy-Schwarz inequality)}\\
=&\E[\Var(e^{-r\Delta T}V_{k+1}|\F_{t_k})]\cdot\text{tr}({\bf{I}}_R)\\
=&\E[\Var(e^{-r\Delta T}V_{k+1}|\F_{t_k})]\cdot R,
\end{split}
\end{equation}
with ${\bf{I}}_R$ being $R\times R$ identity matrix. While the above derivation is similar to the proof of Theorem 1 in \cite{newey1997convergence}, we would emphasize specifically that the tightness of the first occurrence when applying Cauchy-Schwarz inequality is justified by $\vec{\phi}^{P^\text{LSM}}_{R,k}$ and $V_{k+1}$ being agnostic to each other. 

\subsection{Optimized Exercise Boundaries (Opt-EB)}\label{sec:OptimizedExerciseBoundaries}

As alternative independent check for the PV of American-style options, we introduce a brute force method within simulation framework. For ease of illustration but without loss of generality, we focus on Bermudan option only, although this method will work as expected given the payoff is monotonic.  

First of all, we define the variational stopping time function $\tilde{\tau}(\overrightarrow{EB})$, in light of Eqn.\ref{eqn:tauBermuda}, as
\[
\tilde{\tau}(\overrightarrow{EB}):=
\begin{cases}
\inf\{t_k|S_B(t_k)\geq EB_{t_k}\} \text{ for call option;} \\
\inf\{t_k|S_B(t_k)\leq EB_{t_k}\} \text{ for put option,}
\end{cases}
\]
which depends on $\overrightarrow{EB}=\{EB_{t_k}|1\leq k \leq M\}$ defining the exercise boundaries on all exercise dates.

For a given $\overrightarrow{EB}$, the European-style option PV can be obtained by standard Monte Carlo method with ease as $\tilde{PV}(\overrightarrow{EB})=\E[e^{-r\tilde{\tau}(\overrightarrow{EB})}Z(S_B(\tilde{\tau}(\overrightarrow{EB})))]$.

Then the Bermudan option PV under Opt-EB can be computed via external optimization:
\begin{equation}\label{eqn:optEb}
\begin{split}
PV_{Opt-EB} &= \E[\sup_{\tau}[e^{-r\tau}Z(S_B(\tau))]] \\
&= \sup_{\overrightarrow{EB}}\E[e^{-r\tilde{\tau}(\overrightarrow{EB})}Z(S_B(\tilde{\tau}(\overrightarrow{EB})))] \\
&= \sup_{\overrightarrow{EB}}\tilde{PV}(\overrightarrow{EB}). 
\end{split}
\end{equation}
In practice, the result is obtain by optimizing the exercise boundary levels $\overrightarrow{EB}$ to achieve a maximum $\tilde{PV}$ as much as possible. In our implementation, a multi-dimensional Simplex algorithm is used. 

This brute-force variational method is a very time consuming approach, which can take many (up to 100+) iterations before achieving convergent results. However, it would be serving as one of few available benchmarks, or at lease lower (upper) bound estimates for holder (issuer) exercise, especially when full PDE FD solution is unavailable under higher dimensions with $d>2$.

\subsection{Properties of iterative backward sequence: $x_{k}=\sqrt{a_{k} (x_{k+1}^2 + b)} + c_{k} x_{k+1}$}\label{sec:sequenceProperties}

Here we analyze the iterative backward sequence
\begin{equation}\label{eqn:sequenceDefinition}
x_{k}=\sqrt{a_{k}(x_{k+1}^2 + b)} + c_{k} x_{k+1}
\end{equation}
under the constraints \( a_k, c_k \in (0, 1) \) for all \( k \in [1, M] \), the initial condition \( x_M = 0 \). We are interested in the behavior of this sequence as \( b \to 0 \).

\subsubsection{Inductive Proof of Convergence under finite $M$}\label{sec:sequenceFiniteInduction}

We will prove by induction that for all \( k \in [1, M] \), \( x_k \sim O(\sqrt{b}) \) as \( b \to 0 \). We have $x_M = 0\sim O(\sqrt{b})$ a base case to start with. Assuming that \( x_{k+1} \sim O(\sqrt{b}) \) as \( b \to 0 \), it means that there exists a constant \( C_{k+1} > 0 \) such that \( x_{k+1}=C_{k+1} \sqrt{b} \) for sufficiently small \( b \).

By using the inductive hypothesis, $x_k$ can be derived as follows:
\begin{equation}
\begin{split}
x_{k}&=\sqrt{a_{k}(x_{k+1}^2 + b)} + c_{k} x_{k+1}\\
&=\sqrt{a_{k} (C^2_{k+1}b + b)} + c_{k} (C_{k+1} \sqrt{b})\\
&= \sqrt{b} \left( \sqrt{a_{k} (C_{k+1}^2 + 1)} + c_{k} C_{k+1} \right).
\end{split}
\end{equation}

Denoting \( C_k = \sqrt{a_{k}(C_{k+1}^2 + 1)} + c_{k} C_{k+1} \), we have \( x_k = C_k \sqrt{b} \). Since \( a_k, c_k \in (0, 1) \) and \( C_{k+1} \) is a constant, \( C_k \) is also a constant. It concludes
\begin{equation}\label{eqn:sequenceResultsInduction}
x_k \sim O(\sqrt{b})
\end{equation}
for all $k$ by induction, given $M$ is finite.

\subsubsection{Fixed Point Analysis in the limit as $M\to\infty$}\label{sec:sequenceFixedPointLimit}
The asymptotic behavior of $x_k$ in the limit as $k\to 1$ and $M\to\infty$ depends on $a_0=\lim\limits_{\substack{k\to 1 \\ M\to\infty}}a_k$ and $c_0=\lim\limits_{\substack{k\to 1 \\ M\to\infty}}c_k$. 

Applying the limits to both hand sides of Eqn.\ref{eqn:sequenceDefinition}, it gives
\begin{equation}\label{eqn:sequenceLimit}
\lim\limits_{\substack{k\to 1 \\ M\to\infty}}x_k=\sqrt{a_0((\lim\limits_{\substack{k\to 1 \\ M\to\infty}}x_{k+1})^2 + b)} + c_{0}\lim\limits_{\substack{k\to 1 \\ M\to\infty}}x_{k+1}
\end{equation}

Assuming that the sequence reaches a fixed point, i.e., $x_0=\lim\limits_{\substack{k\to 1 \\ M\to\infty}}x_k=\lim\limits_{\substack{k\to 1 \\ M\to\infty}}x_{k+1}$, we have:

\[x_0 = \sqrt{a_0(x_0^2 + b)} + c_0 x_0\].

Solving above for $x_0$, we obtain
\begin{equation}\label{eqn:fixedPointSolution}
x_0 = \sqrt{\frac{b}{\frac{(1 - c_0)^2}{a_0}- 1}}.
\end{equation}

For \(x\) to be a real number as a fixed point to exist and $b>0$, it requires
\begin{equation}\label{eqn:fixedPointCondition}
a_0<(1 - c_0)^2.
\end{equation}
Under this condition, we have $x_0\sim O(\sqrt{b})$.

\bibliographystyle{apalike-refs}
\bibliography{bibliography}

\end{document}